\documentclass{egpubl}

\usepackage[T1]{fontenc}
\usepackage{dfadobe}  

\usepackage{cite}  % comment out for biblatex with backend=biber
\BibtexOrBiblatex
\electronicVersion
\PrintedOrElectronic
\ifpdf \usepackage[pdftex]{graphicx} \pdfcompresslevel=9
\else \usepackage[dvips]{graphicx} \fi

\usepackage{egweblnk} 
\usepackage{braket}
\usepackage{algorithm}
\usepackage{algorithmic}
\usepackage{ascmac}
\usepackage{amssymb, amsmath}
\usepackage{multicol}
\usepackage{vwcol}
\usepackage{xcolor}
\usepackage{soul}

\newcommand{\green}[1]{\colorbox{green}{#1}}
\newcommand{\red}[1]{\textcolor{red}{#1}}

\renewcommand{\hl}[1]{#1}
\renewcommand{\red}[1]{#1}
\renewcommand{\green}[1]{#1}
\title[]%
      {Quantum Coin Method for Numerical Integration\vspace{-0.8cm}}     
\author[]{N. H. Shimada \quad T. Hachisuka \\ The University of Tokyo }

\begin{document}
%--------------------------- Teaser images --------------------------------------------
\teaser{
    \vspace{-1cm}
    \centering
    % \mbox{}
    % \includegraphics[width=0.185\linewidth]{Figures/Grad-1.png}
    % \includegraphics[width=0.185\linewidth]{Figures/Grad-2.png}
    % \includegraphics[width=0.185\linewidth]{Figures/Grad-3.png}
    % \includegraphics[width=0.185\linewidth]{Figures/Grad-4.png}
    % \includegraphics[width=0.185\linewidth]{Figures/Grad-5.png}
    % \includegraphics[width=0.032\linewidth]{Figures/Error-bar-1.png}
    % \includegraphics[width=0.386\linewidth]{Figures/Table--.png}
    % \includegraphics[width=0.185\linewidth]{Figures/Error-1.png}
    % \includegraphics[width=0.185\linewidth]{Figures/Error-2.png}
    % \includegraphics[width=0.185\linewidth]{Figures/Error-3.png}
    % \includegraphics[width=0.032\linewidth]{Figures/Error-bar-2.png}
    % \mbox{}
    \mbox{}
    \newcommand{\varwidth}{0.155}
    \includegraphics[width=\varwidth\linewidth]{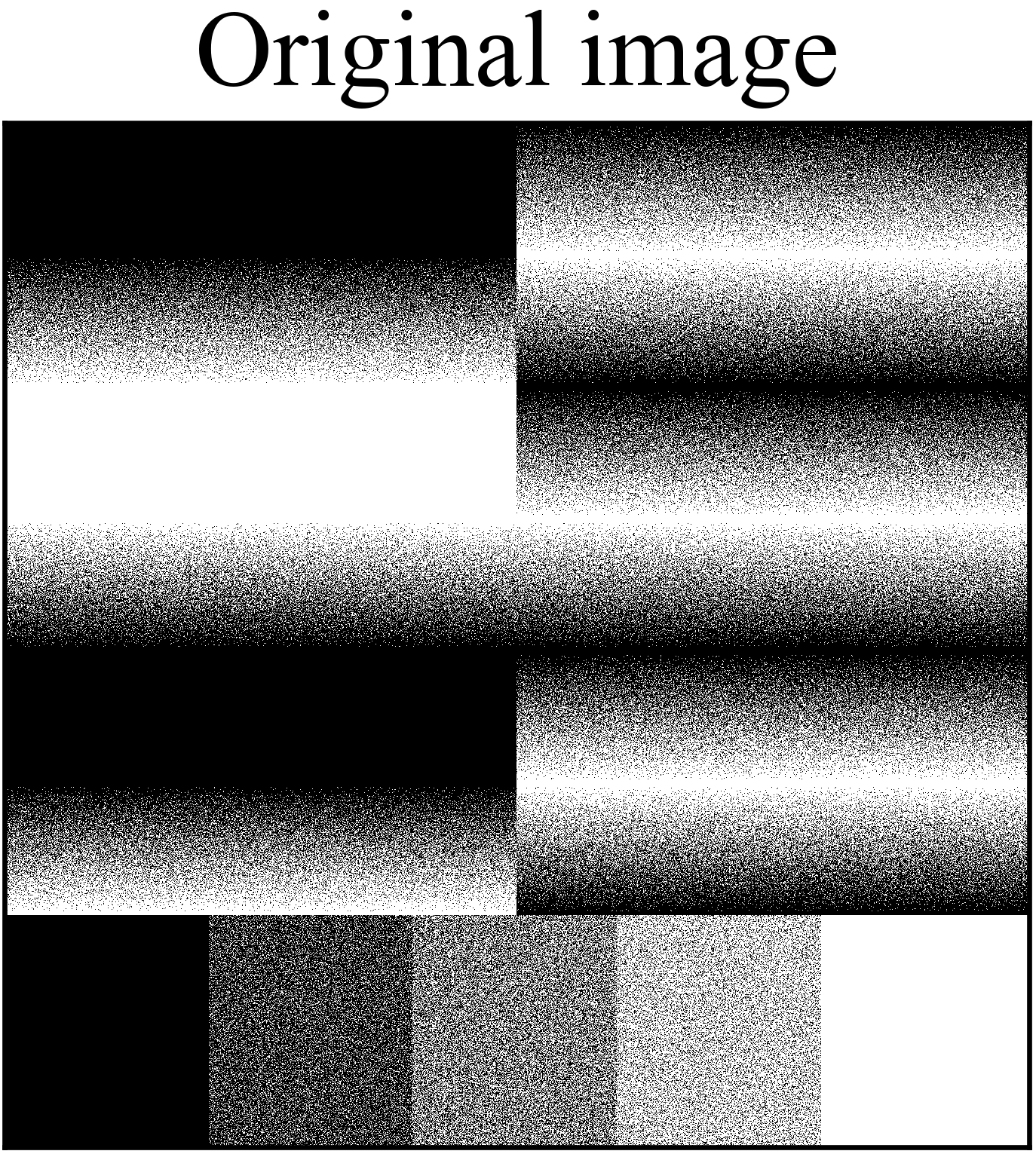}
    \includegraphics[width=\varwidth\linewidth]{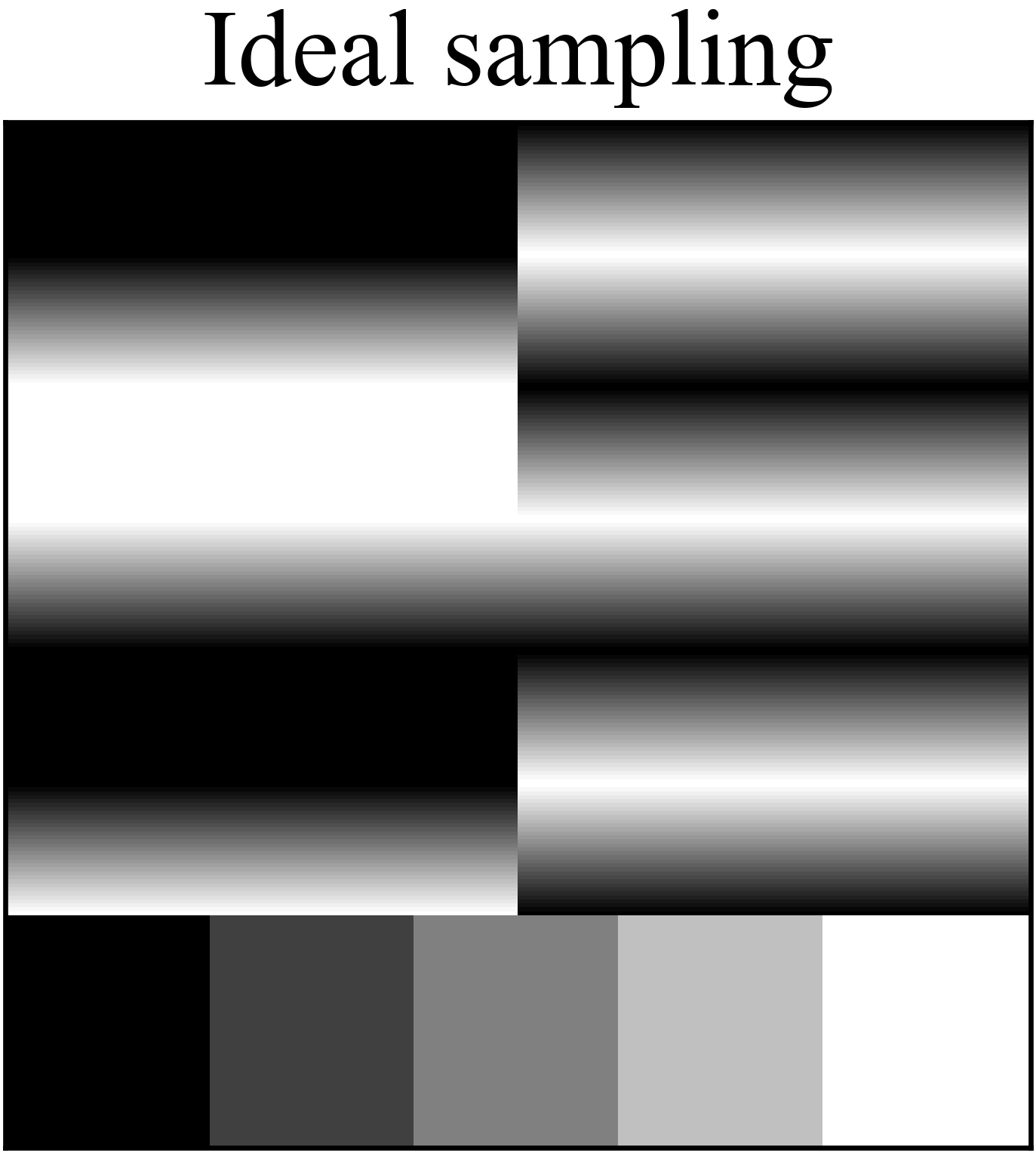}
    \includegraphics[width=\varwidth\linewidth]{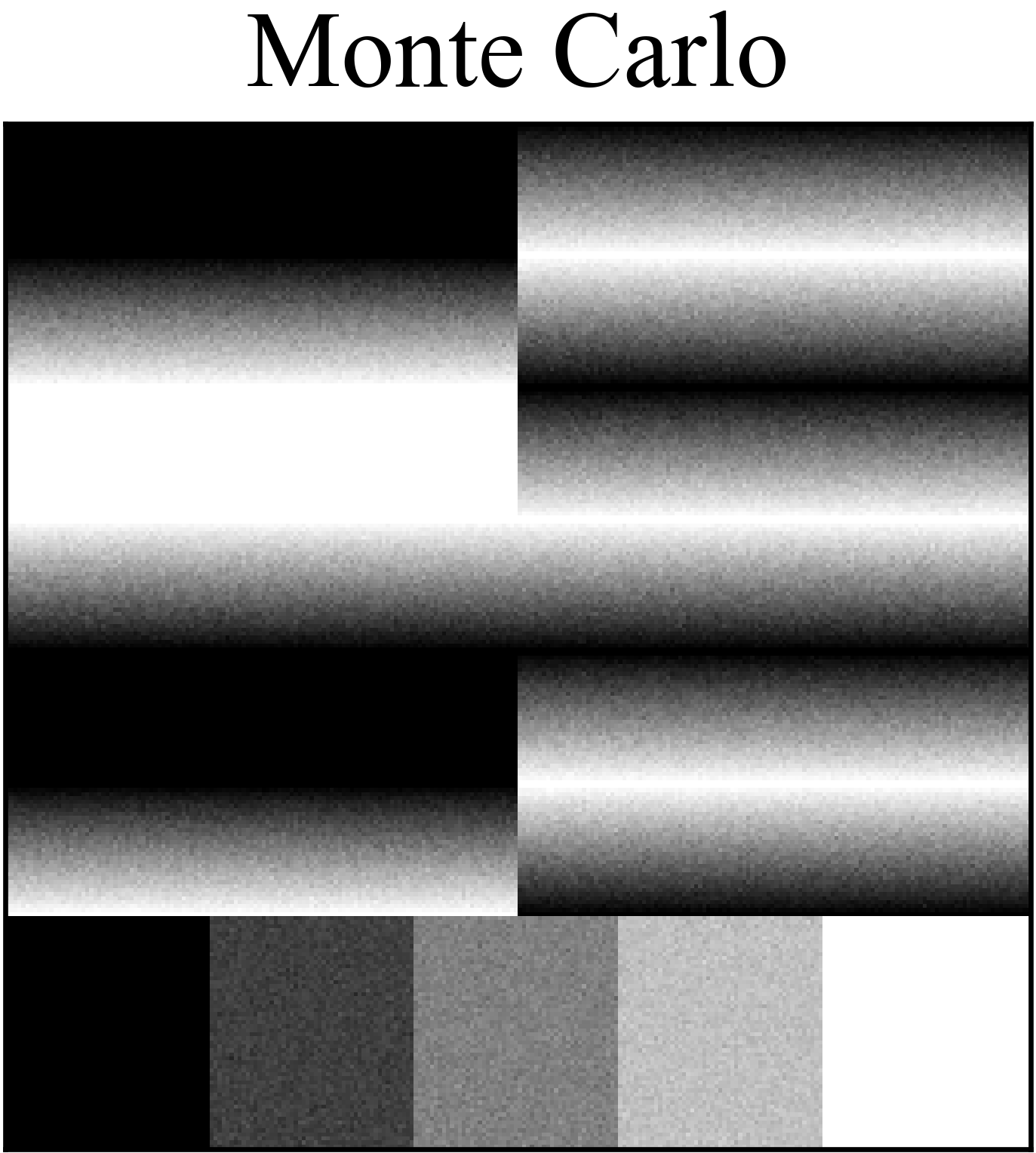}
    \includegraphics[width=\varwidth\linewidth]{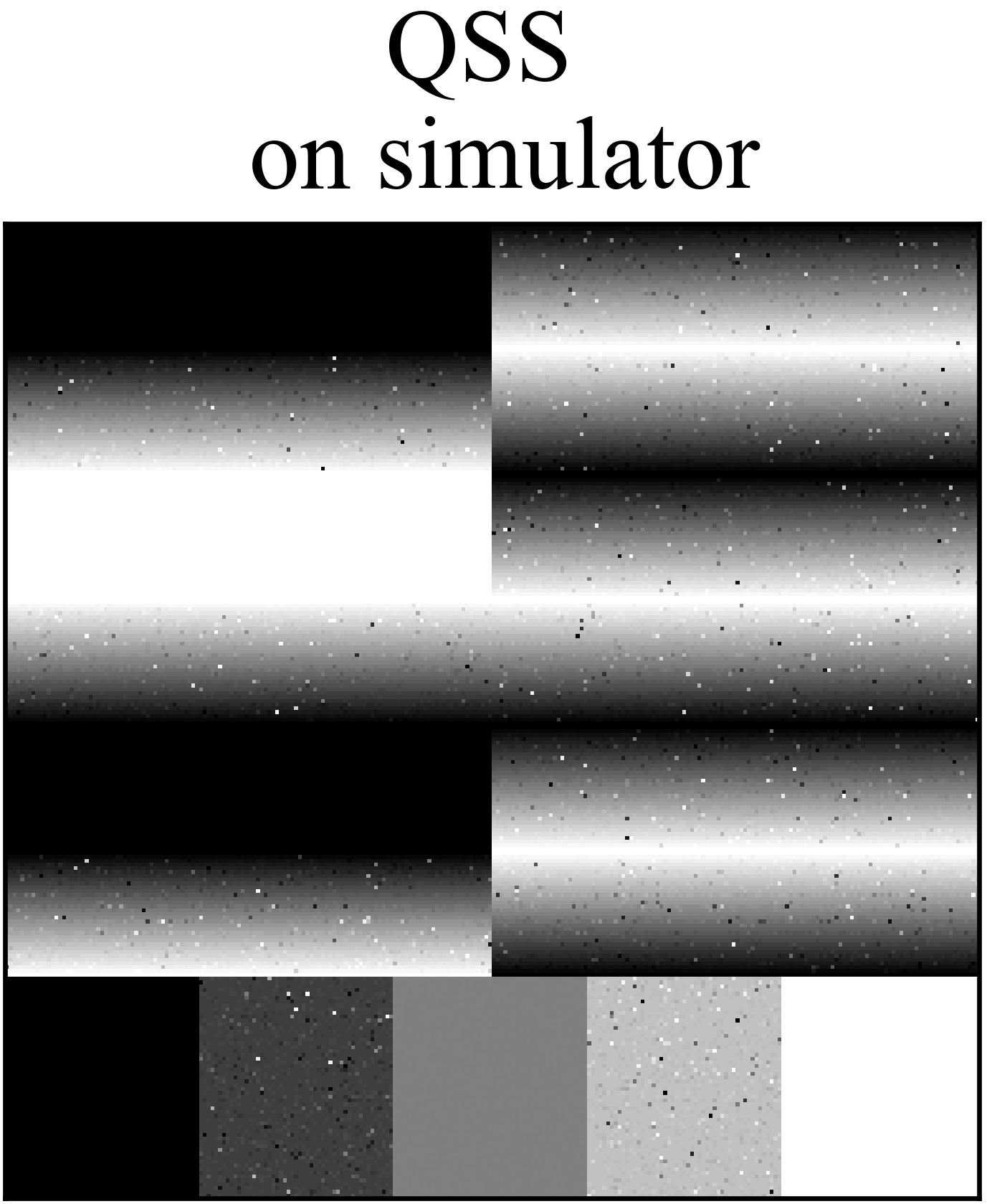}
    \includegraphics[width=\varwidth\linewidth]{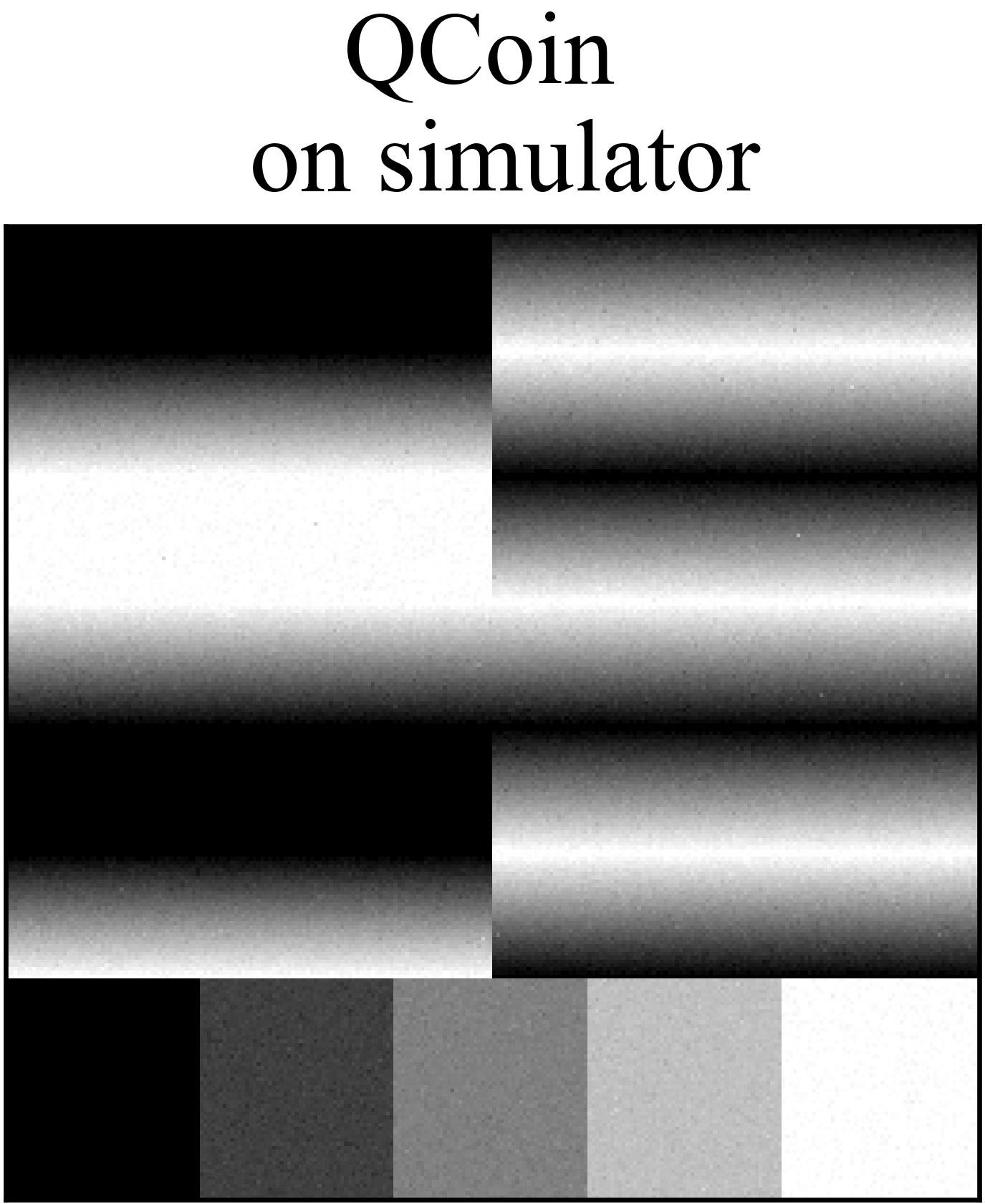}
    \includegraphics[width=\varwidth\linewidth]{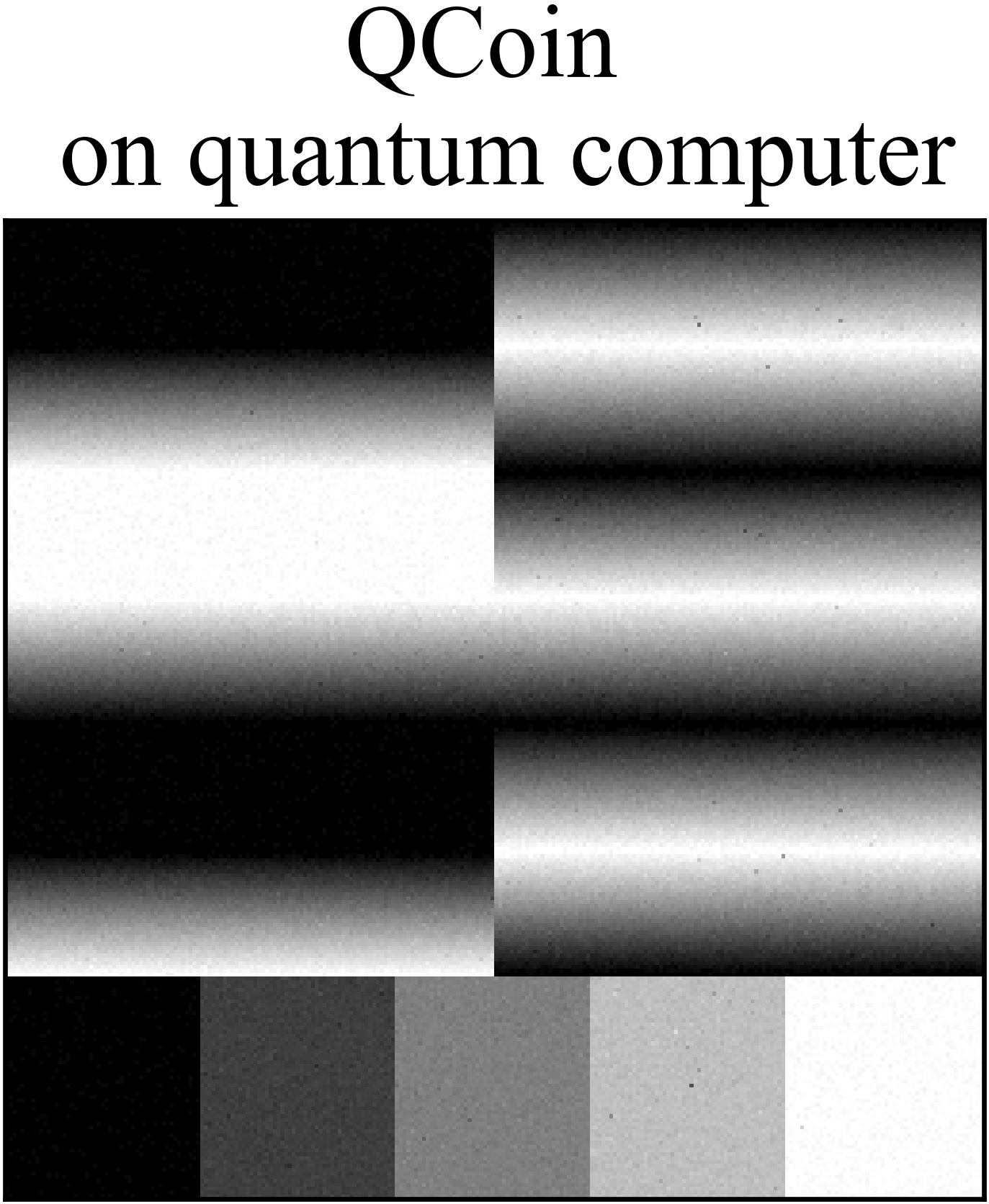}
    \includegraphics[width=0.027\linewidth]{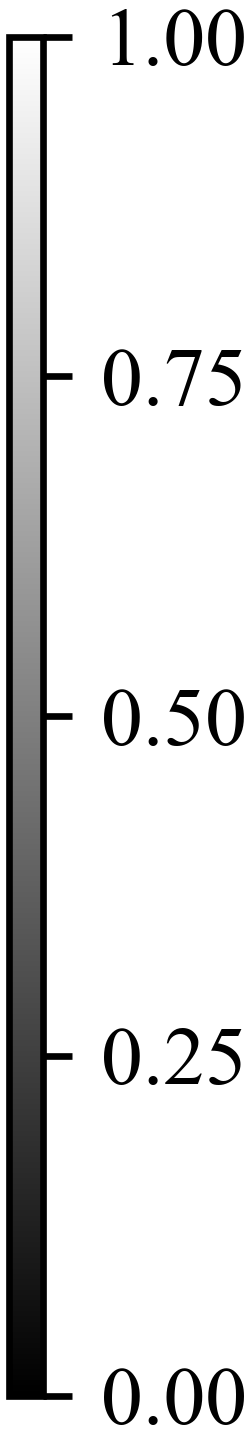}
    \includegraphics[width=0.327\linewidth]{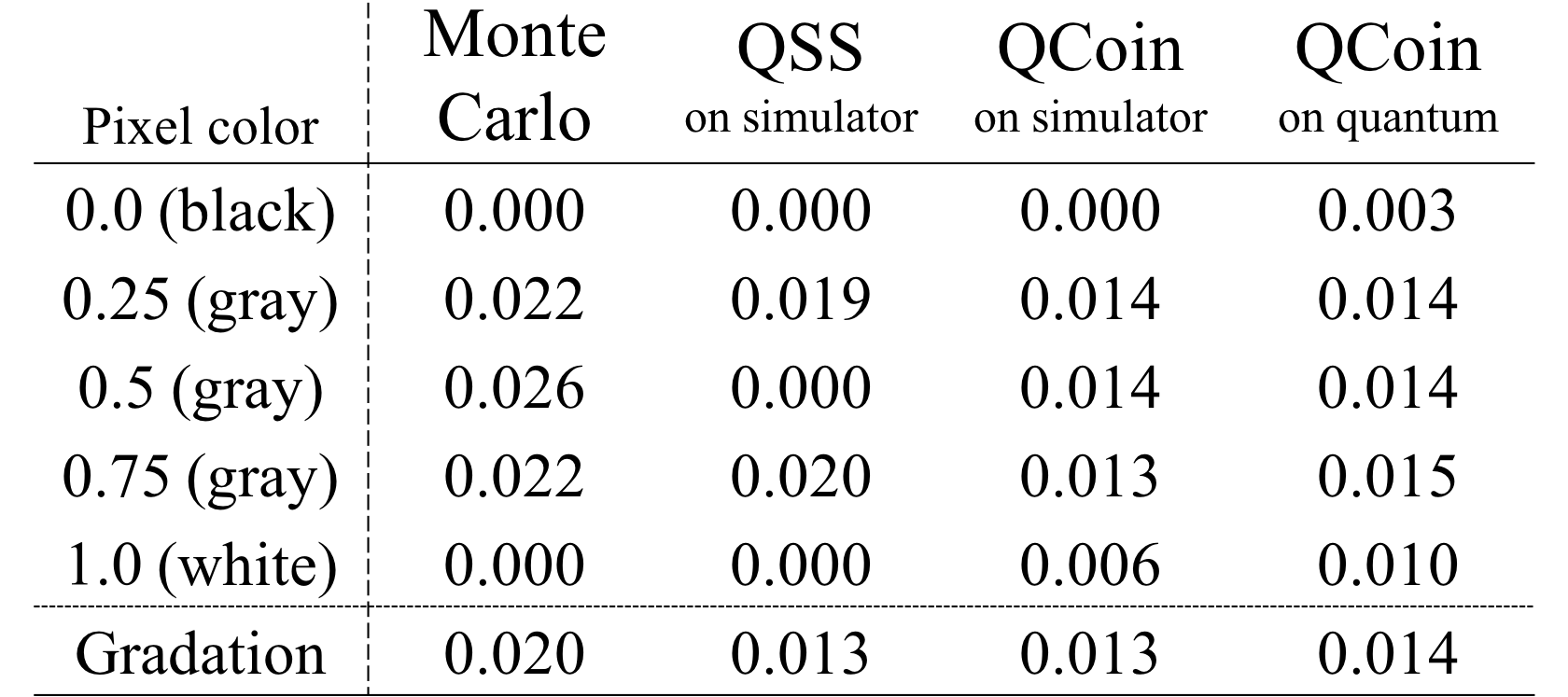}
    \includegraphics[width=\varwidth\linewidth]{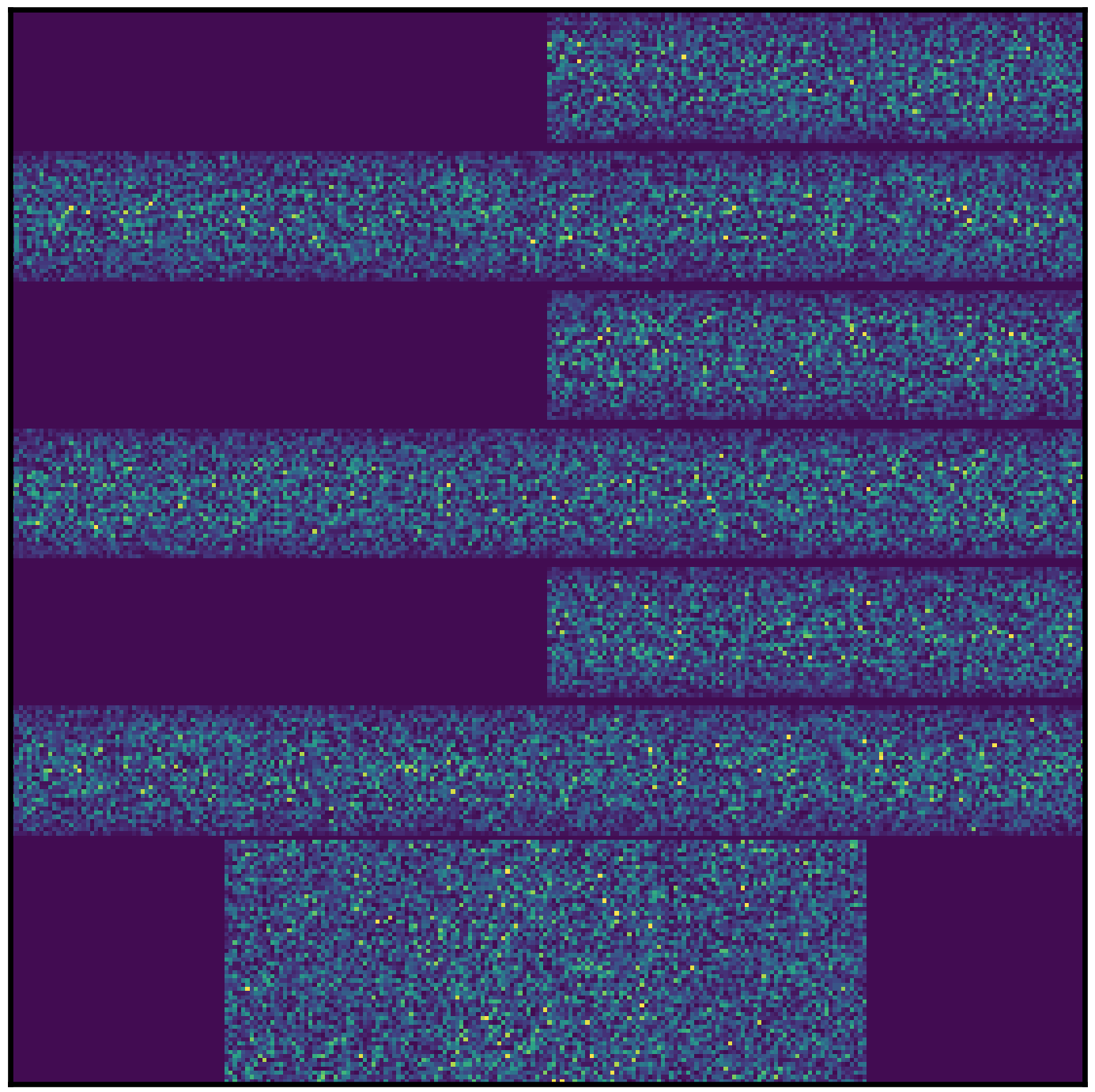}
    \includegraphics[width=\varwidth\linewidth]{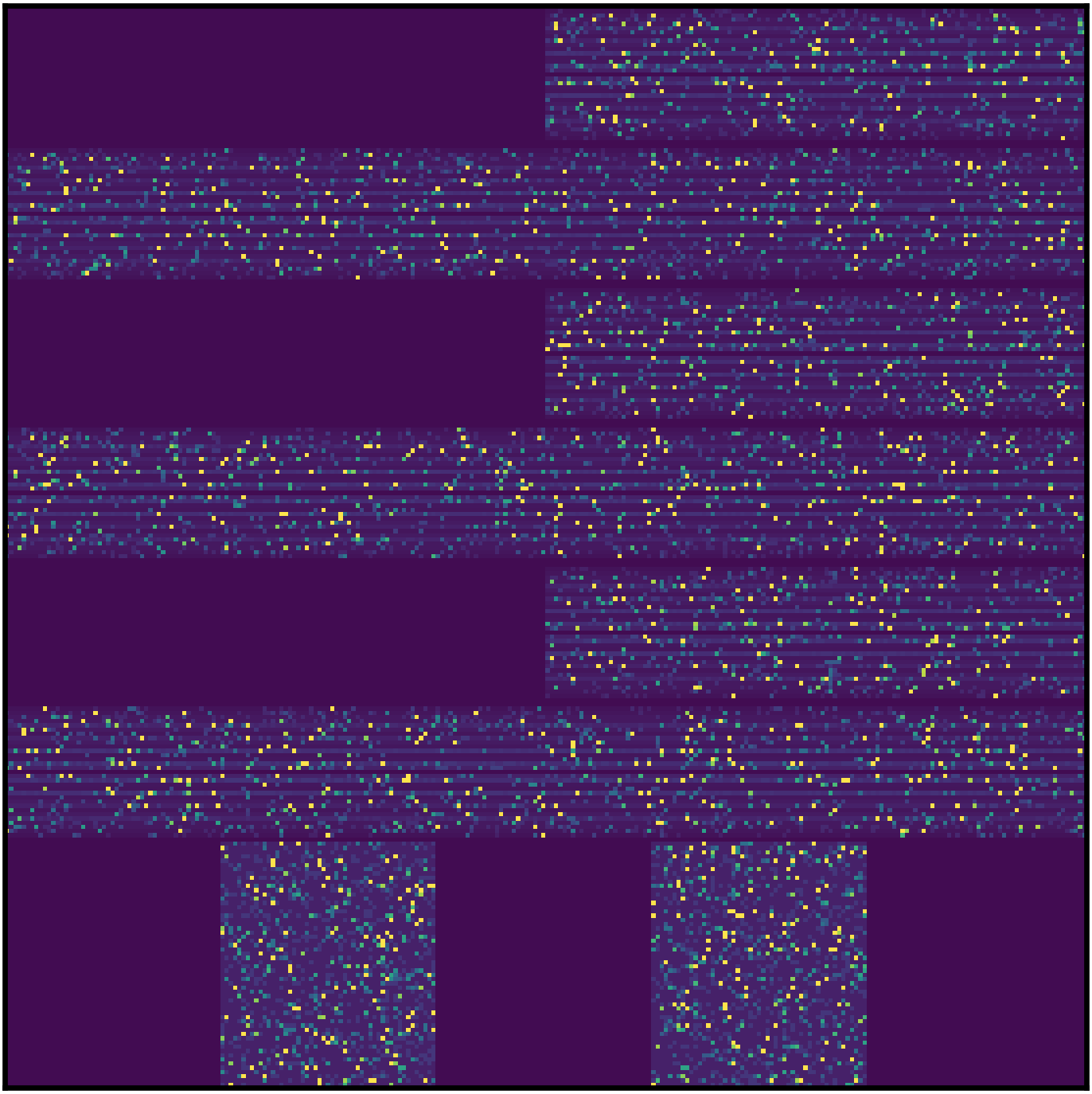}
    \includegraphics[width=\varwidth\linewidth]{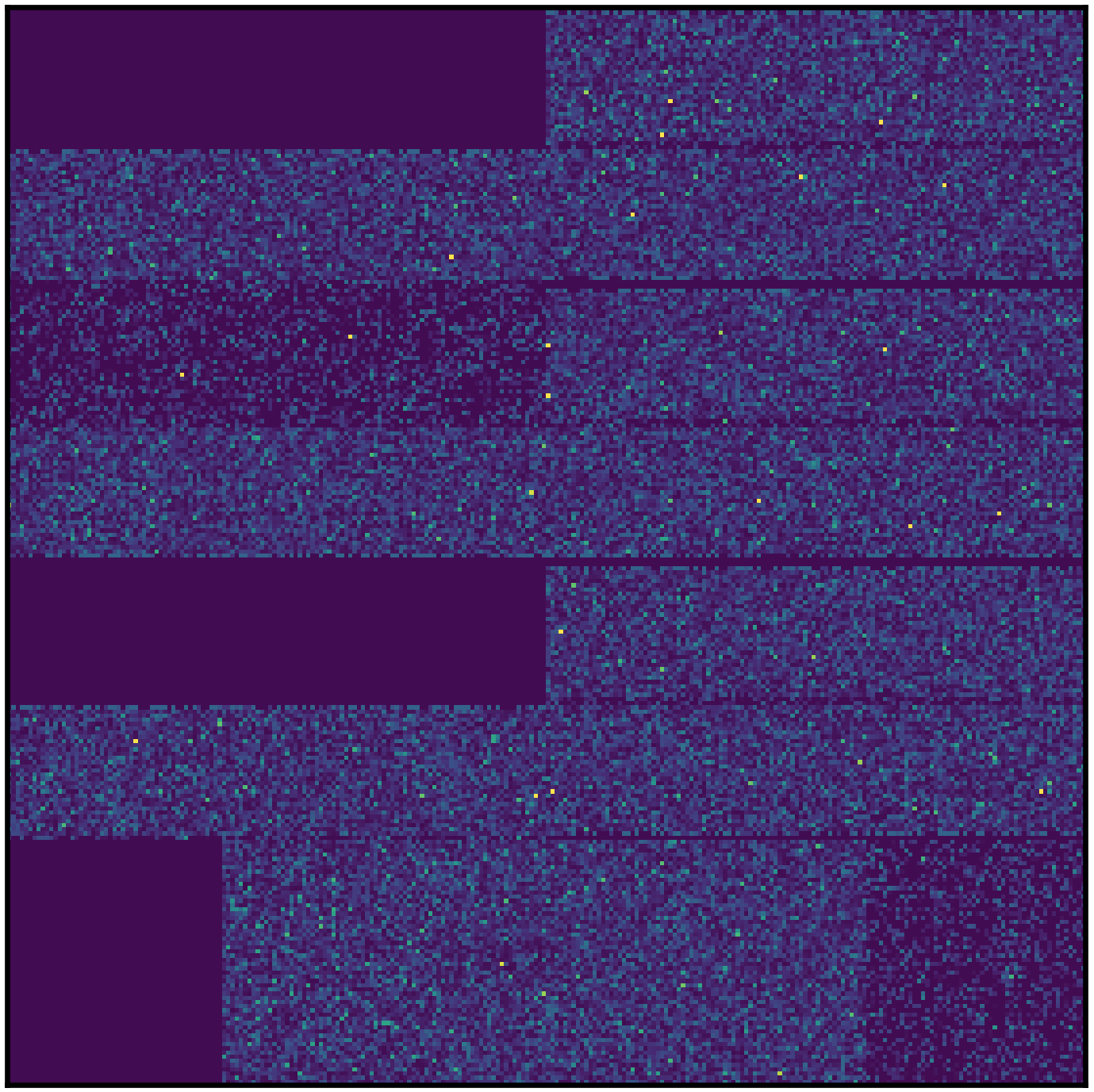}
    \includegraphics[width=\varwidth\linewidth]{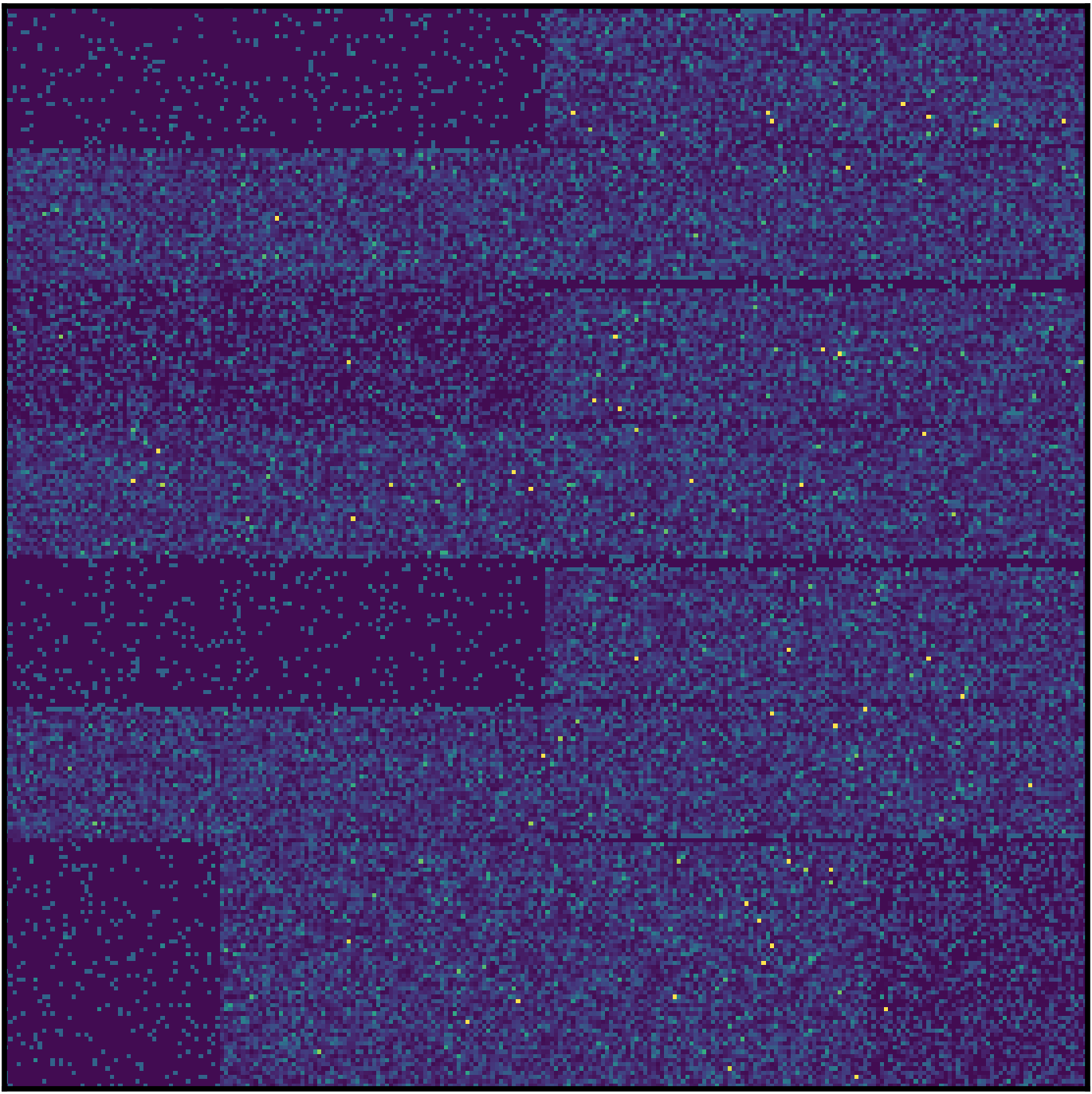}
    \includegraphics[width=0.027\linewidth]{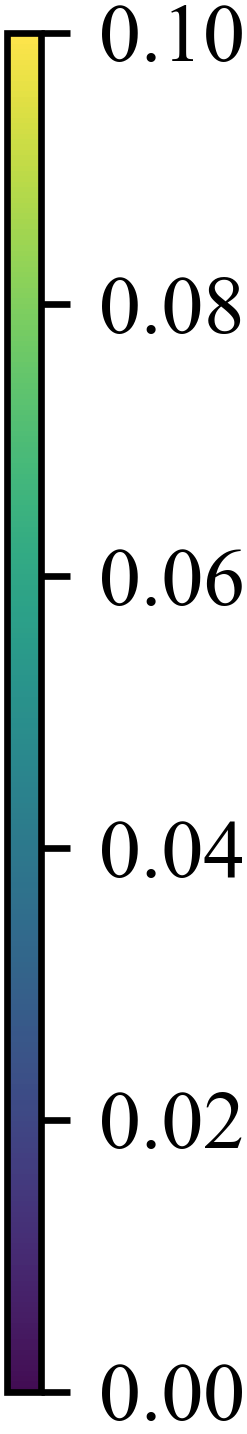}
    \mbox{}
    \vspace{-1.5EM}
    \caption{\red{Experiment with supersampling. We supersample $8\times8$ subpixels of "Original image". "Ideal sampling" shows the ground truth, which takes an average of $8\times8$ subpixels. Monte Carlo and our method (QCoin) take samples from subpixels to approximate this average for each pixel. The images show the results of numerical experiments using Monte Carlo, Quantum Supersampling~\cite{QSS}(QSS) on a noiseless simulator, QCoin on a noiseless simulator, and QCoin on an actual quantum computer with the equal sample counts of 240 queries (only QSS is done by 255 queries). The table shows mean absolute error of 5 colored rectangular regions (black, gray, and white) in the bottom of supersampling images and of gradation regions on the right half of the images. QCoin produces more accurate results than Monte Carlo does because of its accurate estimation using quantum computers. While QSS is comparable to QCoin, it works well only on a noiseless simulator as reported by Johnston~\cite{QSS}. QCoin, on the other hand, works well even on an actual quantum computer for the first time.}
    %Comparison between classical Monte Carlo, quantum supersampling (QSS)~\cite{QSS}, and our quantum coin method (QCoin). \red{QCoin is asymptotically faster than classical Monte Carlo as shown in Figure~\ref{fig:ex2}, and we show one comparison example with 240 queries here (2nd and 3rd figures from the left). Although the performance of QCoin is compatible to QSS on a simulator as shown in Figure~\ref{QCoin-real}, ours is far more robust than QSS on an actual quantum computer (4th and 5th figures from the left) (The number of queries in QSS is limited to 6 queries here due to limitation of the current architecture of quantum computers, and is enough to achieve no error theoretically in this case).}
    }
    \vspace{2EM}
\label{fig:teaser}
}

\maketitle
%-------------------------------- Abstract -----------------------------------------
\begin{abstract}
Light transport simulation in rendering is formulated as a numerical integration problem in each pixel, which is commonly estimated by Monte Carlo integration. Monte Carlo integration approximates an integral of a black-box function by taking the average of many evaluations (i.e., samples) of the function (integrand). For $N$ queries of the integrand, Monte Carlo integration achieves the estimation error of $O(1/\sqrt{N})$. Recently, Johnston~\shortcite{QSS} introduced quantum supersampling (QSS) into rendering as a numerical integration method that can run on quantum computers. QSS breaks the fundamental limitation of the $O(1/\sqrt{N})$ convergence rate of Monte Carlo integration and achieves the faster convergence rate of \hl{approximately} $O(1/N)$ which is the best possible bound of any quantum algorithms we know today~\cite{NayakWu}. We introduce yet another quantum numerical integration algorithm, quantum coin (QCoin)~\cite{QCoin}, and provide numerical experiments that are unprecedented in the fields of both quantum computing and rendering. We show that QCoin's convergence rate is equivalent to QSS's. We additionally show that QCoin is fundamentally more robust under the presence of noise in actual quantum computers due to its simpler quantum circuit and the use of fewer qubits. Considering various aspects of quantum computers, we discuss how QCoin can be a more practical alternative to QSS if we were to run light transport simulation in quantum computers in the future.
% \ccsdesc[100]{Methods and Applications~Monte Carlo Techniques}
% \keywords{Quantum Computing, Monte Carlo}
% \printccsdesc  
\end{abstract}

%---------------------------------- Body ------------------------------------------
\section{Introduction}
The use of quantum computers for computer graphics is a fascinating idea and potentially leads to a whole new field of research. Lanzagorta and Uhlmann~\cite{QuantumRendering} mentioned this idea for the first time and suggested many interesting directions for further research. Their main focus is on Grover's database search algorithm~\cite{Grover}, and they showed how its application could lead to fundamentally more efficient algorithms than those on classical computers for various tasks in rendering, such as rasterization, ray casting, and radiosity. Since  Lanzagorta and Uhlmann, however, there has been little effort put into this direction, mostly due to the limited availability of actual quantum computers at that time.

Recently, Johnston~\shortcite{QSS} introduced a quantum algorithm called \textit{Quantum SuperSampling} (QSS) into computer graphics. Johnston proposed to use this algorithm to perform supersampling of sub-pixels in rendering. This problem is essentially a numerical integration problem in each pixel, which is commonly done by Monte Carlo integration on \hl{classical} computers. Johnston showed that the performance of this quantum algorithm is fundamentally better than classical Monte Carlo integration in terms of time complexity. On the other hand, his experiments on an actual quantum computer are not as successful as the simulated results due to the presence of noise in quantum computers. Since noise is essentially unavoidable in the current architecture of quantum computers, this issue restricts the use of QSS in practice.

We introduce yet another quantum algorithm for numerical integration which runs well also on \textit{actual} quantum computers; the Quantum Coin method (QCoin). We show that the performance of QCoin is equivalent to QSS both theoretically and numerically, including its convergence rate. We discuss the difference between two algorithms in terms of their implementations on a quantum computer. Unlike QSS, QCoin can be regarded as a hybrid of quantum-classical algorithm~\cite{VQE}. Being a hybrid algorithm, we show how \hl{QCoin} is much more practical than QSS in the presence of noise and the various restrictions on actual quantum computers. We tested our QCoin on a real quantum computer and confirmed that QCoin already shows better performance than classical Monte Carlo integration. 
\red{Figure~\ref{fig:teaser} shows one experiment where we compared Monte Carlo and QCoin with the equal sample counts. QCoin achieves more accurate estimation both on a simulator and an actual quantum computer.}
%Thanks to its asymptotic improvement, QCoin achieves more accurate estimation at each pixel, which is demonstrated on both a simulator and an actual quantum computer unlike QSS.}
We also discuss several open problems for running rendering tasks on quantum computers in the future. 
\vspace{2EM}
\section{Background}
Before diving into the details of our method, we first summarize some basic concepts of quantum computing for readers who are not familiar with them. While we do cover the basics that are necessary to understand our method in this paper, for some further details, readers might want to refer to a standard  textbook of quantum computing~\cite{NielsenChang} \hl{or an introductory textbook for readers with computer science background}~\cite{JohnstonBook}.

\paragraph*{Single-qubit and superposition.}
\red{On a classical computer, all the information is stored as a set of \textit{bits} where each bit represents only a binary number $0$ or $1$. We represent a state of a bit having 0 as $\ket0$ and 1 as $\ket1$. The notation of $\ket{}$ is called "\textit{bra-ket}", which is commonly used in the field of quantum computing. On a quantum computer, a single \textit{qubit} can represent a \textit{superposition} of both $\ket 0$ and $\ket 1$. For example, we can represent a superposition state, supposing the real number $a \in [-1, 1]$ as an \textit{amplitude} of $\ket 0$:}
\begin{eqnarray}
    \label{eqn:single-qubit}
    \ket \psi = a\ket0 + \sqrt{1-a^2}\ket1.
\end{eqnarray}
\red{If we \textit{measure} (read out) a qubit, the state converges to either side of $\ket 0$ or $\ket 1$. That is, the only information we can get is either $0$ or $1$ as in a classical case. This process is \textit{probabilistic} and the probability is given by a squared value of its amplitude. For example, in the case of Equation~\ref{eqn:single-qubit}, the measurement of $\ket \psi$ returns $\ket0$ with the probability $a^2$ and $\ket1$ with the probability $1-a^2$.}

%----------------------------
\vspace{1EM}
\paragraph*{Quantum logic gates.}
Just like logic gates for bits on classical computers, there are several known \textit{quantum logic gates} that are used to manipulate qubits. We summarize some of them here.

\setlength{\columnsep}{0.15cm}
\begin{multicols}{2}
%\begin{vwcol}[width={0.45,0.45}]
\begin{itembox}[l]{Identity gate $\hat I$}
    \vspace{-1.5EM}
    \begin{eqnarray}
        \hat I \ket0 &=& \ket0 \nonumber{} \\
        \hat I \ket1 &=& \ket1 \nonumber{}
    \end{eqnarray}
\end{itembox}

\begin{itembox}[l]{Hadamard gate $\hat H$}
    \vspace{-1.5EM}
    \begin{eqnarray}
        \hat H \ket0 &=& \frac{\ket0+\ket1}{\sqrt2} \nonumber{} \\
        \hat H \ket1 &=& \frac{\ket0-\ket1}{\sqrt2} \nonumber{}
    \end{eqnarray}
    \vspace{-1.5EM}
\end{itembox}

\begin{itembox}[l]{Pauli $\hat X, \hat Z$ gates}
    \vspace{-1.5EM}
    \begin{eqnarray}
        &\hat X \ket0 = \ket 1, &\hat Z \ket0 = \ket 0 \nonumber{} \\
        &\hat X \ket1 = \ket 0, &\hat Z \ket1 = -\ket 1 \nonumber{}
    \end{eqnarray}
\end{itembox}

\begin{itembox}[l]{Rotation gate $\hat{U}_\theta$}
    \vspace{-1.5EM}
    \begin{align*}
    & U_\theta \ket0 = \cos \theta \ket0 + \sin \theta \ket1  \nonumber{} \\
    & U_\theta \ket1 = -\sin \theta \ket0 + \cos \theta \ket1 \nonumber{}
    \end{align*}
    \vspace{-0.5EM}
    \centering
    ($\theta$ is a rotation angle)
\end{itembox}
%\end{vwcol}
\end{multicols}

\paragraph*{Multi-qubits.}
We express a multi-qubit state by concatenating single-qubit states. For example, a two-qubits state whose qubits are both $\ket 0$ are expressed as $\ket0 \otimes \ket0$ or $\ket{00}$. The symbol $\otimes$ represents a tensor product which means the concatenation of qubits in this case. In the following, we omit the symbol $\otimes$ for simplicity when it is obvious. In general, a two-qubits state whose qubits are both superposition states as Equation~\ref{eqn:single-qubit} can be written as
\begin{flalign}
    \ket \psi_a &= a_0 \ket0 + a_1 \ket 1, \,\,
    \ket \psi_b = b_0 \ket0 + b_1 \ket1   \nonumber \\
    \rightarrow \,\, \ket\psi_a \otimes \ket\psi_b &= (a_0 \ket0 + a_1 \ket1) \otimes (b_0 \ket0 + b_1 \ket1) \nonumber \\
    &= a_0b_0 \ket{00} + a_0b_1 \ket{01} + a_1b_0 \ket{10} + a_1b_1 \ket{11}. \nonumber
\end{flalign}
Since this explicit binary notation quickly becomes tedious for many qubits, we use another notation $|i)$ for a decimal number $i$ in the binary representation $i_{n-1} \cdots i_1 i_0$ as 
%Suppose that $i$ (decimal) $=$ $i_{n-1} \cdots i_1 i_0$ (binary), $|i)$ is defined by
%
%
\begin{flalign}
    |i) \equiv \ket{i_{n-1}} \otimes \cdots \otimes \ket{i_{1}} \otimes \ket{i_0}. \label{decimal}
\end{flalign}
For example, in the case of 4 qubits, we write as 
\begin{flalign}
|0) = \ket{0000}, |1) = \ket{0001}, |2) = \ket{0010}, \cdots, |15) = \ket{1111}. \nonumber
\end{flalign}

%----------------------------
\vspace{1EM}
\paragraph*{Quantum operation as a tensor product.}
%\cTH{In this paragraph, explain how applying a logic gate can be expressed as a tensor product. Provide a simple example using the Hadamard gate.}

In quantum computing, tensor products are also used to represent logic gate operations. For example, given the initial two-qubits state $\ket{0} \otimes \ket0$, the application of the Hadamard $\hat H$ gate for the first qubit and the Pauli $\hat Z$ gate for the second qubit can be written as 
\begin{flalign}
     (\hat H \otimes \hat Z) (\ket{0} \otimes \ket {0}) = \hat H \ket{0} \otimes \hat Z \ket {0}.
\end{flalign}
If we only operate the $\hat H$ gate for the first qubit and leave the second qubit unchanged, we can use the identity gate $\hat I$:
\begin{flalign}
    (\hat H \otimes \hat I) \ket{0} \otimes \ket {0}.
\end{flalign}
When we apply the same gate to all the qubits, we omit the $\otimes$ symbol and simplify the notation as
%as for the case of the state omitting $\otimes$ symbol, we define that a gate operates for all qubits consisting the state:
%
%
\begin{flalign}
    \hat H \ket{00} \equiv \hat H \ket{0} \otimes \hat H \ket{0}.
\end{flalign}
This notation is also adopted in the case of the decimal representation in Equation~\ref{decimal}. \hl{For the 4 qubits case,}
%the operation of the $\hat H$ gate to %the 4 qubits $|0)$ state is written as
%
%
\begin{flalign}
    \hat H |0) &\equiv \hat H \ket{0000} =  \hat H \ket0 \otimes \hat H \ket0 \otimes \hat H \ket0 \otimes \hat H \ket0 \nonumber \\
    &= \frac{\ket0 + \ket 1}{\sqrt2} \otimes \frac{\ket0 + \ket 1}{\sqrt2} \otimes \frac{\ket0 + \ket 1}{\sqrt2} \otimes \frac{\ket0 + \ket 1}{\sqrt2} \nonumber \\
    &= \frac{1}{\sqrt{2^4}} \left( \ket{0000} + \ket{0001} + \cdots \ket{1111} \right) = \frac{1}{\sqrt{2^4}} \sum_{i=0}^{15} |i). \label{gate-decimal} 
\end{flalign}
%
%
%Note that $\hat H |0)$ results in a superposition of all the $|i)$ states which is often used in quantum algorithms.

%----------------------------------
\vspace{1EM}
\paragraph*{Oracle gate.}
In quantum computing, it is usually assumed that we have a (quantum) circuit which converts the information of an input data for each specific application as a quantum state. This circuit is commonly called an \textit{oracle} gate. 
For example, in a database-search problem~\cite{Grover} with the input data $[a_0,a_1,a_2,a_3]$, the oracle $\hat O$ gate works as using a normalization constant $C$:
%to make sure that the probabilities sum to one:
%
%
\begin{eqnarray}
\hat O \ket{00} \rightarrow \frac{1}{C} \left( a_0 \ket{00} + a_1 \ket{01} + a_2 \ket{10} + a_3 \ket{11} \right)
\end{eqnarray}
which converts the input data into the amplitudes. The exact design of the quantum circuit of an oracle gate is usually omitted in the design each quantum algorithm, but the computational universality~\cite{deutsch1995universality} almost guarantees the existence of such an circuit.

In the context of ray tracing, $\hat O$ can be considered as a \textit{ray trace function}. Given the (sub-)pixel index (i.e., quantized pixel coordinate) $x$, a ray trace function $F(x)$ traces a ray from camera through the pixel $x$ and returns the light throughput along this ray, which can model many rendering algorithms such as path tracing~\cite{kajiya1986rendering}. In advanced algorithms like path tracing, $x$ is defined as a quantized high dimensional coordinate including the pixel coordinate.
%(e.g., a point in the primary sample space~\cite{kelemen2002simple}). 
For $M$ (sub-)pixels, a classical computer needs to repeat this process $M$ times by evaluating the ray trace function for all the (sub-)pixels. On a quantum computer, however, one can evaluate the ray trace function for all the (sub-)pixels in \textit{one shot}:
\begin{equation}
    \hat O |0) \rightarrow \frac{1}{C} \sum_{x=1}^{M} F(x) |x).
    \label{eqn:oracleRay}
\end{equation}
%
%
%It is theoretically possible to construct an arbitrary complex oracle gate by designing a quantum circuit using basic logic gates. The actual design of such a circuit is commonly left as future work in quantum computing since current architecture is not powerful enough. 
We assume the existence of such a ray tracing oracle gate, which is %theoretically possible due to the universality of quantum computation. It is 
equivalent to the fact Monte Carlo integration assumes that one can evaluate the integrand, without specifying how to evaluate.
%
%The design of quantum mean-estimation algorithms (QSS and QCoin) is, however, independent from the actual design of the oracle gate, similarly to the fact that the algorithmic design of Monte Carlo integration is independent from the integrand.

\vspace{1EM}
\paragraph*{Products using the bra-ket notation.}
Under the \textit{bra-ket} notation~\cite{NielsenChang}, a \textit{bra} vector $\bra{A}$ denotes as a complex transpose of \textit{ket} vector $\ket{A}$. For example, when $\ket{A} = \hat U \ket{00...0}$, we have
\begin{equation}
    \bra{A} =\ket{A}^\dagger = (\hat U \ket{00...0})^\dagger = \bra{00...0} \hat{U}^{-1}
\end{equation}
where $\hat U$ is a unitary matrix which represents a gate operation,
%(in quantum computation, all gate operations are an unitary matrix except for a measurement operation). 
and the complex transpose of a unitary matrix is an inverse matrix. One can think of $\bra{A}$ ($\ket{A}$) as a row-vector (column-vector) representation. Using this notation, inner product (scalar) can be expressed as $\langle A|B \rangle$ and outer product (matrix) can be expressed as $\ket{B}\bra{A}$.

\begin{figure}[t]
  \centering
  \includegraphics[width=\linewidth]{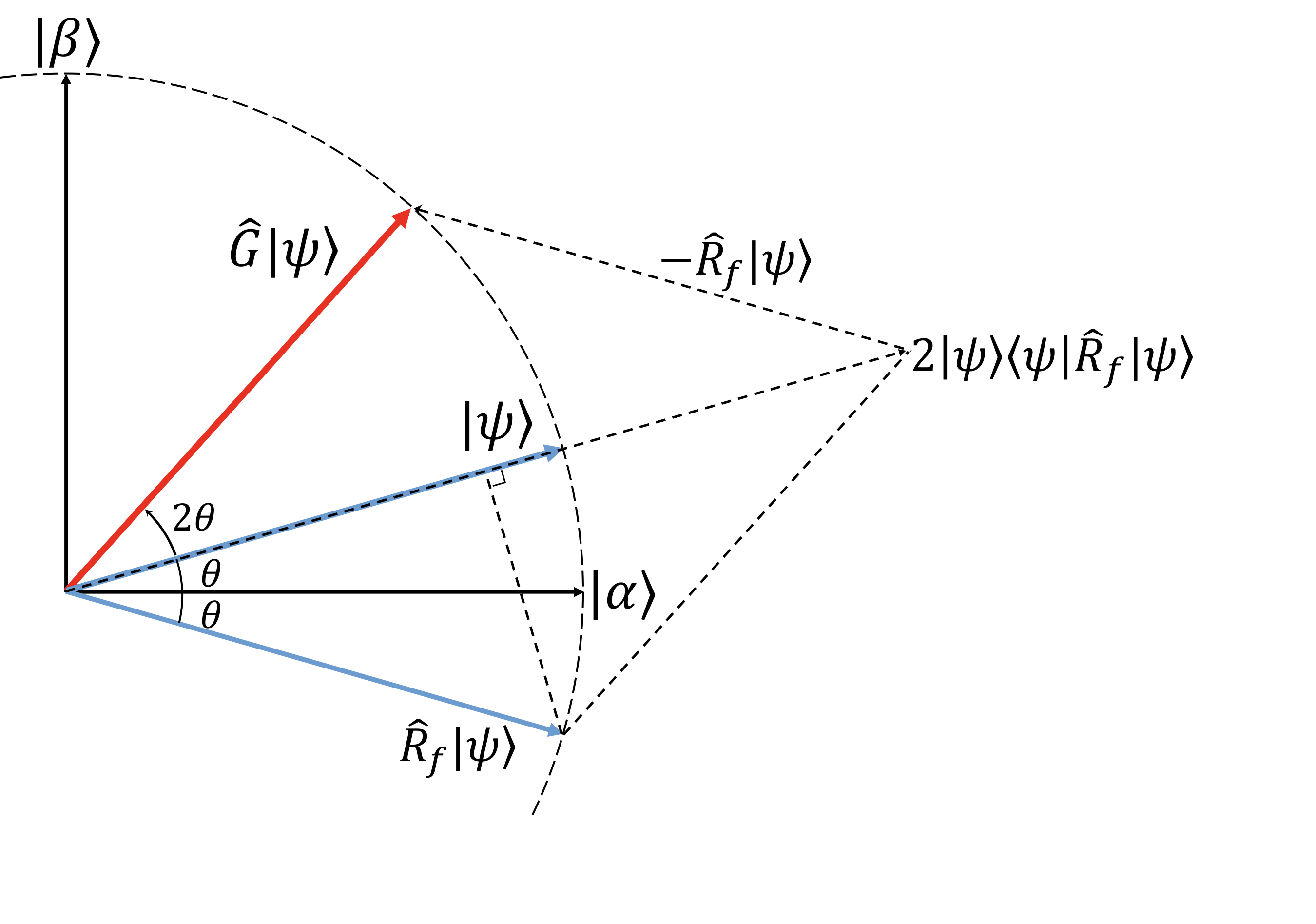}
  \vspace{-3EM}
  \caption{\label{AA} \red{Amplitude amplification from $\ket \psi$ to $\hat G \ket \psi$. The amplitude of $\ket \beta$ is amplified from $\cos \theta$ to $\cos 3\theta$ in the case that $\theta$ is a small value. The initial state $\ket \psi$ is sequentially changed as $\ket \psi \rightarrow \hat{R}_f \ket \psi \rightarrow 2\ket \psi \bra \psi \hat{R}_f \ket \psi \rightarrow \hat G \ket \psi$.} \vspace{-2EM}}
  %Firstly, we flip state against $\ket\alpha$ axis. Secondly, we project to the original vector and scale its magnitude by two. Finally, we subtract another vector to return on the circumference. The rhombus is composed of vectors, hence the angle between $\ket \psi$ and $\hat G \ket \psi$ is the same as one between $\ket \psi$ and $\hat R_f \ket \psi$; $2\theta$. As a result, the initial state $\ket \psi$ is rotated by $2\theta$ degrees and approaches toward the axis of $\ket\beta$.\vspace{-1EM}}
\end{figure}
\newpage
\section{Quantum Mean Estimation}
Let us consider the problem of computing the \hl{mean} of $F(x)$ in Equation~\ref{eqn:oracleRay}. This problem corresponds to supersampling $M$ sub-pixels (or $M$ quantized bins in high-dimensional integrands) in the context of ray tracing. When $M$ is large, a popular algorithm on a classical computer is Monte Carlo integration; we randomly sample multiple subpixels and use their average as the estimate of the correct average. The estimation error of Monte Carlo integration is $O(1/\sqrt{M})$ for $M$ samples.

On a quantum computer, we can \hl{evaluate} $F(x)$ at \textit{all the possible $x$} in one-shot using the oracle gate. As we explain later, it is also trivial to transform the resulting state into another state whose amplitude is the correct average value $f \equiv \frac{1}{M} \sum_{x=1}^{M} F(x)$ as:
\begin{eqnarray}
\ket \psi = \sqrt{1-f^2} \ket 0 + f \ket 1.
\label{eqn:mean}
\end{eqnarray}
Unlike classical computers, it does not fundamentally matter how large $M$ is on quantum computer since all the $M$ values are computed in one shot. The remaining problem, however, is to estimate the amplitude $f$ using this state.

One naive solution to this problem is to simply prepare $N$ instances of $\ket \psi$ by querying the oracle $N$ times and measure all of them (we cannot simply copy $\ket \psi$ $N$ times just by querying the the oracle $1$ time at the beginning, due to the no-cloning theorem~\cite{park1970concept}). We then count the number of measured states \red{belonging to $\ket 1$} and deduce the value of $f$ from that. This naive solution is essentially classical Monte Carlo integration, hence the convergence rate for $N$ queries (i.e., samples) is $O(1/\sqrt{N})$, and does not provide any benefit compared to classical Monte Carlo integration. 
It is thus important to design \hl{a} more efficient estimation algorithm which outperforms the classical calculation.
We focus on two quantum algorithms in this paper: QSS and QCoin, which almost achieve $O(1/N)$ error with $N$ queries. They use two other basic quantum algorithms called \textit{amplitude amplification} and \textit{quantum Fourier transformation}.

%-----------------------------------------------------------------------
\subsection{Amplitude Amplification}
The idea of amplitude amplification (AA) was first introduced in the context of a quantum database-search algorithm which is commonly known as \textit{Grover's algorithm}~\cite{Grover}.
%The algorithm finds a target quantum state by amplifying its amplitude and thus its observation probability.
We consider an oracle $\hat O$ which results in
\begin{eqnarray}
\ket{\psi} = \hat O \ket{00...0} = \cos\theta \ket\alpha + \sin\theta \ket\beta \label{3-1-1}
\end{eqnarray}
where $\ket\beta$ is a \red{set of target states} and $\ket\alpha$ is \red{a set of the other states}. 
The state $\ket{\psi}$ is represented as a vector $(\cos \theta, \sin \theta)$ within a plane spanned by $\ket\alpha$ and $\ket\beta$ as shown in Figure~\ref{AA}.
The goal \red{of AA} is to increase the \red{small} probability of observing the target state \red{$\ket \beta$}. The idea is to rotate $\ket{\psi}$ counter-clockwise \red{as $\ket \psi \rightarrow \hat G \ket \psi$ in Figure~\ref{AA}}. \red{Note that AA, despite its name, does not necessarily amplify the amplitude when $\theta$ is larger than $\pi/2$, and thus one can treat AA just as a rotation operator.}

\red{As detailed in Figure~\ref{AA}}, we first apply a flip operation $\hat{R}_f$ which flips the state $\ket \psi$ against the $\ket\alpha$ vector. It can be realized by flipping the sign of target states as $\ket\beta \rightarrow -\ket\beta$. We then project the resulting flipped state $\hat{R}_f \ket \psi$ onto the original $\ket \psi$, and multiply the length of the projected vector $\ket \psi \bra \psi \hat{R}_f \ket \psi$ by two. Finally, we subtract $\hat{R}_f \ket \psi$ from it. The resulting state \red{is}
\begin{eqnarray}
\ket {\psi_{\mathrm{result}}} = \cos 3\theta \ket\alpha + \sin 3\theta \ket\beta\red{.} \label{3-1-2}
\end{eqnarray}
\red{The formula of AA operation $\hat G$ is derived as}:
\begin{align}
    \ket {\psi_{\mathrm{result}}} = \hat G \ket \psi
    &\equiv \left(2\ket \psi \bra \psi \right) \hat {R}_f \ket \psi - \hat {R}_f \ket \psi \nonumber \\
    &= \left(2\ket \psi \bra \psi - \hat I \right)  \hat {R}_f \ket \psi \label{3-1-3} \\
    &= \left(2\hat O \ket{00...0} \bra{00...0} \hat O^{-1} - \hat I \right) \hat {R}_f \ket \psi \nonumber \\
    &= \hat O \left(2\ket{00...0} \bra{00...0} - \hat I \right) \hat O^{-1} \hat {R}_f \ket \psi. \label{3-1-4}
\end{align}
The $\left( 2\ket{00...0} \bra{00...0} - \hat I \right)$ operation corresponds to flipping the amplitude of all the states except the state $\ket{00...0}$. Since $\hat G$ includes two oracle gates ($\hat O$ and $\hat O^{-1}$), the AA algorithm makes two queries (i.e., $\hat O$ is called two times) to perform one $\hat G$ operator. Note that AA does not need to know the actual value of $\theta$.
%
%To summarize, AA rotates a quantum state vector $\ket \psi$ to $\ket{\psi_{\mathrm{result}}}$ (i.e., change $\theta$ to $3\theta$) without knowing the actual value of $\theta$ using two queries of the oracle gate $\hat O$. 

%----------------------------------------------------------
\subsection{Quantum Fourier Transformation}
Quantum Fourier transformation (QFT) can be thought as an analogy to classical discrete Fourier transformation. Given a data set $\{ a_0, a_1, a_2, \cdots ,a_{N-1} \}$, classical Fourier transformation $\{ a_k \,\,|\,\, 0 \le k \le N-1 \} \rightarrow \{ b_j \,\,|\,\, 0 \le j \le N-1\}$ conducts the calculation as 
%
%
%\begin{eqnarray}
$
b_j = \frac{1}{\sqrt{N}} \sum_{k=0}^{N-1} e^{-i\frac{2\pi}{N}j k} a_k.
$
%\end{eqnarray}
%
%
The resulting set $\{ b_i \}$ is a set of frequency components of the input data series $\{ a_i \}$, and one can view that Fourier transform is an algorithm which converts $\{ a_i \}$ into $\{ b_i \}$.
In QFT, the input data series is given by the amplitudes:
\begin{eqnarray}
\ket \psi = a_0 |0) + a_1 |1) + \cdots + a_{N-1} |N-1).
\end{eqnarray}
The idea of QFT is to turn this input quantum state into a superposition of frequency components $\{ b_i \}$ as 
\begin{eqnarray}
\ket {\psi_{\mathrm{QFT}}} = b_0 |0) + b_1 |1) + \cdots + b_{N-1} |N-1).
\end{eqnarray}
We will not explain the detailed process of QFT in this paper as it is not important for our discussion. Interested readers can refer to a textbook of quantum computing~\cite{NielsenChang}.
%
%The important difference from the classical algorithm is that we cannot directly access frequency components $b_i$. Instead, it is only that the observation probability of each state is proportional to $b_i^2$. %It is thus impossible to directly read out individual coefficients $b_i$ from a given single $\ket {\psi_{\mathrm{QFT}}}$.

\begin{figure}[t]
  \centering
  \includegraphics[width=1.0\linewidth]{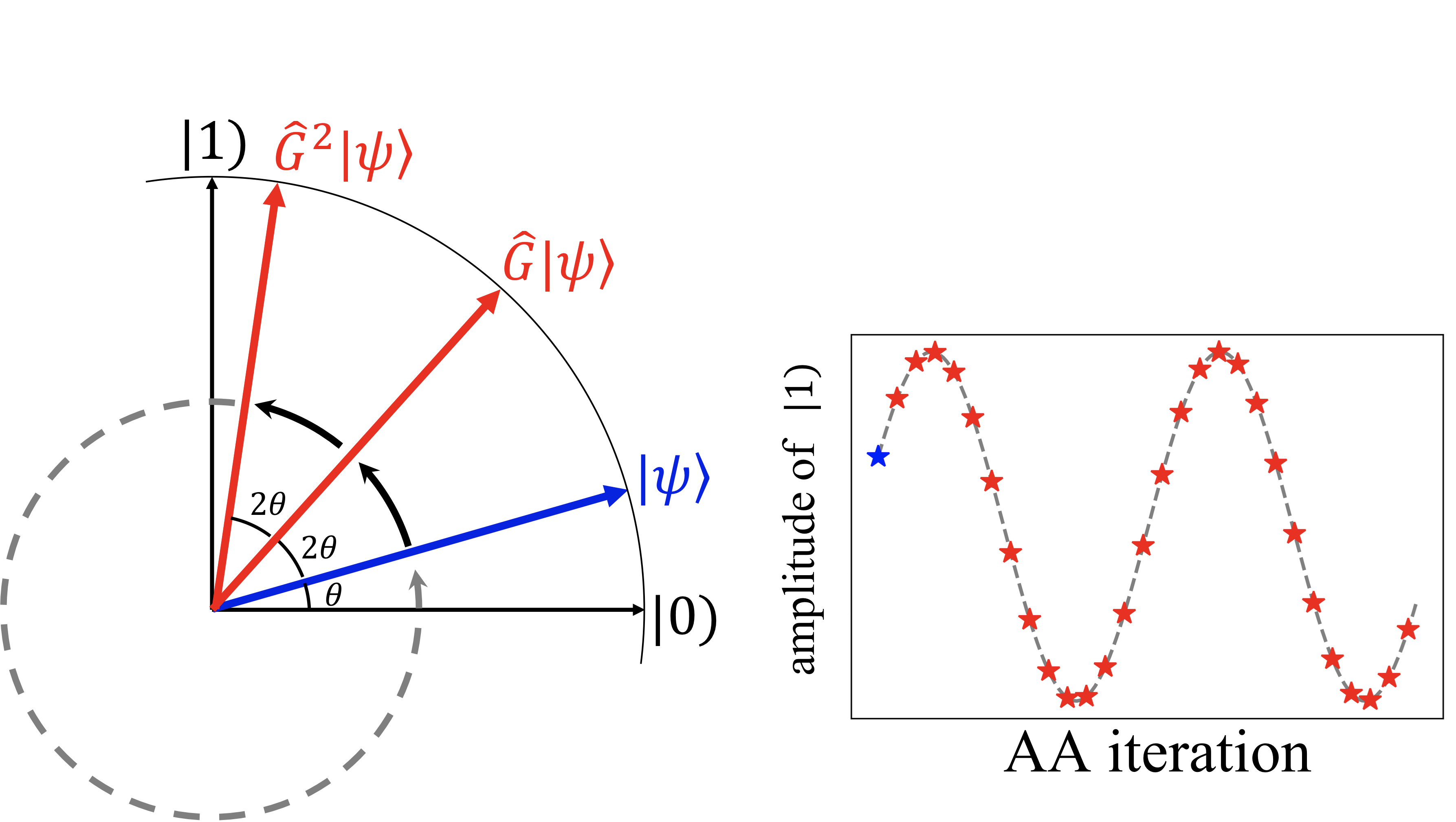}
  \caption{\label{QSS}Example of repeated AA operations. $\theta$ is initially defined as $f = \sin \theta$. The degree of state vector evolves as $\theta \rightarrow 3\theta \rightarrow 5\theta \cdots $ (left). Therefore, the trace of $f$ values tracks a sin curve (right).%cTH{Maybe use a color gradation on arrows and points to illustrate the correspondences between them.}
  \vspace{-1EM}}
\end{figure}

%-----------------------------------------------------------
\subsection{Quantum Supersampling}
Grover~\cite{Grover2} was the first to introduce \hl{a} quantum algorithm for estimating a mean $f = \frac{1}{M} \sum_{i=1}^M F(i)$. The idea is to combine AA with QFT as we explain later. Many theoretical developments have followed since then~\cite{QuantumCounting,Brassard}, but few numerical experiments using simulation have been done so far~\cite{Imai}. Johnston~\cite{QSS} implemented this original idea by Grover to conduct numerical experiments in the context of rendering. The problem addressed there is supersampling of an image, which can be seen as a mean estimation per pixel. We explain QSS by Johnston in the following, to contrast it to our QCoin. In the original work by Johnston~\cite{QSS}, the values of $F(i)$ are assumed to be binary $\{0, 1\}$. We modified it to be able to handle continuous values of $F(i)$. Since the algorithm essentially stays the same, we refer to our modified QSS simply as QSS in the following.

\paragraph*{Main idea.}
The main idea of QSS is to exploit the existence of a periodic cycle when we keep applying amplitude amplification on $\ket \psi$. As we explained before, AA rotates the state within the plane spanned by $\ket\alpha$ and $\ket\beta$, thus the state actually rotates fully after sufficiently many AA operations. It turns out that there is a unique periodic cycle to each corresponding $\theta$ value. Figure~\ref{QSS} shows the movement of the state vector $\ket{\psi}$ (left) and the trace of the amplitude value of $\ket1$ (right).
Applying QFT on the history of rotated $\ket \psi$, we can extract the frequency of this periodic cycle, which then allows us to calculate the corresponding $\theta$ (and therefore $f$).

%---------------------------
\paragraph*{Problem setting.}
In QSS, given a black-box function
%
%\begin{eqnarray}
$
F(a) : a \rightarrow [0,1] \label{3-3-prob-1}
$
%\end{eqnarray}
%
and a quantum oracle operator
\begin{eqnarray}
\hat Q_{F} : \ket0 \otimes |i) \rightarrow \left( \sqrt{1-F(i)} \ket0 + \sqrt{F(i)}\ket1 \right) \otimes |i),  \label{3-3-prob-2}
\end{eqnarray}
the objective is to get the average $f$ of $F(a)$ with $N (=2^n)$ samples:
\begin{eqnarray}
f \equiv \frac{1}{N} \sum_{i=0}^{N-1} F(i). \label{3-3-prob-4}
\end{eqnarray}

\newpage
%---------------------------
\paragraph*{Algorithm.}
In QSS, we use the oracle $\hat Q_F$ and make a superposition state $\ket {\psi_0}$ from the initial state whose all qubits (= register, target, and input \hl{qubits)} are $\ket0$, where the numbers of qubits for each are $\log_2 P$, 1, $\log_2 N$. We thus write the initial state as 
\begin{eqnarray}
\ket{\underbrace{0\cdots0}_{\log_2 P}} \otimes \ket{0} \otimes \ket{\underbrace{0\cdots0}_{\log_2 N}} = |0) \otimes \ket0 \otimes |0)
\end{eqnarray}
\red{We generate a superposition state $\ket {\psi_0}$} as
\begin{eqnarray}
\ket{\psi_0} &=& \hat Q_F (\hat H \otimes \hat I \otimes \hat H) |0) \otimes \ket0 \otimes |0) \label{3-3-algo-1} \\
&=& \frac{1}{\sqrt{PN}} \sum_{m=0}^{P-1} \sum_{i=0}^{N-1} |m) \otimes \hat Q_F (\ket 0 \otimes |i) ) \label{3-3-algo-2} \\
&=& \frac{1}{\sqrt{PN}} \sum_{m=0}^{P-1} \sum_{i=0}^{N-1} |m) \otimes \left( \sqrt{1-F(i)} \ket0 + \sqrt{F(i)}\ket1 \right) \otimes |i), \nonumber{} \\
\label{3-3-algo-3}
\end{eqnarray}
\red{where $\hat Q_F$ in Equation~\ref{3-3-algo-1} operates the latter two states.} The total measurement probability of $\frac{1}{\sqrt N} \sum_{i=0}^{N-1} \sqrt{F(i)} \ket1 \otimes |i)$ states is $\sum_{i=0}^{N-1} \sqrt{\frac{F(i)}{N}}^2 = f$. If we define $\ket0' \equiv \frac{1}{\sqrt{N}}\sum_{i=0}^{N-1} \ket0 \otimes |i)$ and $\ket1'$ in the same manner, the amplitude of $\ket1'$ is $\sqrt f$:
\begin{eqnarray}
\ket{\psi_0} &=& \frac{1}{\sqrt{P}} \sum_{m=0}^{P-1} |m) \otimes \left( \sqrt{1-f} \ket0' + \sqrt{f}
\ket1' \right). \nonumber
\end{eqnarray}
We can define  $\cos \theta$ and $\sin \theta$ as $\sqrt{1-f}$ and $\sqrt{f}$, and $\ket{\psi_0}$ is
\begin{eqnarray}
\ket{\psi_0} &=& \frac{1}{\sqrt{P}} \sum_{m=0}^{P-1} |m) \otimes \left( \cos \theta \ket0' + \sin \theta
\ket1' \right). \label{3-3-algo-4}
\end{eqnarray}
We then apply AA to the $\left( \cos \theta \ket0' + \sin \theta \ket1' \right)$ state for \red{$P$ times:}
\begin{equation} 
\ket{\psi_1} = \frac{1}{\sqrt{P}} \sum_{m=0}^{P-1} |m) \left( \cos (2m+1)\theta \ket0'  + \sin (2m+1)\theta \ket1' \right). \label{3-3-algo-5}
\end{equation}
We then \red{measure the target qubits}. We assume that the state is converged to $\ket1'$:
\begin{eqnarray}
\ket{\psi_2} = \frac{1}{C} \sum_{m=0}^{P-1} \sin{(2m+1)\theta} \,\, |m) \ket1' \label{3-3-algo-6}
\end{eqnarray}
Finally, we perform QFT on $\ket{\psi_2}$. With a sufficiently large probability~\cite{QuantumCounting}, the result of measurement after QFT will be
\begin{eqnarray}
t \simeq \frac{P \theta}{\pi}, \frac{P (\pi-\theta)}{\pi}. \label{3-3-algo-7}
\end{eqnarray}
If the measured and converged state is $\ket0'$, we get the same result. Therefore, we can deduce the estimated average $f'$ by
\begin{eqnarray}
f \approx f' = \sin^2 \left(\frac{t \pi}{P}\right).
\label{3-3-algo-8}
\end{eqnarray}
Since $\frac{t \pi}{P}$ can be determined by the precision $O(1/P)$ in this process, $f'$ also has the precision of $O(1/P)$. Johnston~\shortcite{QSS} proposed to use a precomputed table instead of the analytical expression in Equation~\ref{3-3-algo-8} by considering only discrete values of $f$.
The estimation error $|f-f'|$ is inversely proportional to the number of AA operations $P$. Since AA uses two queries per operation, we perform $O(N)$ queries to achieve $O(1/N)$ error. Note that this convergence rate is faster than $O(1/\sqrt{N})$ of Monte Carlo integration.

%---------------------------
\paragraph*{Example.}
We show how the whole process works for the 5 qubits case where $P=4$ and $N=4$. The initial state of 5 qubits is
$
 (\ket 0 \otimes \ket 0) \otimes \ket 0 \otimes (\ket 0 \otimes \ket 0) = |0) \otimes \ket0 \otimes |0).
 $
At first, we apply $\hat H \otimes \hat I \otimes \hat H$ as in Equation \ref{3-3-algo-1}:
\begin{flalign}
 &\left( \hat H \otimes \hat I \otimes \hat H \right) |0) \otimes \ket0 \otimes |0) \nonumber \\
 &= \left( \frac{|0)+|1)+|2)+|3)}{\sqrt{4}}\right) \otimes \ket 0 \otimes \left( \frac{|0)+|1)+|2)+|3)}{\sqrt{4}}\right). \nonumber
\end{flalign}
This transformation is as Equation~\ref{gate-decimal}. Then, the oracle $\hat Q_{F}$ works as \red{(omitted the register qubits)}:
\begin{flalign}
 &\hat{Q}_F \ket 0 \otimes \left( \frac{|0)+|1)+|2)+|3)}{\sqrt{4}}\right) \nonumber \\
&= \frac{1}{\sqrt4} \left( \sqrt{1-F(0)}\ket0 + \sqrt{F(0)}\ket1 \right) \otimes |0) \nonumber \\
&+ \frac{1}{\sqrt4} \left( \sqrt{1-F(1)}\ket0 + \sqrt{F(1)}\ket1 \right) \otimes |1) \nonumber \\
&+ \frac{1}{\sqrt4} \left( \sqrt{1-F(2)}\ket0 + \sqrt{F(2)}\ket1 \right) \otimes |2) \nonumber \\
&+ \frac{1}{\sqrt4} \left( \sqrt{1-F(3)}\ket0 + \sqrt{F(3)}\ket1 \right) \otimes |3). \nonumber
\end{flalign}
The probability of observing $\ket1$ is calculated as
\begin{flalign}
 &\left|\sqrt{\frac{F(0)}{4}}\right|^2 + \left|\sqrt{\frac{F(1)}{4}}\right|^2 + \left|\sqrt{\frac{F(2)}{4}}\right|^2 + \left|\sqrt{\frac{F(3)}{4}}\right|^2 = f \nonumber 
 %&= \frac{F(0)+F(1)+F(2)+F(3)}{4} = f, \nonumber 
\end{flalign}
hence the total amplitude of $\ket1$ is $\sqrt f$. By grouping a set of states with $\ket1$ as $\ket1'$ (and those with $\ket0$ as $\ket0'$) for brevity, the resulting state vector can be written as
%
%\begin{flalign}
$
 \sqrt{1-f} \ket0' + \sqrt f \ket1' = \cos\theta \ket0' + \sin\theta \ket1' \nonumber
$
%\end{flalign}
%
where we write $\sqrt f = \sin \theta$. The state $\ket{\psi_0}$ is
\begin{flalign}
 \ket{\psi_0} = \left( \frac{|0)+|1)+|2)+|3)}{\sqrt{4}}\right) \otimes \cos\theta \ket0' + \sin\theta \ket1'. \nonumber
\end{flalign}
We perform AA operations ($\hat G$) corresponding to the decimal number of the register qubits' state
\begin{align}
 \ket{\psi_1} 
 &= \frac{1}{\sqrt4} |0) \otimes \hat G^{0} \left(\cos \theta \ket0' + \sin \theta \ket1' \right) \nonumber \\
 &+ \frac{1}{\sqrt4} |1) \otimes \hat G^{1} \left(\cos \theta \ket0' + \sin \theta \ket1' \right) \nonumber \\
 &+ \frac{1}{\sqrt4} |2) \otimes \hat G^{2} \left(\cos \theta \ket0' + \sin \theta \ket1' \right) \nonumber \\
 &+ \frac{1}{\sqrt4} |3) \otimes \hat G^{3} \left(\cos \theta \ket0' + \sin \theta \ket1' \right) \nonumber \\
 &= \,\,\,\, \frac{1}{\sqrt4} |0) \otimes \left(\cos \theta \ket0' + \sin \theta \ket1' \right)\,\,\,  \nonumber \\
 &+ \frac{1}{\sqrt4} |1) \otimes \left(\cos 3\theta \ket0' + \sin 3\theta \ket1' \right) \nonumber \\
 &+ \frac{1}{\sqrt4} |2) \otimes  \left(\cos 5\theta \ket0' + \sin 5\theta \ket1' \right) \nonumber \\
 &+ \frac{1}{\sqrt4} |3) \otimes  \left(\cos 7\theta \ket0' + \sin 7\theta \ket1' \right). \nonumber
\end{align}
We then measure the target qubit $\ket1'$ to obtain
%
%\begin{flalign}
$
 \ket{\psi_2} = \frac{1}{C} (\sin \theta |0) + \sin 3\theta |1) + \sin 5\theta |2) + \sin 7\theta |3) ) \otimes \ket1' \nonumber
$
%\end{flalign}
%
which allows us to estimate $\theta$ (thus $f$) value using QFT.
\section{Quantum Coin Method}
We introduce another mean-estimation quantum algorithm, which we call as the \textit{quantum coin method} (QCoin). While the theory of QCoin was introduced by Abrams and Williams 20~years ago~\cite{QCoin}, its actual implementation was not discussed and no numerical experiment has been done so far. We provide the first practical implementation of this algorithm by identifying practical issues and performed the first set of numerical experiments. 
%

%------------------------------------------
\paragraph*{Quantum coin.}
QCoin uses a \textit{quantum coin} as its core. \red{A quantum coin is a quantum state as described in Equation~\ref{eqn:mean}, which returns} the target state $\ket1$ ("head") with the probability of $f^2$, and other states $\ket0$ ("tail") with the probability $1 - f^2$. By counting the number of "heads" out of the total number of trials, we can estimate $f^2$ (and $f$) with $\delta$ error with $O(1/\delta^2)$ queries. As we discussed before, this process alone is equivalent to Monte Carlo integration, thus it will not provide any benefit.

\begin{figure}[t]
\centering
\includegraphics[width=.9\linewidth]{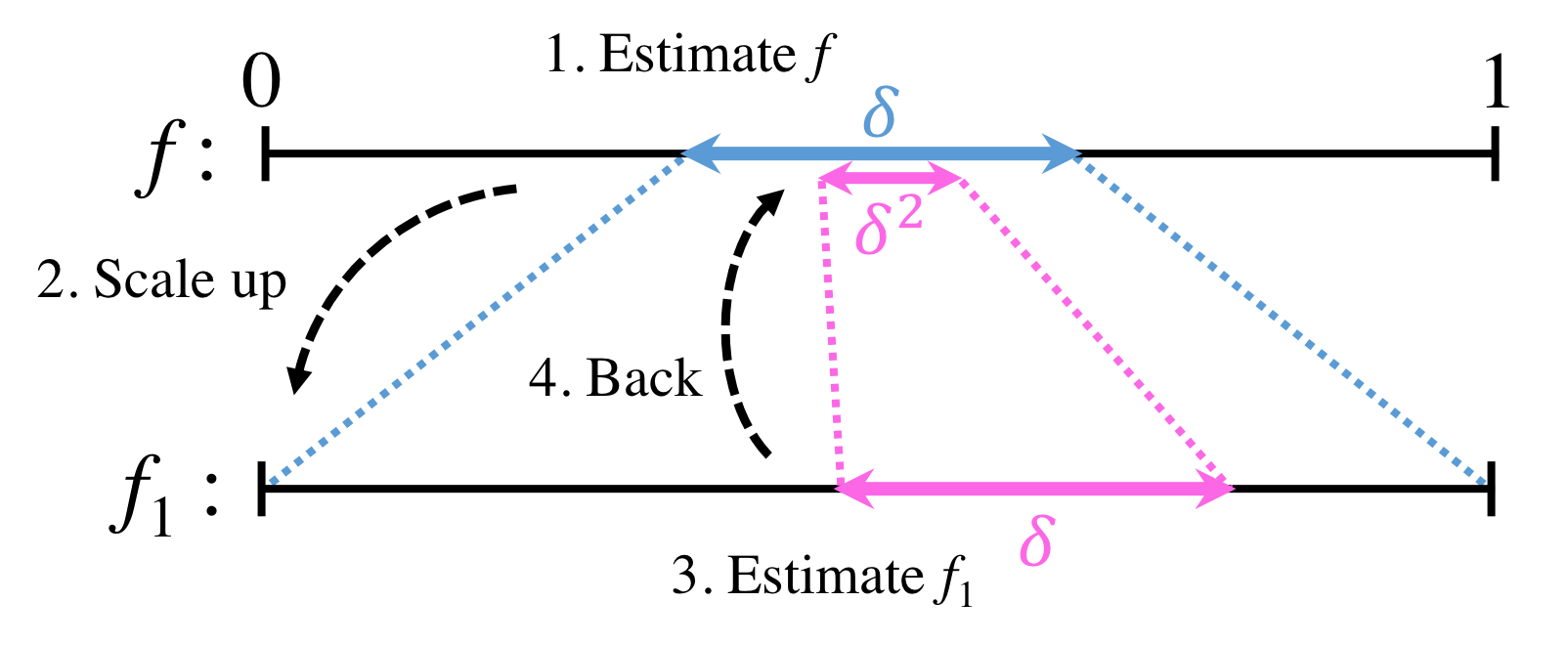}
\caption{Shifting-Scaling process of QCoin: 1. We estimate the value of $f$ and decide bounded-error range: $[f'-\frac{\delta}{2},f'+\frac{\delta}{2}]$. 2. We scale up quantum coin to [0,1]. 3. Now the target value $f$ is changed to $f_1$, we can estimate $f_1$ with $\delta$ error. 4. We can estimate $f$ with $\delta^2$ error via calculating back from estimated $f_1$ value.
\vspace{-1EM}}
\label{QCoin}
\end{figure}

%------------------------------------------
\paragraph*{Main idea.}
Suppose that we have a rough estimate $f'$ by running Monte Carlo integration using a quantum coin as described above with $N$ queries. According to the error analysis of Monte Carlo integration, with a certain confidence probability, one can say that the actual value of $f$ is in the interval of $\left[ f'-\frac{\delta}{2}, f'+\frac{\delta}{2} \right]$ where $\delta = O(1/\sqrt{N})$. The idea of QCoin is to repeatedly shrink this interval by shifting and scaling it using quantum computation until we are sufficiently close to $f$. Figure~\ref{QCoin} illustrates this process.

\paragraph*{Problem setting.}
QCoin considers a black-box function
\begin{eqnarray}
    F(a) : a \rightarrow [0,1] \label{4-prob-1}
\end{eqnarray}
and a quantum oracle operator $\hat Q_{F,E}$ which includes the function $F(a)$ and the offset (shifting) parameter $E$:
\begin{eqnarray}
    \hat Q_{F,E} \ket0 \otimes |i) \rightarrow \left( \sqrt{1-\left(F(i)-E\right)^2} \ket0 + \left(F(i)-E\right) \ket1 \right) \otimes |i). \nonumber{} \\ \label{4-prob-2}
\end{eqnarray}
Our goal is to estimate the average value $f$ similar to QSS.
%
%\begin{eqnarray}
%   f \equiv \frac{1}{N} \sum_{i=0}^{N-1} F(i)
%\end{eqnarray}
%

%----------------------------------------
\paragraph*{Algorithm.}
%\cTH{The following is not good for a paper. Revise it.} \cNS{I choosed to add the contents of supplementary in this article.}
%\deleted{In this part, the development of equations might be complicated. If necessary, please refer to the detailed example of 3 qubits' case in Supplementary2.} 
%
%
For the first step, using oracle $\hat Q_{F,0}$, we make the initial superposition state \red{(the number of qubits of input is $\log_2 N$)}:
\begin{eqnarray}
    \ket{\psi_0} &=& \hat Q_{F,0} (\hat I \otimes \hat H) \ket 0 \otimes |0) \label{4-algo-1} \nonumber \\
    &=& \hat Q_{F,0} \sum_{i=0}^{N-1} \ket 0 \otimes |i) \nonumber \\
    &=& \frac{1}{\sqrt{N}} \sum_{i=0}^{N-1}  \left( \sqrt{1-F(i)^2} \ket0 + F(i) \ket1 \right) \otimes |i) \label{4-algo-2}
\end{eqnarray}
The construction of a quantum coin is in fact simple; we perform $\hat H$ operators for all the qubits after the oracle gate operation. After this process, each state is distributed with $\frac{1}{\sqrt{N}}$ amplitude to a |0) state and any amplitude to all the other states:
\begin{eqnarray}
\ket{\psi_0}' &=& (\hat I \otimes \hat H) \ket{\psi_0} \label{4-algo-3} \nonumber \\
&=& \frac{1}{N} \sum_{i=0}^{N-1}F(i) \ket1 \otimes |0) + \cdots %\label{4-algo-4} \\
= f \ket1 \otimes |0) + \cdots \label{4-algo-5}
\end{eqnarray}
%

%---------------------------

%---------------------------
We show the construction of a quantum coin for the 3 qubits case. The initial state of 3 qubits is 
$
 \ket0 \otimes \left( \ket0 \otimes \ket0 \right) = \ket0 \otimes |0).
$
Applying $\hat I \otimes \hat H$ operation, we have
\begin{flalign}
 \left( \hat I \otimes \hat H \right) \left( \ket0 \otimes |0) \right)
 = \ket0 \otimes \left( \frac{|0)+|1)+|2)+|3)}{\sqrt4} \right). \nonumber
\end{flalign}
We get $\ket{\psi_0}$ in Equation~\ref{4-algo-2} using $\hat Q_{F,0}$:
\begin{flalign}
\ket{\psi_0} 
&=\hat Q_{F,0} \ket0 \otimes \left( \frac{|0)+|1)+|2)+|3)}{\sqrt4} \right) \nonumber \\
&= \frac{1}{\sqrt4} \left(\sqrt{1-F(0)^2}\ket0 + F(0)\ket1\right) \otimes |0) \nonumber \\
&+ \frac{1}{\sqrt4} \left(\sqrt{1-F(1)^2}\ket0 + F(1)\ket1\right) \otimes |1) \nonumber \\
&+ \frac{1}{\sqrt4} \left(\sqrt{1-F(2)^2}\ket0 + F(2)\ket1\right) \otimes |2) \nonumber \\
&+ \frac{1}{\sqrt4} \left(\sqrt{1-F(3)^2}\ket0 + F(3)\ket1\right) \otimes |3). \nonumber \;\;\;\;\;\;\;\;\;\;\;\;\;\;\;\;\;\;\;\;\; %(\ref{4-algo-2})'
\end{flalign}
Now, if we operate $\hat H$ on the input qubits, the states are changed to
\begin{flalign} 
\hat H |0) 
&= \frac{|0) + |1) + |2) + |3)}{\sqrt4} , \;\;
\hat H |1) %&= \hat H \ket 0 \otimes \hat H \ket 1 \nonumber \\
%&= \left( \frac{\ket 0 + \ket 1}{\sqrt2} \right) \otimes \left( \frac{\ket 0 - \ket 1}{\sqrt2} \right) \nonumber \\
= \frac{|0) - |1) + |2) - |3)}{\sqrt4} \nonumber \\
\hat H |2) &= \frac{|0) + |1) - |2) - |3)}{\sqrt4}, \;\;
\hat H |3) = \frac{|0) - |1) - |2)+ |3)}{\sqrt4}. \nonumber
\end{flalign}
All states are distributed to $|0)$ with $+\frac{1}{\sqrt4}$ amplitude, hence
\begin{flalign}
 \ket{\psi_0}'  
 &= \left( \hat I \otimes \hat H \right) \ket{\psi_0} \nonumber \;\;\;\;\;\;\;\;\;\;\;\;\;\;\;\;\;\;\;\;\;\;\;\;\;\;\;\;\;\;\;\;\;\;\;\;\;\;\;\;\;\;\;\;\;\;\;\;\;\;\;\;\;\;\;\;\;\;\;\;\;\;\;\;\;\;\;\;\; \\%(\ref{4-algo-3})' \\
 &= \left( \frac{F(0) + F(1) + F(2) + F(3)}{4} \right) \ket 1 \otimes |0) + \cdots  \,\,\,\,\,\,\,\,\,\,\,\,\,\,\,\,\,\,\,\,\,\,\,\, \nonumber \\%(\ref{4-algo-4})' \nonumber \\
 &= f \ket 1 \otimes |0) + \cdots . \;\;\;\;\;\;\;\;\;\;\;\;\;\;\;\;\;\;\;\;\;\;\;\;\;\;\;\;\;\;\;\;\;\;\;\;\;\;\;\;\;\;\;\;\;\;\;\;\;\;\;\;\;\;\;\;\;\;\;\;\;\;\;\;\;\; \nonumber%(\ref{4-algo-5})' \nonumber
\end{flalign}
\begin{algorithm}                      
\caption{\label{alg:qcoin}Our implementation of Qcoin $(F, k, L)$}         
\label{alg1}   
\begin{algorithmic}
\STATE // 1st step
\STATE $f_0 \leftarrow 0$
\FOR{$i=1$ to $L$}
\STATE make QCoin : $\hat{Q}_{\sqrt{F},0} \ket0|0)$
\IF{Measure(QCoin) $==\ket1$}
\STATE $f_0$ += $1$
\ENDIF
\ENDFOR
\STATE $f_0$ /= $L$
\STATE

\STATE // The other steps
\STATE $E_{-} \leftarrow 0.0, \; E_{+} \leftarrow 1.0$
\FOR{$i=1$ to $k$}
\STATE $\delta \,\,\,\,\,\,\,\, \leftarrow \sin (\pi/2^{i+1})$ \,\,\,\,\,\,\,\, \,\,\,\,\,\,\,\, \, // hypothetical error
\STATE $E_{-} \leftarrow \mathrm{Max}(f_{i-1}-\frac{\delta}{2}, E_-)$ \, // lower bound of error range
\STATE $E_{+} \, \leftarrow \mathrm{Min}(f_{i-1}+\frac{\delta}{2}, E_+)$ \,\, // upper bound
\STATE $f_i \leftarrow 0$
\FOR{$j=1$ to $L$}
\STATE make QCoin : $\hat{G}_{F,E_{-}}^{2^{i-1}} \ket0|0)$
\IF{Measure(QCoin) $==\ket1|0)$}
\STATE $f_i$ += $1$
\ENDIF
\ENDFOR
\STATE $f_i$ /= $L$
\STATE $f_i \leftarrow$ Min $\left( E_{-} + \sin \left( \frac{\mathrm{asin} (f_i)}{2^{i}} \right), E_{+} \right)$
\ENDFOR

\STATE
\STATE // Output
\PRINT $f_k$

\end{algorithmic}
\end{algorithm}
Note that the amplitude of $\ket 1 \otimes |0)$ is equal to $f$. This $\ket{\psi_0}'$ state thus can be regarded as a quantum coin.
We use $\ket{\psi_0}'$ to perform a rough estimate of $f$ within $\delta$ error using $O(1/\delta^2)$ queries just like Monte Carlo integration. Suppose that the estimated value is $f_0$, then we can say that the correct value $f$ is in the interval $[f_0-\frac{\delta}{2}, f_0+\frac{\delta}{2}]$ with a certain high probability (1st process in Figure \ref{QCoin}).

%\cTH{The blue part below needs a significant revision. I recommend you to ask someone to proofread those parts. I will try to revise them later as it will take some time for me to grasp the ideas behind them and rewrite them.}
For the next step, we set $E \equiv f_0-\frac{\delta}{2}$ and the oracle gate as $\hat{Q}_{F,E}$. We make the quantum coin $\ket{\psi_1}'$ using $\hat{Q}_{F,E}$ as above:
\begin{eqnarray}
\ket{\psi_1}' = \left(f-E\right) \ket1 \otimes |0) + \cdots. \label{4-algo-6}
\end{eqnarray}
Now, \red{the amplitude of $\ket 1 \otimes |0)$} is $f-E$. This value is in the interval $[0, \delta]$. If we define $\sin \theta \equiv f - E$,
\begin{eqnarray}
\ket{\psi_1}' \equiv \sin \theta \ket 1 \otimes |0) + \cdots. \label{4-algo-7}
\end{eqnarray}
We then operate AA for $O(1/\delta)$ times to make the error range from $[0,\delta]$ to $[0,1 - \epsilon]$ (upper limit is not always precisely $1$). It can be done without knowing the exact value of $f$. If we conduct $m$ times AA operations, the state is changed as:
\begin{eqnarray}
\ket{\psi_1}'' = \sin (2m+1) \theta \, \ket 1 \otimes |0) + \cdots. \label{4-algo-8}
\end{eqnarray}
This corresponds to the 2nd process in Figure~\ref{QCoin}. Now, we can estimate the value of $\sin (2m+1) \theta$ within $\delta$ error measuring the state for $O(1/\delta^2)$ times (3rd process in Figure \ref{QCoin}).

We assume the estimated value is $f_1$. Then, we can easily calculate back to the original scale: calculate the value of $\theta$ from $m$ and $\sin (2m+1) \theta$ values, and $f$ is calculated by the relation $``f = \sin\theta + E"$. As a result, we get to estimate $f$ value with the error range $\delta^2$ (4th process in Figure \ref{QCoin}). 
If this step is repeatedly for $k$ times, we achieve the error $\delta^{k+1}$.

%----------------------------------
\paragraph*{Convergence rate.}
In the case of $k=1$ step as above, the estimation error is $\delta^2$, and the total number of queries is calculated as:
\begin{eqnarray}
O(1/\delta^2) + \left(1+2O(1/\delta)\right) \times O(1/\delta^2) = O(1/\delta^3)
\end{eqnarray}
The convergence rate is improved from "$\delta$ error with $O(1/\delta^2)$ queries" to "$\delta^2$ error with $O(1/\delta^3)$ queries". For comparison, assuming the numbers of queries are both $N_{\mathrm{all}}$, the estimation error is reduced from $O\left(\frac{1}{{N_{\mathrm{all}}}^{0.5}}\right)$ to $O\left(\frac{1}{{N_{\mathrm{all}}}^{0.66\cdots}}\right)$. 
%\cTH{I guess you are trying to say that $O(1/\delta^{k+2}) = O(1 / N^{1 - 1/(k+2)})$? I think we should mention the relationship between $\delta$ and $N$ somewhere before this.}

As for the case of $k \gg 1$, we show the convergence rate here. If we use $M$ queries in the Monte Carlo integration part of QCoin, we achieve $O(1/ \sqrt{M})$ as the error value of $\delta$ (equivalently, $\delta = O(1/ \sqrt{M})$). For a total of $k-1$ iterations, QCoin achieves the final error value $O(\delta^k)$ as described above. On the other hand, the total number of queries $N_{\mathrm{all}}$ in this case is asymptotically defined as
\begin{eqnarray}
    N_{\mathrm{all}} = M \cdot O\left(1 + M^{1/2} + \cdots + M^{(k-1)/2} \right) = O\left( M^{k/2} \right)
\end{eqnarray}
for a large enough $k$. Given that we have $\delta = O(1/ \sqrt{M})$, we can conclude that QCoin achieves the final error value of $O(\delta^k) = O(1 / \sqrt{M}^k) = O(1 / N_{\mathrm{all}})$ using $N_{\mathrm{all}}$ queries using QCoin.

\begin{figure*}[ht]
  \centering
  \includegraphics[width=0.95\linewidth]{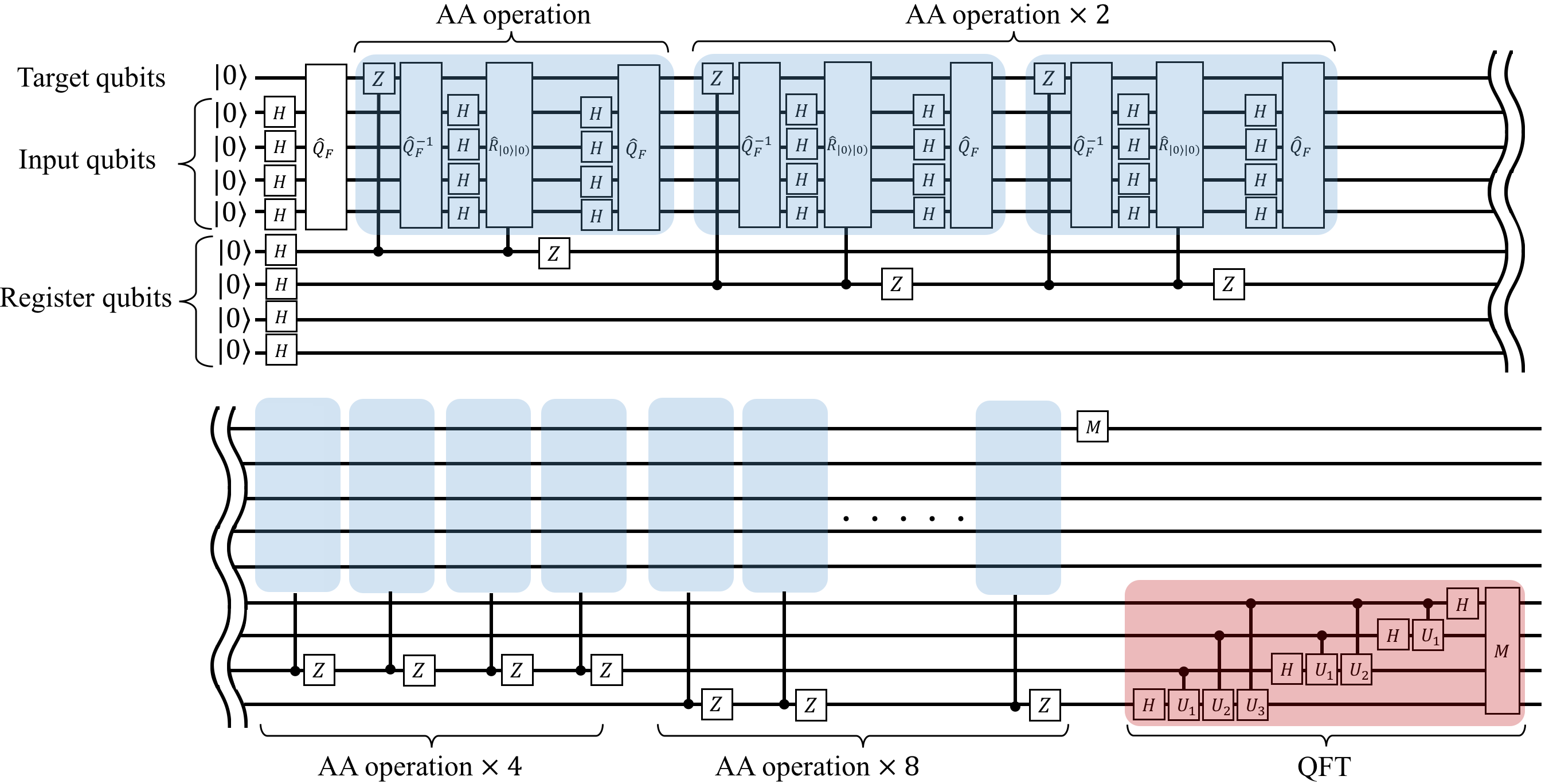}
  \caption{\label{Qcircuit-QSS}%
           Example of the quantum circuit of QSS with 4 input qubits and 4 register qubits case. ($X,Z$: Pauli gates, $\hat H$: Hadamard gate, $\hat Q_F$ and ${\hat{Q}_F}^{-1}$: oracle gate and inverse oracle gate, $M$: measurement gate, $R_{\ket0 |0)}$: phase flip gate only for the state $\ket0 |0)$, $U_n (n=1,2,3)$: $e^{i\frac{\pi}{2^n}}$ phase shift gate for $\ket1$ state.)}
\end{figure*}

\begin{figure*}[t]
  \centering
  \includegraphics[width=0.78\linewidth]{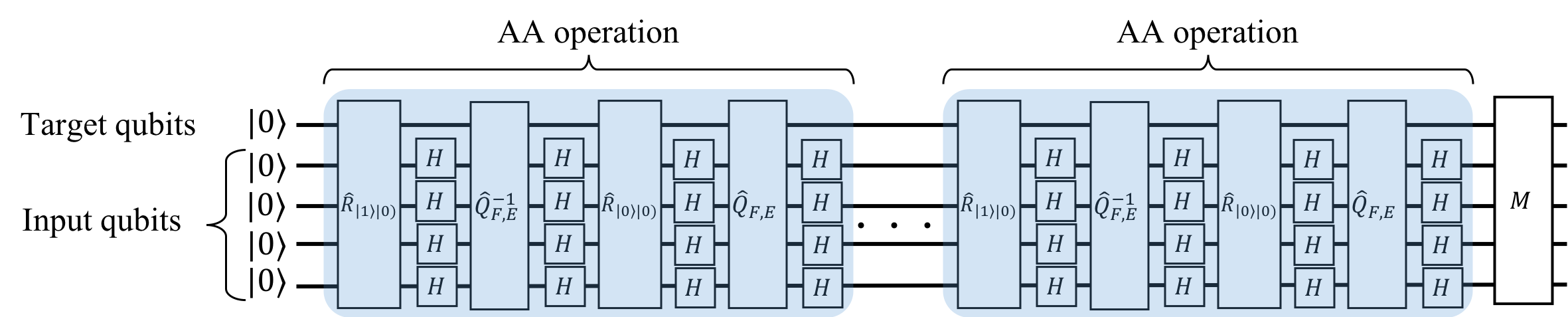}
  \caption{\label{Qcircuit-QCoin}%
           Example of the quantum circuit of QCoin with 4 input qubits. ($\hat Q_{F,E}$ and ${\hat{Q}_{F,E}}^{-1}$: oracle gate and its inverse, $R_{\ket0 |0)}$ and $R_{\ket1 |0)}$: phase flip gate only for the state $\ket0 |0)$ and $\ket1 |0)$ respectively.)}
\end{figure*}

\paragraph*{Our contributions over Abrams and Williams.}
Compared to the original work by Abrams and Williams~\cite{QCoin}, our work provides the following contributions.
\begin{itemize}
    \item We conducted numerical experiments to clarify the followings:
    \begin{itemize}
        \item While Abrams and Williams~\cite{QCoin} showed that the convergence rate of QCoin approaches to $O(1/N)$ as $k$ increases, it has not been clear how the convergence rate changes for a finite (practical) $k$ as we have done.
        \item Similar to classical Monte Carlo integration, there is non-zero possibility that $f$ resides outside the estimated interval at each step. Its influence is difficult to investigate just by looking at the theory, which we have shown by numerical experiments.
        \item In the QCoin algorithm, Monte Carlo estimates are done by estimating $f^2$ (i.e., the probability of "heads") and then taking its square-root, making its estimation more error-prone toward $f \approx 0$. This causes the fluctuation of a estimation error in accordance with the observed values. How this \hl{influences the efficiency of the algorithm} is unknown.
    \end{itemize}
\end{itemize}
\begin{itemize}
    \item We redesigned and implemented the whole algorithm of QCoin as shown in Algorithm~\ref{alg:qcoin}. Some technical points we implemented are as follows:
    \begin{itemize}
        \item We showed how to deal with the cases where Monte Carlo estimation at each step is outside the range $[0, 1]$ which has been ignored so far, yet certainly happens in practice.
        \item We proposed to directly estimate $f$ for the first estimate using a different quantum coin than the rest. %\cTH{I think this claim asks us to show a comparison with and without this modification. Maybe we cannot claim it.}
        \item We fixed the number of scaling-shifting \hl{operations} per step. %\cTH{I am not sure if it should be listed.}
    \end{itemize}
\end{itemize}
\begin{itemize}
    \item We pointed out the superiority of QCoin against QSS, for the first time, in terms of usefulness on an actual quantum computer.
    \begin{itemize}
        \item We pointed out that QCoin belongs to a class of hybrid quantum-classical algorithms~\cite{VQE} and demonstrate its usefulness on actual quantum computers in the presence of noise (shown and discussed later).
    \end{itemize}
\end{itemize}

\begin{figure*}[htb]
  \centering
  \mbox{}
  \includegraphics[height=0.25\linewidth]{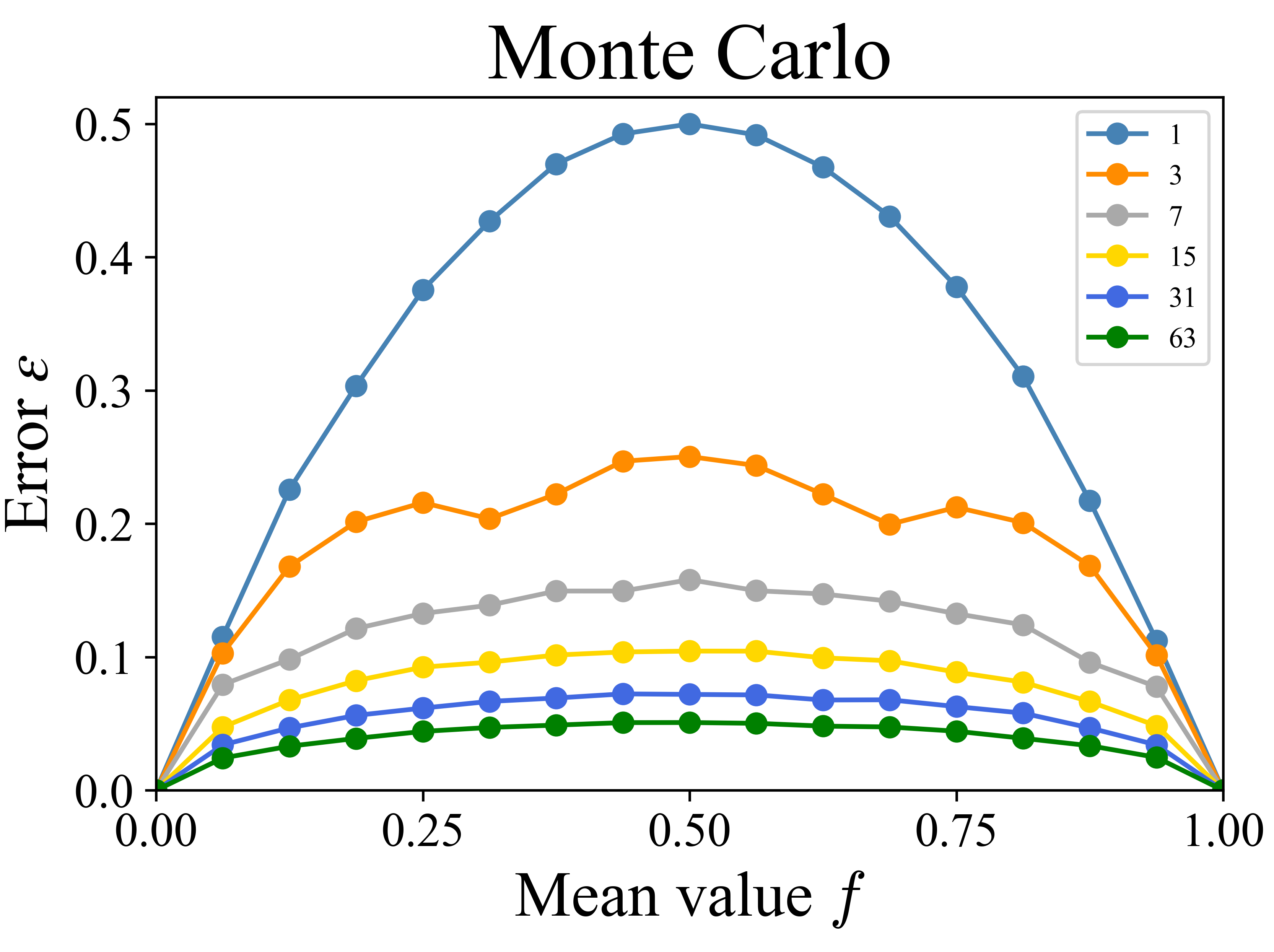}
  \includegraphics[height=0.25\linewidth]{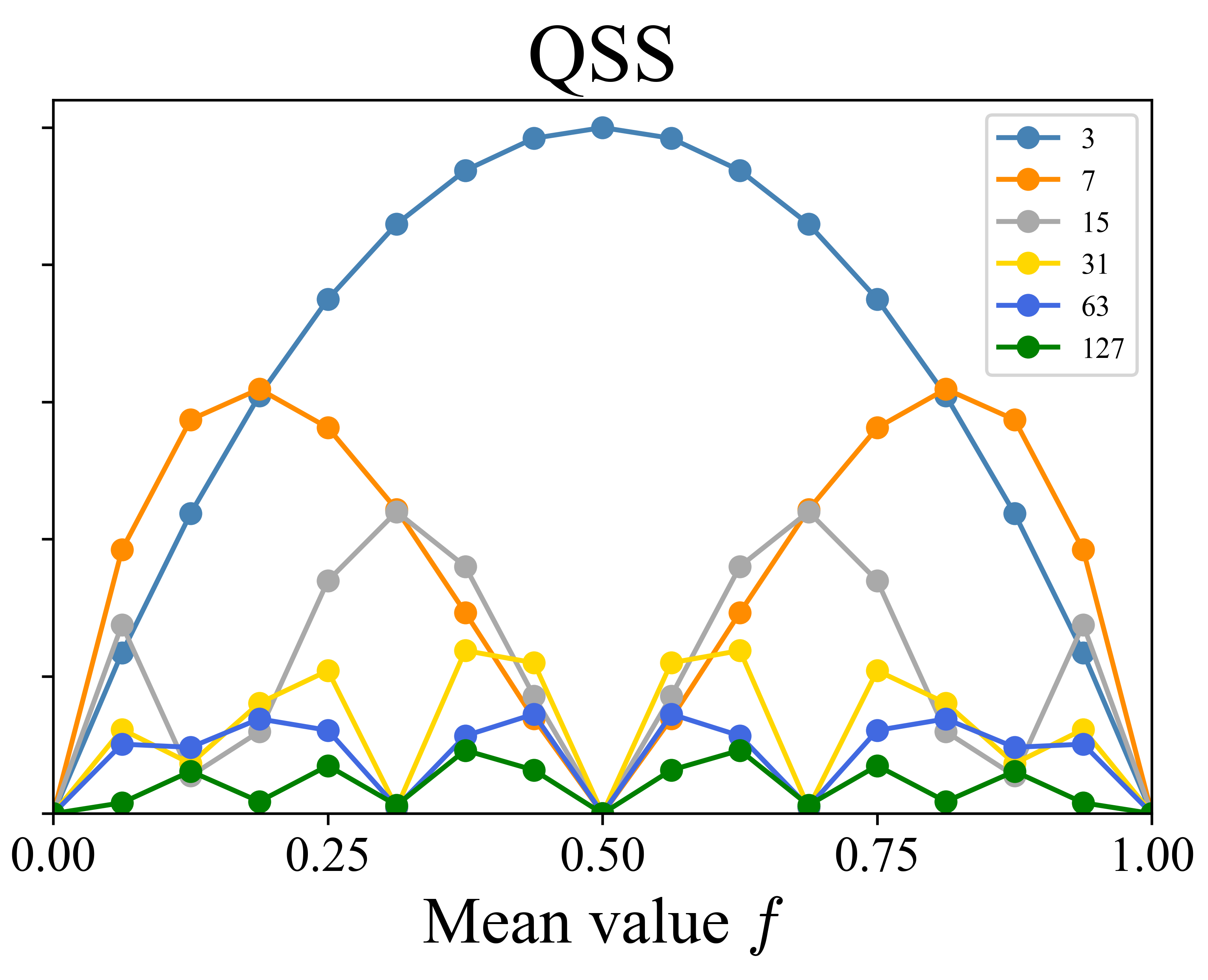}
  \includegraphics[height=0.25\linewidth]{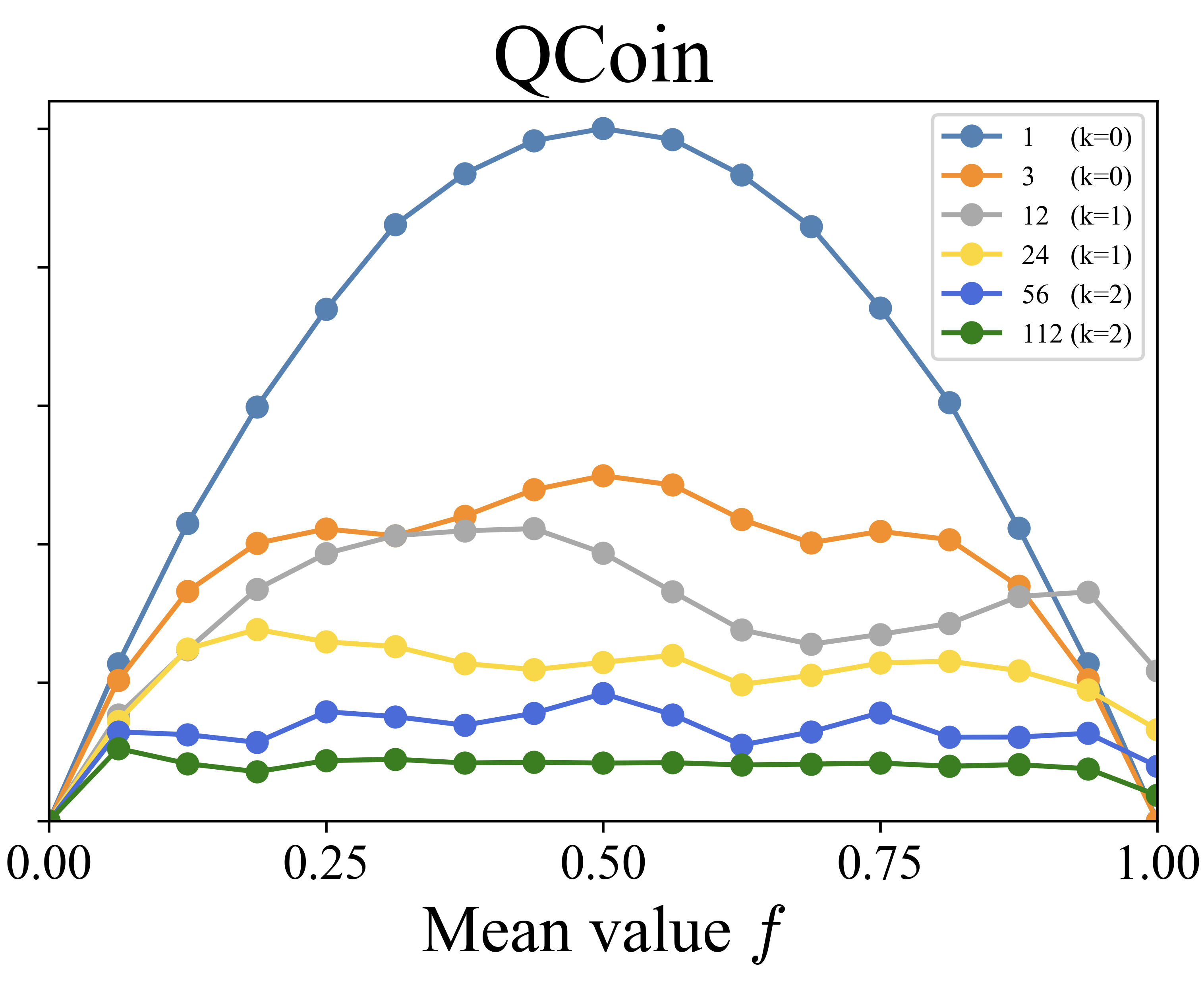}
  \mbox{}
  \caption{\label{fig:ex1}%
           Error plots against query times (represented by colors indicated in the legends) with various target mean $f$ in three methods; Monte Carlo, QSS, and QCoin. }
\end{figure*}

\begin{figure*}[tbp]
\centering
  \mbox{}
  \includegraphics[height=0.325\linewidth]{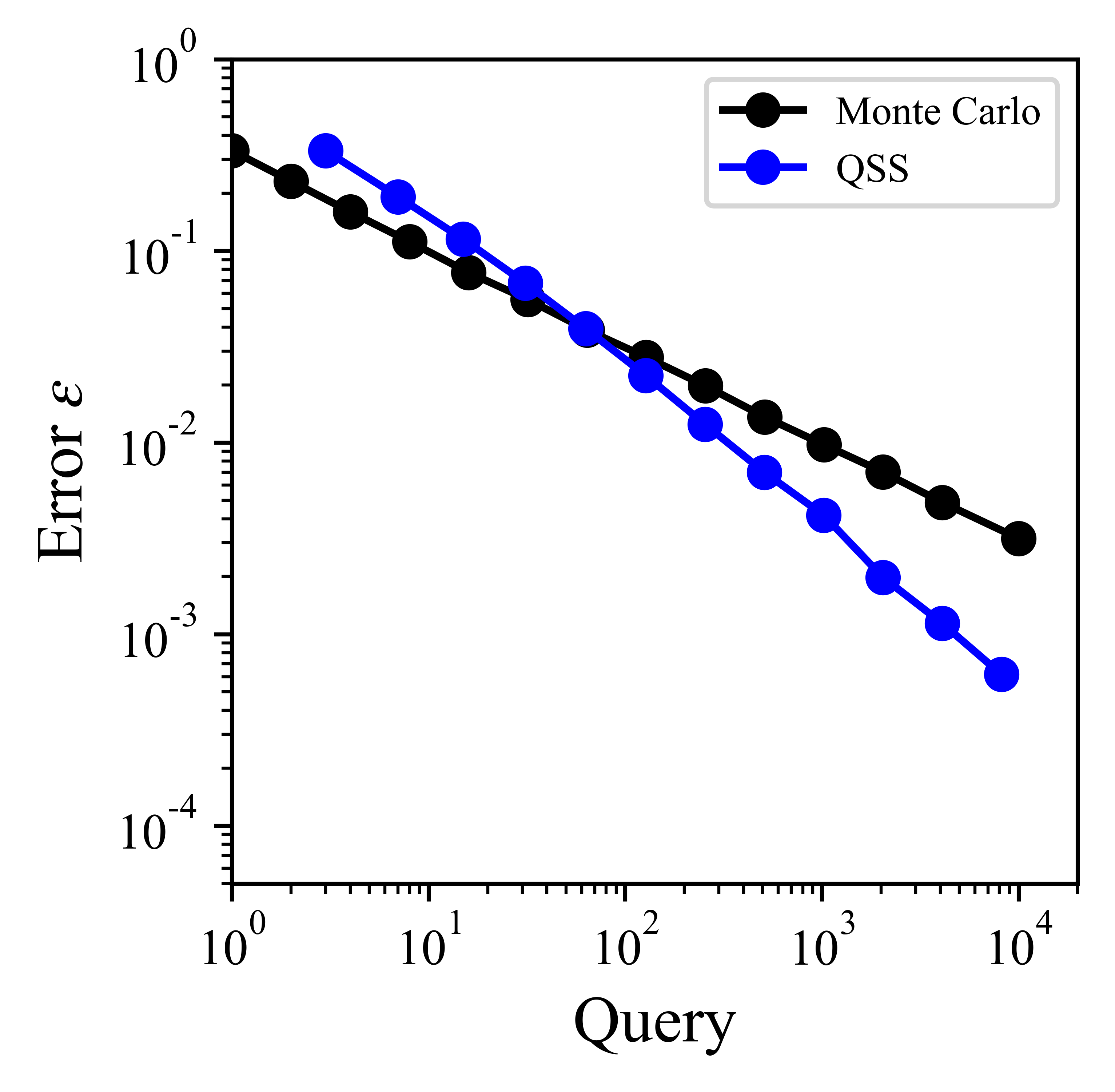}
  \includegraphics[height=0.325\linewidth]{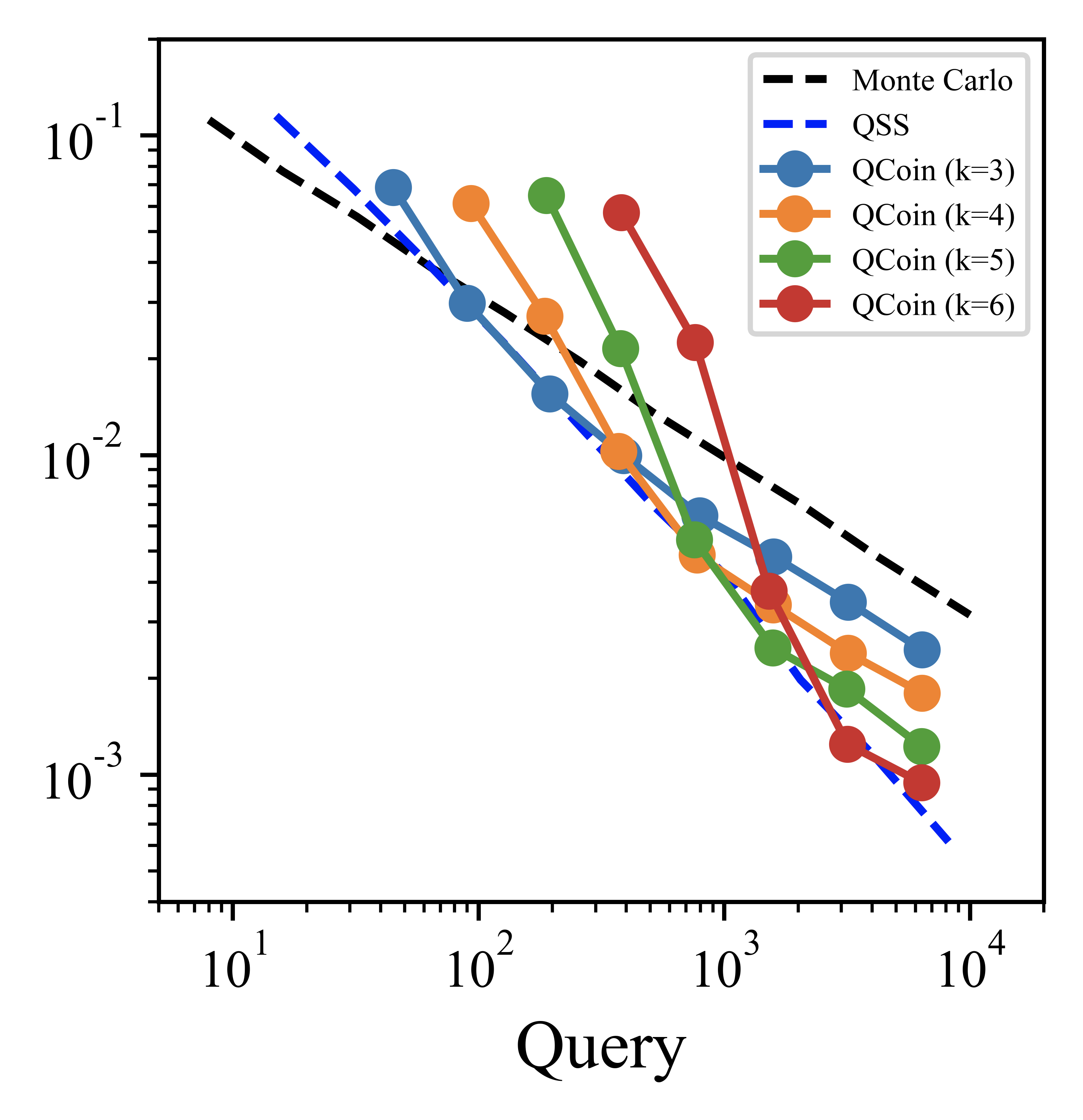}
  \includegraphics[height=0.325\linewidth]{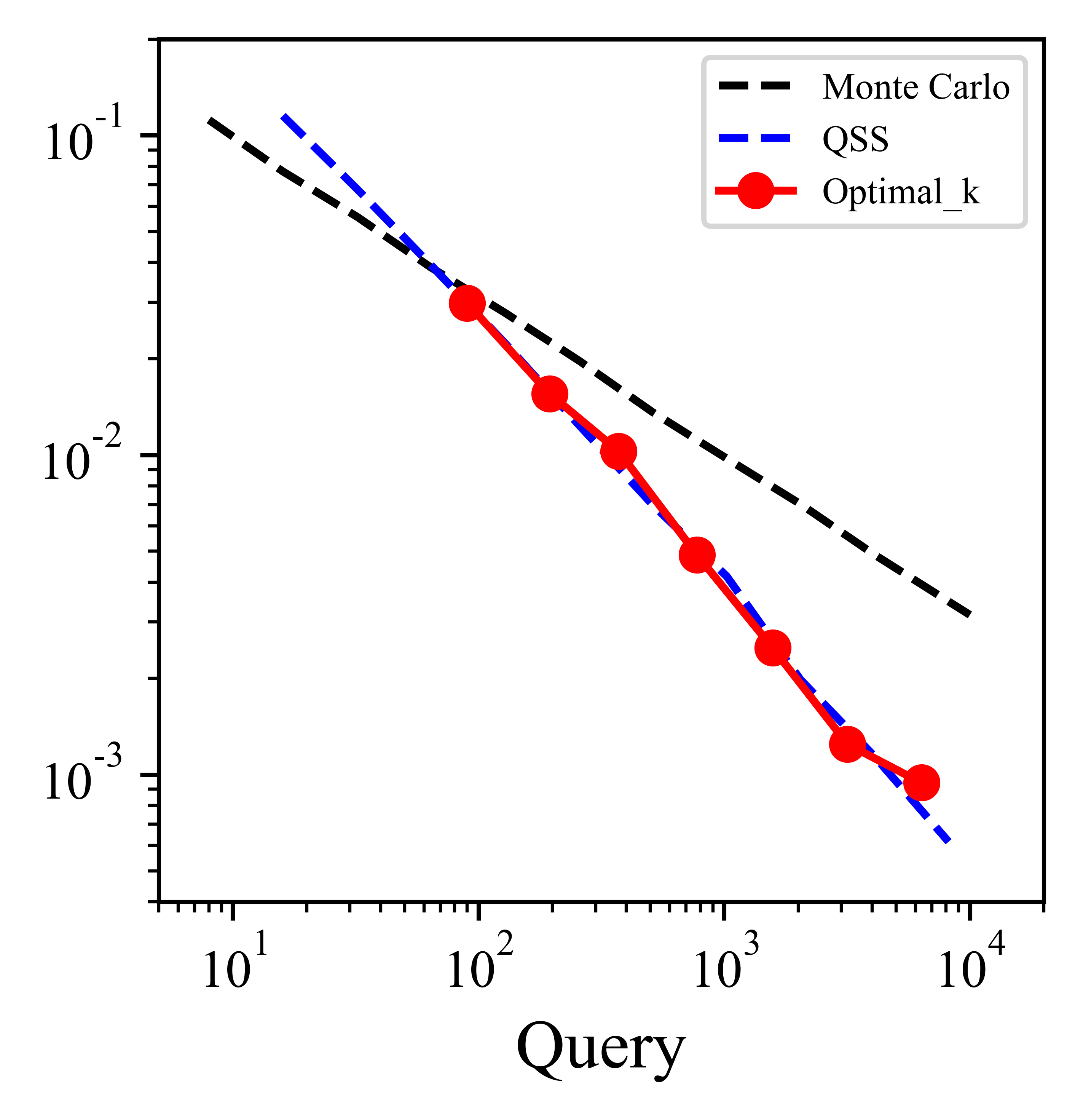}
  \mbox{}
  \caption{\label{fig:ex2}%
           (Left and Center) Mean absolute error plots with query times in Monte Carlo, QSS, and QCoin($k=3,4,5,6$). (Right) The best performance of QCoin with selected optimal $k$ values is plotted.}
\end{figure*}
\section{On a Simulator}
We now explain our implementations of QSS and QCoin on a simulator of quantum computing using Microsoft Q\#~\cite{Qsharp}. \green{Our source code is available on Github~\cite{Github}.}

\paragraph*{QSS.}
The AA operation $\hat{G}_F$ using from Equation \ref{3-3-algo-4} to Equation \ref{3-3-algo-5}  in QSS is defined as Equation \ref{3-1-3}:
\begin{eqnarray}
\hat G_F \equiv (2\ket{\psi_0} \bra{\psi_0} - \hat I) \, \hat Z. \label{5-1}
\end{eqnarray}
 $\ket{\psi_0}$ can be decomposed as Equation \ref{3-3-algo-1}:
 \begin{eqnarray}
\ket{\psi_0} &=& \hat{Q}_F (\hat I \otimes \hat H) \, \ket0|0) \label{5-2} \\
\bra{\psi_0} &=& \red{\bra0(0|} \, (\red{\hat I \otimes \hat H}) {\hat{Q}_F}^{-1}. \label{5-3}
\end{eqnarray}
We substitute Equation \ref{5-2} and \ref{5-3} to Equation \ref{5-1}:
\begin{eqnarray}
\hat G_F = \hat{Q}_F (\hat I \otimes \hat H) \, \left(2\ket0|0)\red{\bra0(0|} - \hat  I\right) \, (\red{\hat I \otimes \hat H}) \hat{Q}_F^{-1} \, \hat Z \nonumber \\
= - \hat{Q}_F (\hat I \otimes \hat H) \, \left( \hat I - 2\ket0|0)\red{\bra0(0|}\right) \, (\red{\hat I \otimes \hat H}) \hat{Q}_F^{-1} \, \hat Z, \nonumber
\end{eqnarray}
where we omit irrelevant register qubits here. If $(\hat I - 2\ket0|0)\red{\bra0(0|} )$ is defined to be represented by $\hat R_{\ket0 |0)}$, we have
\begin{eqnarray}
\hat G_F = -\hat{Q}_F (\hat I \otimes \hat H) \, \hat R_{\ket0 |0)} \, (\hat I \otimes \hat H) \hat{Q}_F^{-1} \, \hat Z. \label{5-4}
\end{eqnarray}
\red{The example of an quantum circuit of QSS using $\hat G_F$ is shown in Figure~\ref{Qcircuit-QSS} where the number of input qubits is 4.} Each blue-colored region of the circuit corresponds to Equation~\ref{5-4}. The operators in Equation~\ref{5-4} are lined up in the reverse order in the circuit (operators are like matrix operations, hence they are indeed conducted from the back of an equation). AA operations are controlled by register qubits, which allows us to store the history of the rotating state vector; only if a control-register qubit is $\ket1$, $\hat G_F$ is run and the state vector rotates.
%In the case of 4 register qubits as in Figure~\ref{Qcircuit-QSS}, the number of operations $\hat G_F$ for each register qubits' state is
%
% \begin{eqnarray}
% {\hat{G}_F}^0 : &\ket{0} \otimes \ket{0} \otimes \ket{0} \otimes \ket{0} = |0)& \nonumber \\
% {\hat{G}_F}^1 : &\ket{0} \otimes \ket{0} \otimes \ket{0} \otimes \ket{1} = |1)& \nonumber \\
% {\hat{G}_F}^2 : &\ket{0} \otimes \ket{0} \otimes \ket{1} \otimes \ket{0} = |2)& \nonumber \\
% &\vdots& \,\,\,\, \nonumber \\
% {\hat{G}_F}^{15} : &\ket{1} \otimes \ket{1} \otimes \ket{1} \otimes \ket{1} = |15).& \nonumber
% \end{eqnarray}
%
Finally, the red region of the circuit performs QFT, which operates on the register qubits and extracts the period.

%---------------------------
\paragraph*{QCoin.}
\red{The quantum circuit of QCoin is described in Figure~\ref{Qcircuit-QCoin}, where the number of input qubits is 4.}
The exact operator $\hat G_{F, E}$ of AA for Qcoin $\ket{\psi_1}'$ in Equation \ref{4-algo-7} is defined as the equation \ref{3-1-3}:
\begin{eqnarray}
\hat G_{F, E} \equiv (2\ket{\psi_1}' \bra{\psi_1}' - \hat I) \, \hat R_{\ket 1 |0)}, \label{5-QCoin-1}
\end{eqnarray}
where operator $\hat R_{\ket 1 |0)}$ flips the amplitude of $\ket 1 |0)$. $\ket{\psi_1}'$ is decomposed by $\ket{\psi_1}$ and some gate operations like Equation \ref{4-algo-3}:
\begin{eqnarray}
 \hat G_{F, E} &=& (\hat I \otimes \hat H) \, (2\ket{\psi_1} \bra{\psi_1} - \hat I) \, (\red{\hat I \otimes \hat H}) \, \hat R_{\ket 1 |0)} \nonumber{} \\
 &=& (\hat I \otimes \hat H) \ (2\ket{\psi_1} \bra{\psi_1} - \hat I) \, (\red{\hat I \otimes \hat H}), \,  \hat R_{\ket 1 |0)} \label{5-QCoin-2}
\end{eqnarray}
and $\ket{\psi_1}$ is also deconstructed from Equation~\ref{4-algo-1}:
\begin{eqnarray}
2\ket{\psi_1} \bra{\psi_1} - \hat I &=& \hat Q_{\hat F,E} (\hat I \otimes \hat H)\, (2\ket0|0)(0|\bra0 - \hat I) \, (\red{\hat I \otimes \hat H}) \hat Q_{\hat F,E}^{-1} \nonumber \\
&=& \hat Q_{\hat F,E} (\hat I \otimes \hat H)\, \hat R_{\ket 0 |0)} \, (\red{\hat I \otimes \hat H}) \hat Q_{\hat F,E}^{-1}, \label{5-QCoin-3}
\end{eqnarray}
where $(2\ket0|0)(0|\bra0 - \hat I)$ is represented by $\hat R_{\ket 0 |0)}$ for simplicity. Substituting Equation~\ref{5-QCoin-3} to \ref{5-QCoin-2}, we get the explicit form of $\hat G_{F, E}$:
\begin{flalign}
 &\hat G_{F, E} = (\hat I \otimes \hat H) \hat Q_{\hat F,E} (\hat I \otimes \hat H) \, \hat R_{\ket0 |0)} \, (\red{\hat I \otimes \hat H}) \hat Q_{\hat F,E}^{-1} (\red{\hat I \otimes \hat H}) \, \hat R_{\ket 1 |0)}. \nonumber{} \\ \label{5-QCoin-4}
\end{flalign}
%

%-------------------------------------------------------------------------
\subsection{Results}

\begin{figure*}[htb]
  \centering
  \includegraphics[width=0.85\linewidth]{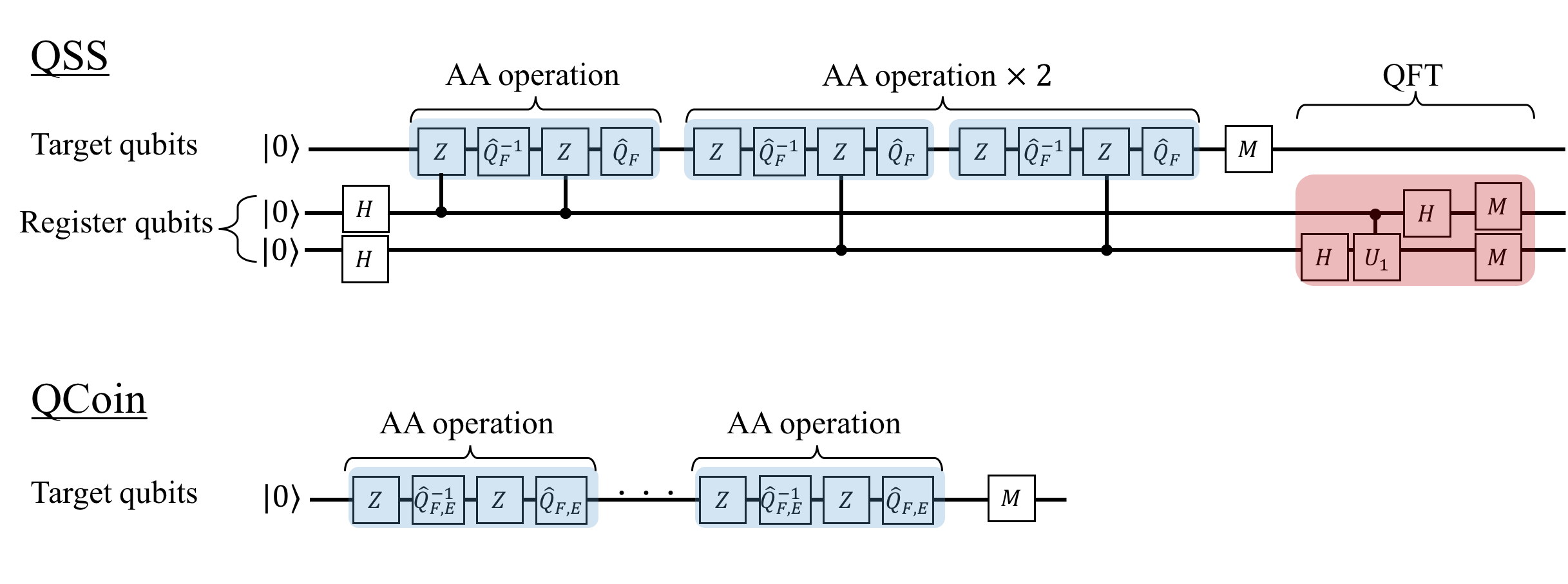}
  \caption{\label{SimpleCir} (Top) QSS's quantum circuit for 6 queries in minimum setting; no input qubit and two register qubits. (Bottom) QCoin's quantum circuit in minimum setting; no input qubit.}
\end{figure*}

\paragraph*{Convergent behavior against target value.}
Figure~\ref{fig:ex1} shows the behaviors of estimation error against the increasing number of queries with various target mean values $f$ in three methods: Monte Carlo, QSS, and QCoin. We conducted numerical experiments with 3000 samples for each point in Monte Carlo and QCoin, and calculated its theoretical error for QSS. In Monte Carlo, all the reduction rates of errors are almost uniform regardless of $f$, while QSS returns almost zero error at specific $f$. This distinctively different behavior is also showed by Johnston, which arises from QFT. Fourier transformation extracts a period of data series, therefore it can definitely detect the frequency of wave whose period just \hl{matches} the data length. %\deleted{the $O(1/P)$ resolution of QSS with $O(P)$ times of AA operations} \cTH{I don't get "$O(1/P)$ resolution of QSS with $O(P)$ times of AA operation".}\cNS{It's wrong. I edited.} 
In QCoin, however, we see almost uniform reduction of error just like Monte Carlo integration. One minor difference occurs at $f=1.0$ where QCoin has non-zero error while Monte Carlo integration has zero error. This difference arises from the fact that the QCoin algorithm scales the bounded-error of quantum coin $[0,\delta]$ to $[0,a]$ ($a$ is not always 1). %\cTH{This reasoning does not sound logical to me. Why is it related to the error at $f=1.0$?} 
If $a$ is always 1, the QCoin with $f=1.0$ only returns $\ket 1$ at any steps.
\hl{This experiment shows that the estimation error of QCoin has a similar characteristic as that of Monte Carlo. On the other hand, QSS behaves quite differently from Monte Carlo. As we discuss later, this similarity between MC and QCoin may allow us to use the existing error reduction methods (e.g., denoising) with QCoin.}
%\hl{It is discussed how the different behaviors of QSS and QCoin result in the difference of abilities to be combined with denoising methods used for Monte Carlo in Discussions section.}

\paragraph*{Convergent behavior against the number of queries.}
Figure~\ref{fig:ex2} shows the mean error of QCoin with random $f$ samplings for each $k$ step. Figure~\ref{fig:ex2} (Left) plots the results of Monte Carlo and QSS. Monte Carlo integration took 10000 samples ($f$ is randomly selected for each sample) for each point, and calculate theoretical error of QSS with uniformly selected 200 $f$ values for each point. In Monte Carlo, the slope of the curve is $-0.50$ in the logarithmic scale which matches the theoretical convergence rate of $O(1/\sqrt{N})$ with $O(N)$ queries. In QSS, the slope is $-0.85$, hence it achieves $O(1/N^{0.85})$ error with $O(N)$ queries. This result is close to the theoretical rate of $O(1/N)$.
Figure~\ref{fig:ex2} (Center) shows the results of QCoin with $k=3,4,5,6$ cases. For all the $k$ values, the error of few queries is large because the trials of a quantum coin in each step is too small for the estimation value be reasonably accurate for the succeeding shifting-scaling operations. Other than that, the slope for the same $k$ value first quickly becomes close to $-1.0$, but asymptotically approaches to  $-0.5$ after many queries while fixing $k$. We can thus observe that there is an optimal number of shifting and scaling operations $k$ for a given total number of queries. 
Figure~\ref{fig:ex2} (Right) plots the results of QCoin with the those optimal $k$ values for each number of queries. This optimal $k$ results in almost the same performance as QSS, and we use this optimal $k$ for the remaining experiments.
\hl{This experiment thus demonstrates that QCoin performs as well as QSS for a finite number of queries on a noiseless simulator. While its theoretical $O(1/N)$ convergence predicts that QCoin asymptotically outperforms Monte Carlo, we are the first to numerically verify its performance for a finite $N$.}

\paragraph*{Supersampling.}
\red{Figure~\ref{fig:teaser} shows an application of our method to a rendering task. The task is supersampling which estimates the average of subpixel values. One can think of this task as a numerical integration problem where the integrand is a function of subpixel values. This experiment is inspired by similar experiments done by Johnston~\cite{QSS}. In our experiment, each pixel contains $8 \times 8$ subpixels. The ground truth image ("Ideal sampling") is computed by simply taking the average of all $64$ subpixels. We used Monte Carlo, QSS on a simulator, QCoin on a simulator, and QCoin on an actual quantum computer (later discussed for the last one) with the same 240 queries ($k=3$ for QCoin) (255 queries only for QSS) by considering subpixels as the values of the integrand. The table in Figure~\ref{fig:teaser} shows mean absolute error of the five rectangular regions at the bottom of each and of the gradation parts at the top right of the images, which highlight errors for particular pixel values.}

\red{In the gradation part of the images, the mean absolute error of "QCoin on simulator" is almost the same as that of "QSS on simulator", and about half as much as Monte Carlo's, which is consistent with the error plots in Figure~\ref{fig:ex2}. However, we can see a striking difference in the images. Although Monte Carlo and QCoin show uniform reduction of error in the region, QSS produces more error in some pixels and less error in other pixels, which is consistent with the convergence behavior seen in Figure~\ref{fig:ex1}. This is also confirmed by colored rectangular regions; in QSS, some regions of particular pixel-color (0.0,0.5,1.0) show no error result, but the other parts indicate more estimation error than in QCoin.}

\section{On an Actual Quantum Computer}

\begin{figure*}[htb]
  \centering
  \mbox{}
  \includegraphics[height=0.325\linewidth]{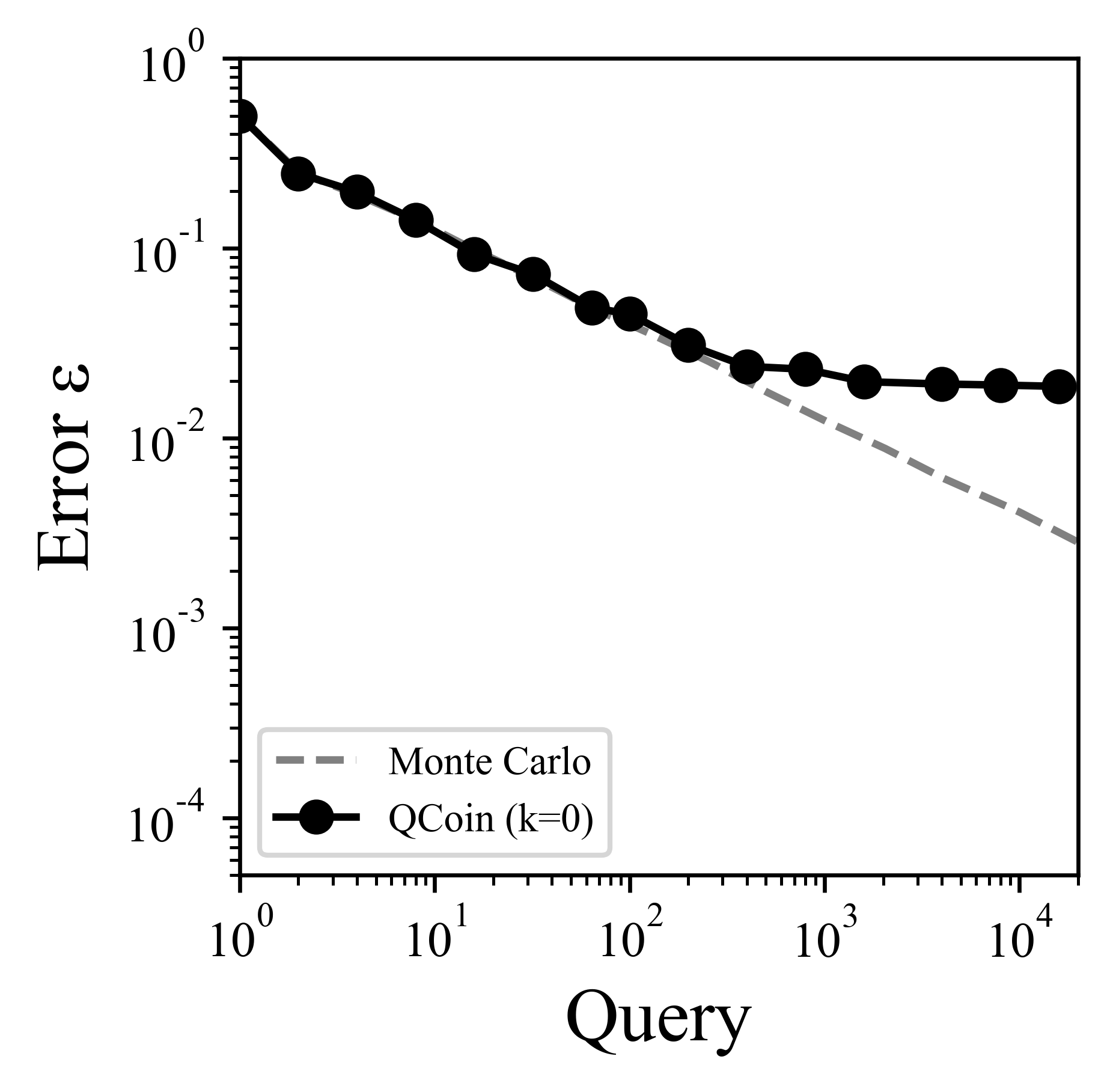}
  \includegraphics[height=0.325\linewidth]{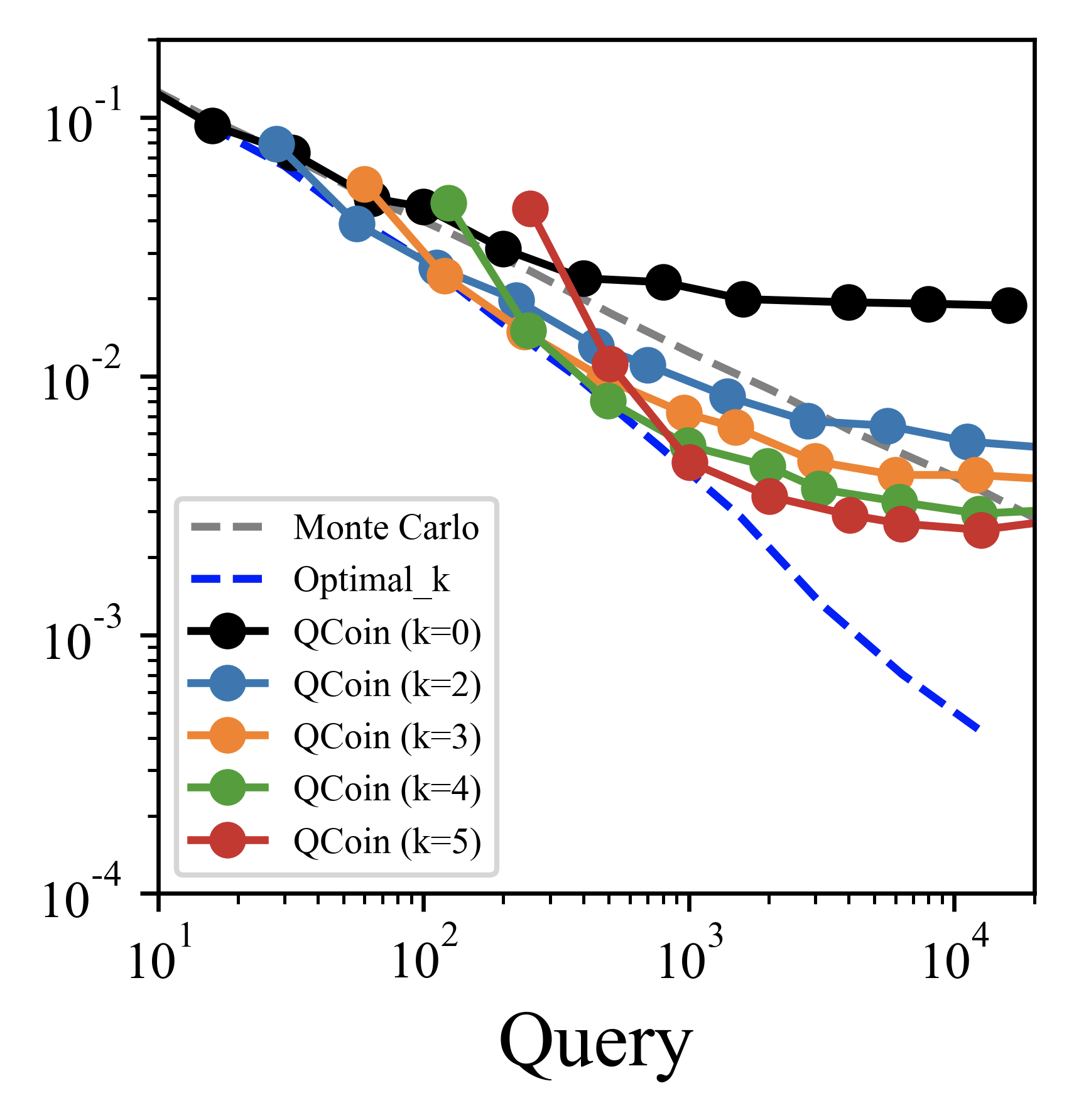}
  \includegraphics[height=0.325\linewidth]{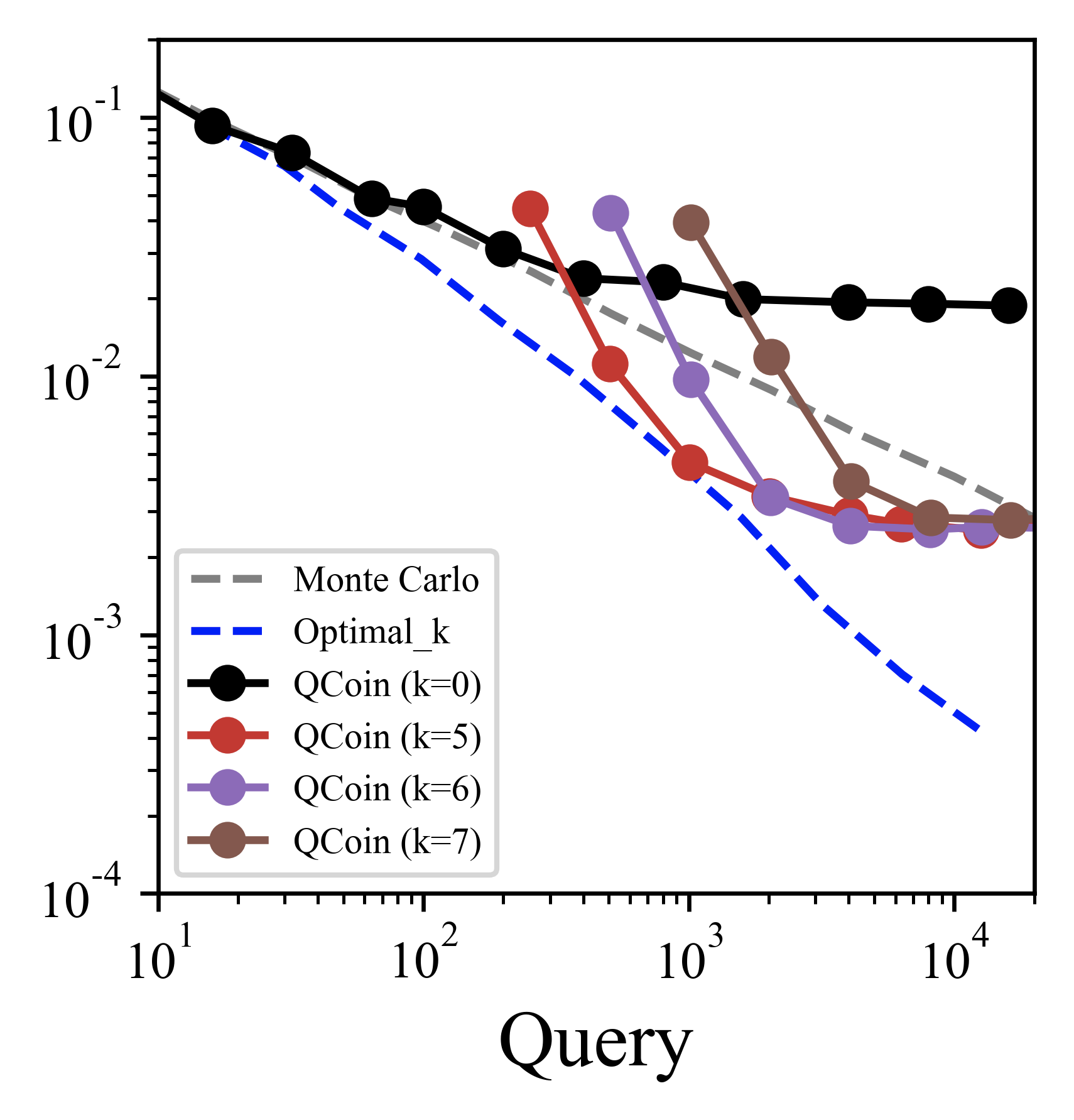}
  \mbox{}
  \caption{\label{QCoin-real}%
    Results of QCoin with $f=0.50$ on an actual quantum computer. We plot $k=0$ on the left, additionally $k=2,3,4,5$ in the center, and $k=5,6,7$ on the right. Monte Carlo plot data points are calculated with 3000 samples, and Optimal\_$k$ data is plotted with the data points calculated by 500 samples on simulator.}
\end{figure*}

We use IBM Q5 Yorktown~\cite{IBMQ} and Qiskit~\cite{Qiskit} to run QSS and QCoin on an actual quantum computer. The hardware resources are very limited, hence we simplified settings for both QSS and QCoin. \green{Our source code is available on Github~\cite{Github}.}

\paragraph*{QSS.}
To implement QSS, we removed the circut for input qubits and assumed that the oracle $\hat Q_{F}$ operates as:
\begin{eqnarray}
\hat Q_{F} \ket0 = \sqrt{1-f}\ket0 + \sqrt{f}\ket1. \nonumber{}
\end{eqnarray}
The AA operation $\hat G$ is expressed as
\begin{eqnarray}
\hat G &=& \hat Q_{F} (2\ket0 \bra0 - \hat 1) \hat Q^{-1}_{F} \hat R_f \\
&=& \hat Q_{F} \hat Z \hat Q^{-1}_{F} \hat Z. \nonumber{}
\end{eqnarray}
In one qubit case, the $R_f$ flip gate counterparts to a Pauli $Z$ gate, and the $(2\ket0 \bra0 - \hat 1)$ flip operation also does to $\hat Z$ gate. Despite its very simple implementation, we can create at most 3 qubits circuit (expressed in Figure~\ref{SimpleCir}) on the IBM Q5 quantum computer due to its requirement for the architecture of qubits as discussed in the previous section.

\paragraph*{QCoin.}
In QCoin, we also simplified the quantum circuit by eliminating the circuit for input qubits and set the oracle as:
\begin{eqnarray}
\hat Q_{F,E} \ket0 = \sqrt{1-f^2}\ket0 + f\ket1. \nonumber{}
\end{eqnarray}
In this case, the oracle can be regarded as the one including the process of making a quantum coin. AA operation $\hat G$ is almost the same as QSS:
\begin{eqnarray}
\hat G &=& \hat Q_{F,E} (2\ket0 \bra0 - \hat 1) \hat Q^{-1}_{F,E} \hat R_f \\
&=& \hat Q_{F,E} \hat Z \hat Q^{-1}_{F,E} \hat Z. \nonumber{}
\end{eqnarray}
The quantum circuit is shown in Figure~\ref{SimpleCir}.

%-------------------------------------------------------------------------
\subsection{Results}

\paragraph*{QCoin.}
Figure~\ref{QCoin-real} shows the error performance of QCoin on the quantum computer with $f=0.50$. All the data points are calculated by 300 simulations. Figure~\ref{QCoin-real} (left) shows the results of Monte Carlo integration and QCoin of $k=0$ (equivalent to Monte Carlo integration) 
%\cTH{I think Monte Carlo integration corresponds to $k=0$ (no AA operation at all). Why $k=1$?} \cNS{I will change $k=0$ and correct all graphs and sentences including $k$.}
on the quantum computer. The convergence rate is almost the same at small number of queries ($\lesssim 100$), while the reduction of error stops in the range of more query. This error seems to mainly arise from the readout error of qubits. Hence we cannot improve this error by more trials. 
%\cTH{I don't understand the following sentences. I think what exactly is "a scale-up" is never explained.} \deleted{However, we are able to overcome this limitation by using a scale-up quantum coin. The error which occurs in measurement of enlarged quantum coin is reduced in the process of calculation back to estimate original $f$.}
However, we are able to overcome this limitation by using the AA steps of quantum coin. We scale-up the bounded error and measure the enlarged quantum coin in each step as shown in Algorithm 1. We only need to estimate $f$ value roughly for each step, hence the influence of readout error becomes relatively smaller than Monte Carlo method.
Figure~\ref{QCoin-real} (center) shows $k=2,3,4,5$ cases of QCoin. We confirm that error performances are better than Monte Carlo and compatible to QCoin's on simulator even on real quantum computer in the range of rather small number of queries. For larger queries, the convergences of error reduction are seen, which seem to be also due to readout error. 
%\cTH{I don't understand why 16 AA corresponds to $k=5$. Isn't $k$ the number of AA operations done for the whole process?} \cNS{As shown in Algorithm 1, the bounded error is $\frac{\pi}{2} \cdot \frac{1}{2^5}$ in the case of $k=5$. The number of AA is $2^5 \cdot \frac{1}{2} = 16$ to change $[0,\frac{1}{2^5}]$ to $[0,1]$.}
%
Figure~\ref{QCoin-real} (right) additionally shows the plots of $k=6,7$ cases. The error does not reduce with $k=6,7$ compared to $k=5$, though the scale-up steps of a quantum coin increases. This means that the scaling-up process over almost 16 AA iterations (corresponds to the $k=5$ case) becomes meaningless due to accumulation of decoherence and gate error in large circuit calculation.
%U3(θ~0.01)レベルの精度で回転できない??
%回路深すぎてdecoherence、 AAstepがmeaninglessになっている

\paragraph*{Supersampling.}
\red{We also conducted the experiment of supersampling for QCoin on an actual quantum computer (the right image in Figure~\ref{fig:teaser}). Note that since we cannot prepare the oracle gates which convert all sub-pixels value into quantum states due to hardware limitaion, we now set and use the oracle gate which directly have a target value as Equation~\ref{eqn:mean}. For gray-colored regions (where the pixel color is 0.25, 0.50, or 0.75), QCoin on IBMQ produces similar results as QCoin on a simulator, which is consistent with the results in Figure~\ref{QCoin-real}. On the other hand, in the black and white regions, QCoin on IBMQ shows a rather large estimation error than the simulation. These regions are sensitive to the noise of an actual quantum computer since the integrand in those regions should be either strictly zero or one, which can be easily corrupted by the noise. This disadvantage in the limited narrow region doesn't decrease overall performance of QCoin's algorithm so much, as we can see the mean absolute error of the gradation part is not almost reduced compared with QCoin on simulator.
}

\begin{figure*}[t]
  \centering
  %\mbox{}
  \includegraphics[width=0.19\linewidth]{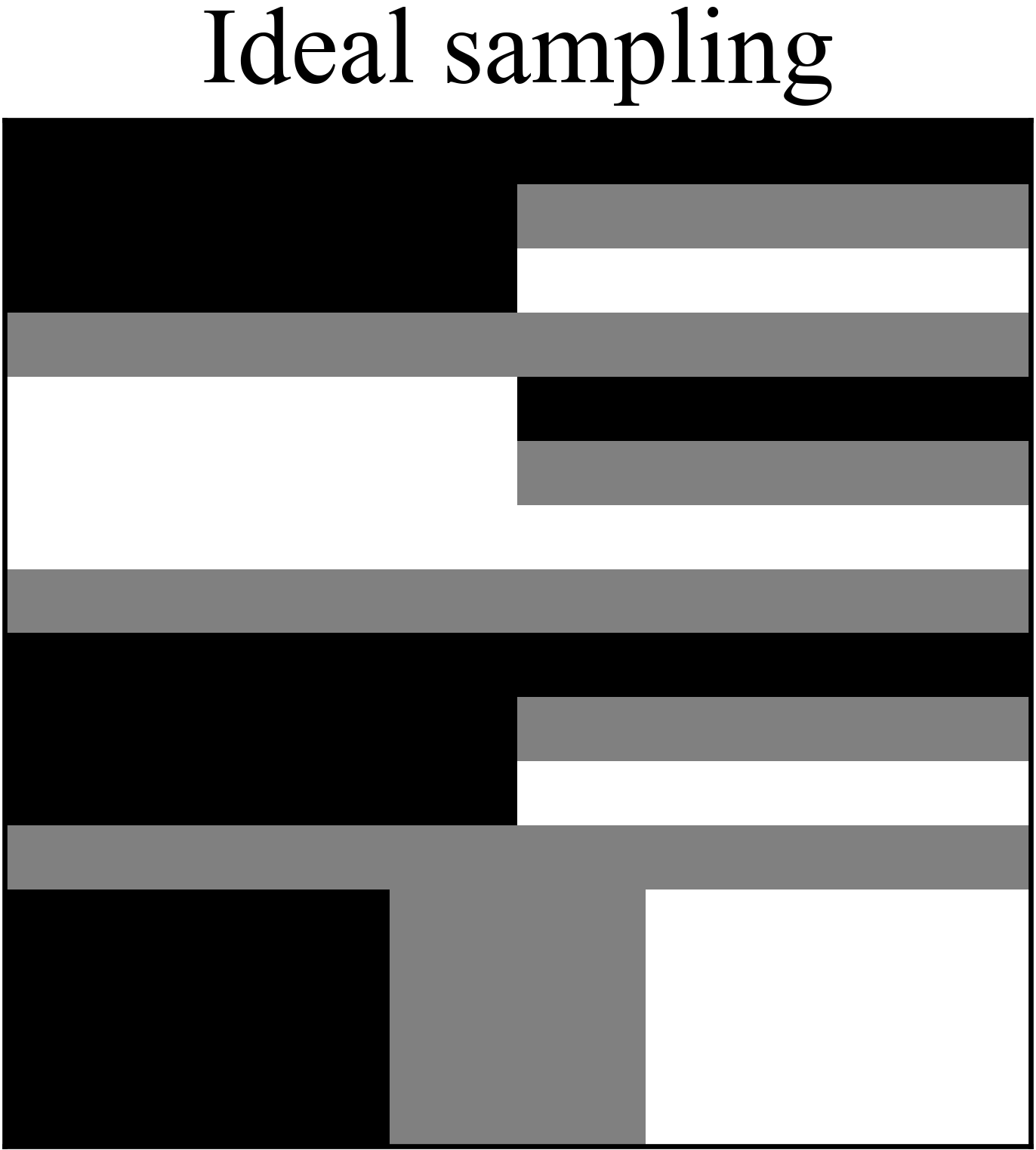}
  \includegraphics[width=0.19\linewidth]{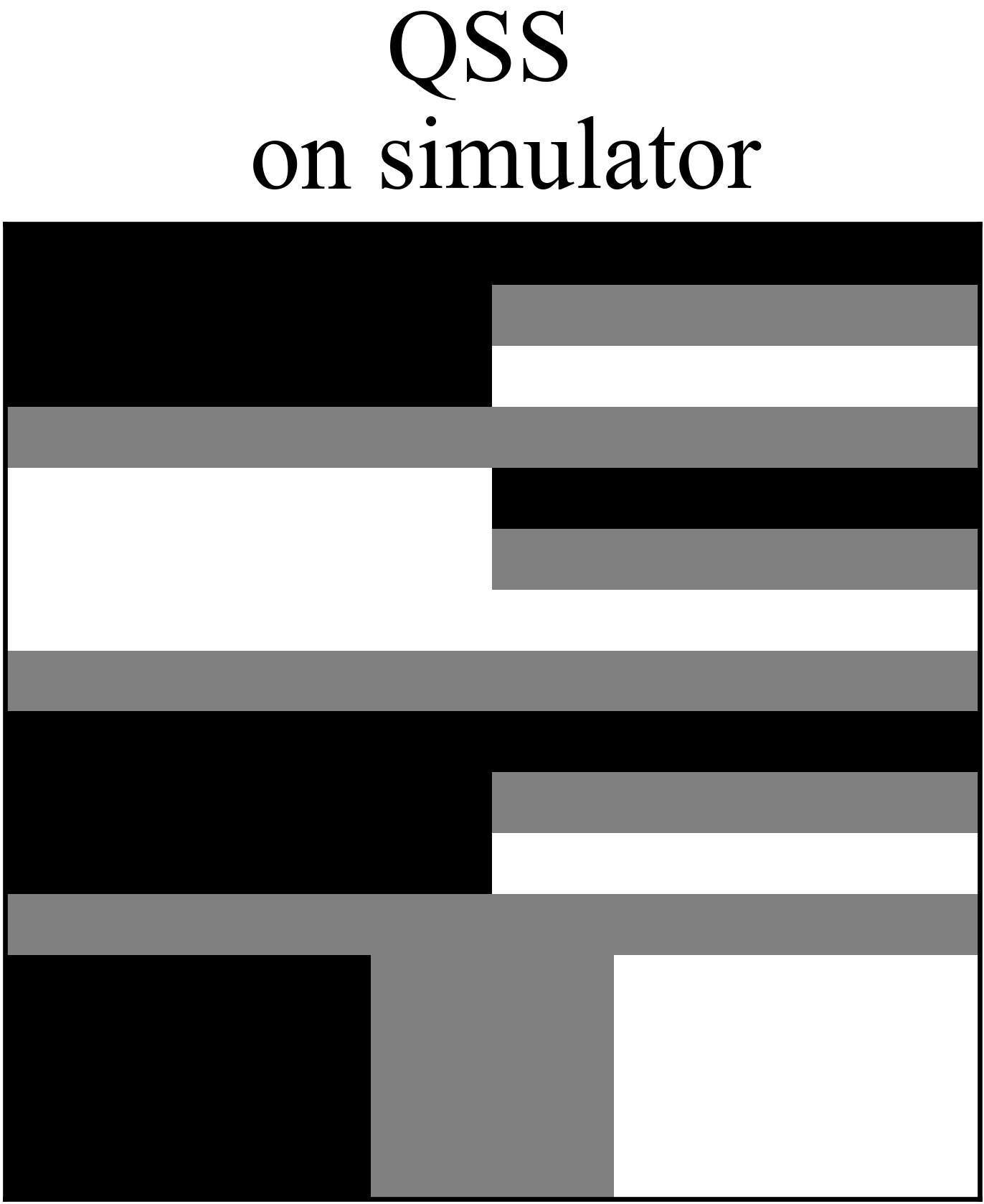}
  \includegraphics[width=0.19\linewidth]{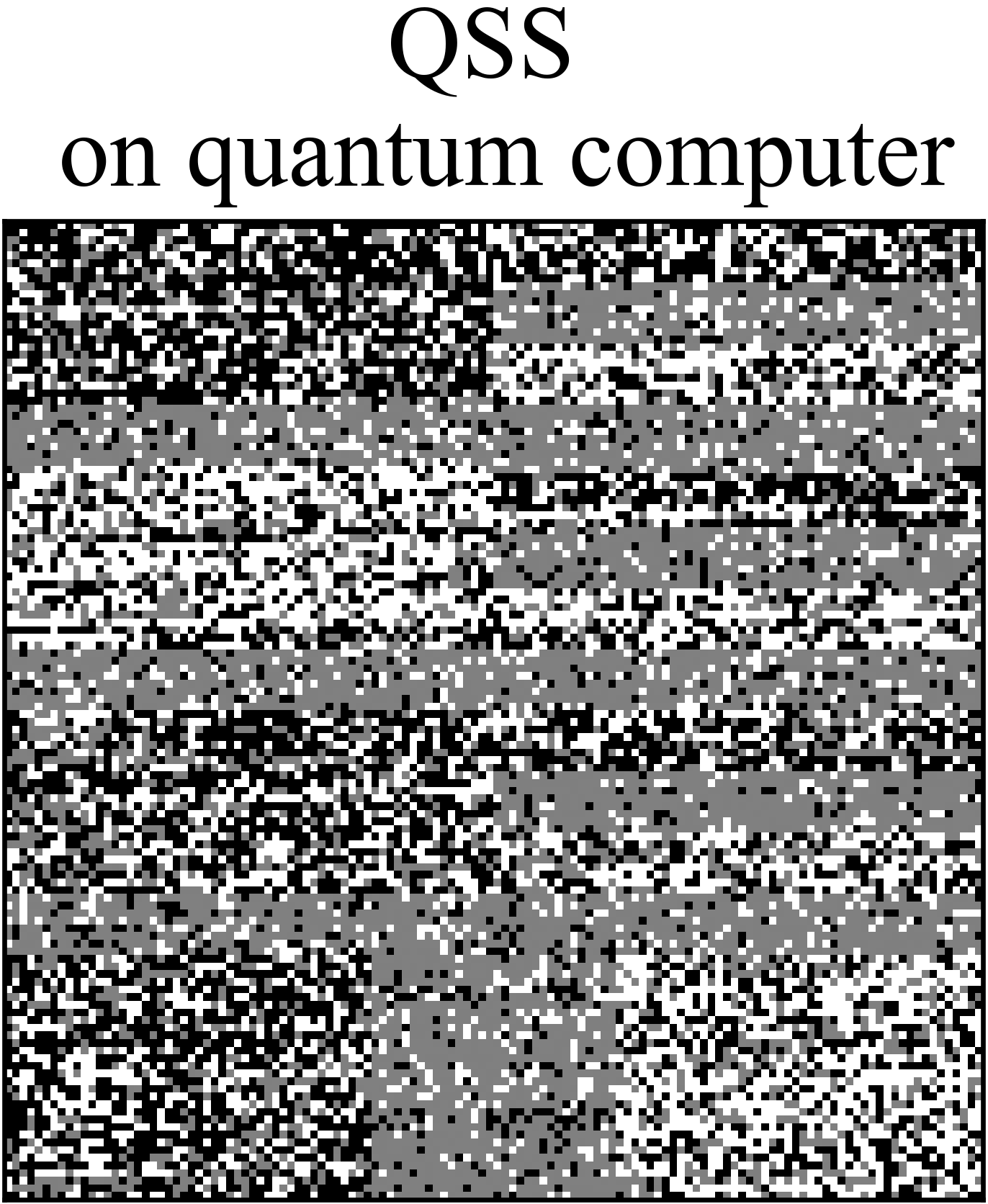}
  \includegraphics[width=0.19\linewidth]{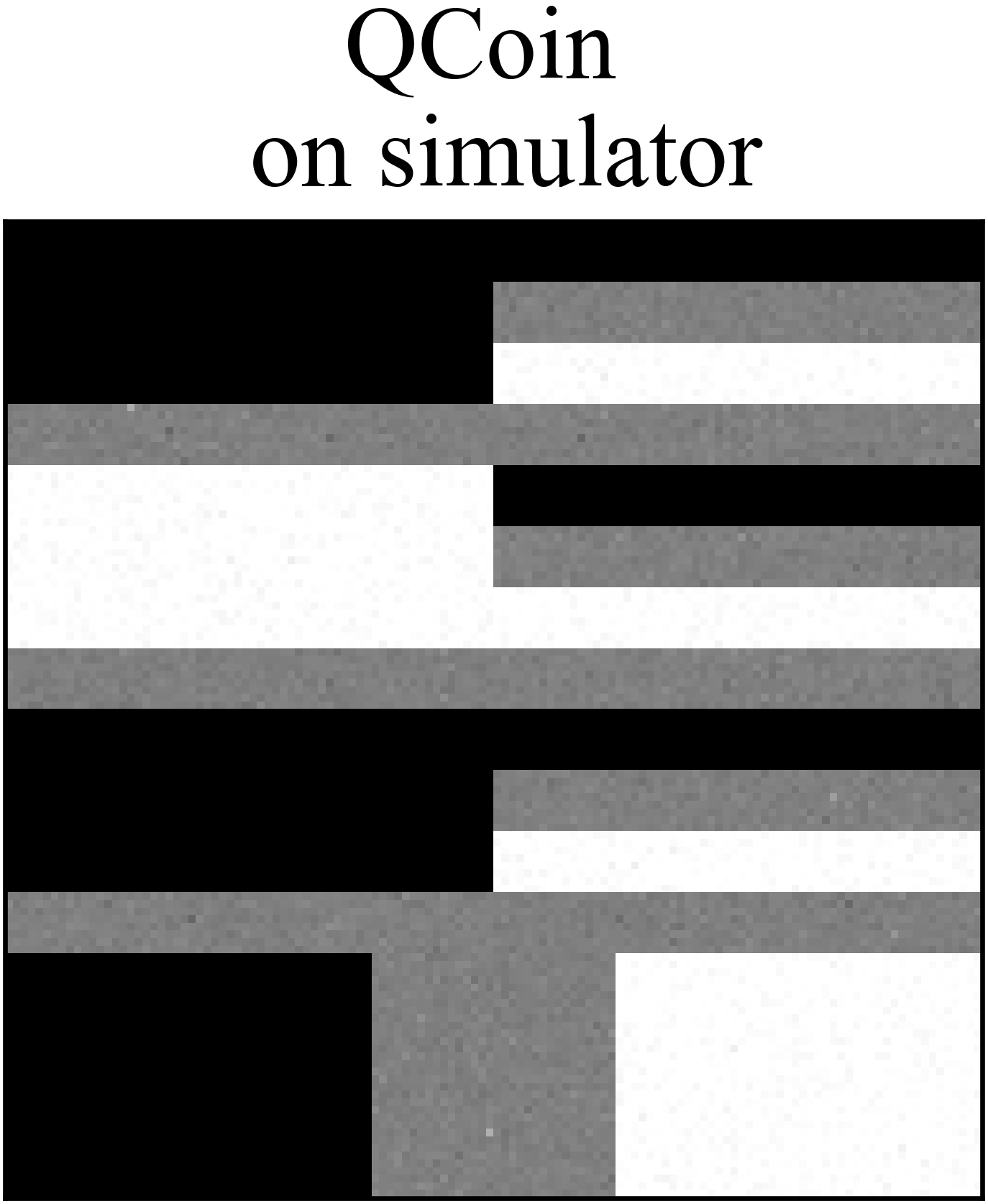}
  \includegraphics[width=0.19\linewidth]{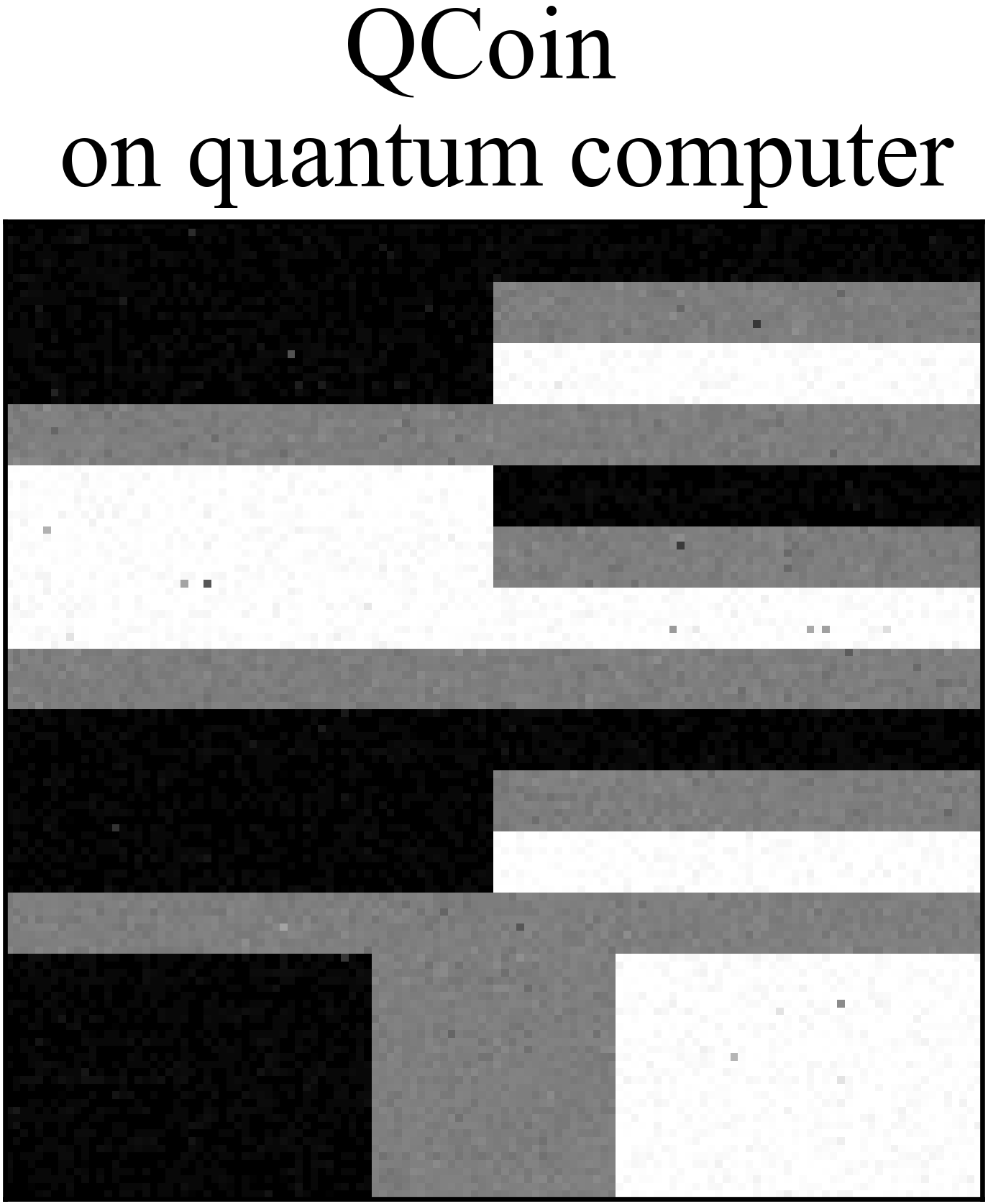}
  %\mbox{}
  %\vspace{-1EM}
  \caption{\label{QSS-real}%
    \red{Supersampling with QSS on an actual quantum computer using 7 queries (3rd form the left). Ideal sampling image is shown in 1st column, and QSS on simulator produces no error result as a 2nd image in this case of the limited pixel colors. QCoin's results are also shown in the right, and the settings of experiments are the same as Figure~\ref{fig:teaser}.}}
    \vspace{-2EM}
\end{figure*}

\red{We also show another supersampling image via QSS in Figure~\ref{QSS-real}. We cannot conduct QSS's algorithm with such large number of queries as in Figure~\ref{fig:teaser} because of the hardware limitation of IBMQ. Then we now treat the case where the target pixel color is limited to 0.0, 0.5, and 1.0 (seeing "Ideal sampling" in the Figure), and the number of queries for QSS is limited to seven. We, however, note that seven queries are enough to estimate the specific pixel colors of this case with no error on a simulator, which is confirmed by seeing QSS's result in Figure~\ref{fig:ex1}. As was also observed by Johnston~\shortcite{QSS}, the result of QSS on IBMQ is significantly influenced by the noise in an actual quantum computer, and the mean absolute error increases $0.30$.
}
The dominant error in QSS on an actual quantum computer seems to be a multi-qubit gate error. Multi-qubit operations are more difficult to execute than single-qubit ones because they must additionally operate controlled functions. All the gate errors are publicly disclosed~\cite{IBMQ}, and the multi-qubit gate errors are almost $5$\%. Therefore, even if we operate it for only a few times, accumulation error quickly reaches to the visible level and thus appears as noise in the image. Thanks to its simpler circuit design, QCoin does not suffer from this issue. One can thus see a striking difference between QSS and QCoin when we run both on an actual quantum computer.

%%\newpage

\section{Discussions}

\paragraph*{Ray tracing oracle gate.}
One might argue that our numerical experiments (inspired by Johnston~\shortcite{QSS}) are too simple compared to the actual use cases of ray tracing, thus it is not really demonstrating the applicability of QCoin in rendering. While we admit that it is not as complex as the actual use cases, the basic idea of QCoin is readily applicable to arbitrary complex integrands just like Monte Carlo integration. The remaining challenge is to design an oracle gate which efficiently performs ray tracing. While it is theoretically possible to design such an oracle gate, we found it infeasible to perform any numerical experiment (\textit{even on a simulator}) at this moment for two major reasons.

First, the maximum number of qubits we can have at the moment is about $50$ qubits~\cite{50qubits}. It is incorrect to assume that we can get away from this limitation on a simulator. Simulation of 50~qubits already takes roughly 16~peta byte of RAM if we store the full quantum states, and a 64 qubits simulator (on a cluster of 128 nodes) is possible only by limiting the complexity of quantum circuits~\cite{chen201864}. Given that even \textit{one} floating point number consumes 32~bits, we concluded that it is currently infeasible to conduct any numerical experiment (both on actual and simulated quantum computers). Note that computation such as square root of floating point numbers adds up to the required number of qubits. While we should be able  to theoretically design such a quantum circuit, if we were to perform numerical experiments, even for \textit{simple} cases like ray tracing of a sphere, we need either a better hardware or a simulator, which are both out of the scope in this paper.

Second, there is currently no research done on how to \textit{appropriately} represent typical data for ray tracing. For example, due to the limited number of input qubits, it will not be immediately possible to handle triangle meshes on a quantum computer. While Lanzagorta and Uhlmann~\cite{QuantumRendering} mentioned a theoretical possibility of using Grover's method for ray tracing, implementation of this idea for any practical scene configurations is currently impossible. We thus believe that further research on a suitable data representation for rendering on quantum computers is deemed necessary, and this topic alone can lead to a series of many research questions and thus cannot be a short addendum in our paper. We focused on a numerical integration algorithm which serves as a basic building block for ray tracing on quantum computers. Our work should be useful as a stepping stone to conduct further research along this line.

\paragraph*{Error distribution over the image}
A unique requirement of solving many integrals on the image place in rendering highlights another important difference between QSS and QCoin. In QCoin, the distribution of errors over the image is essentially noise due to random sampling, which is the same as Monte Carlo integration. Being a hybrid quantum-classical method as we explain later, one can easily apply many exiting tools developed for Monte Carlo integration, such as denoising via image-space filtering~\cite{zwicker2015recent}, to QCoin. QCoin can thus directly replace Monte Carlo integration while being asymptotically faster. In QSS, however, erroneous pixels appear as completely wrong pixel values~\shortcite{QSS} which cannot be easily recovered or identified by the existing tools for Monte Carlo integration. For example, denoising for Monte Carlo rendering would not work as-is for rendered images via QSS. We thus believe that QCoin is more readily applicable to rendering than QSS.

\begin{figure}[t]
  \centering
  \mbox{}
  \includegraphics[height=0.47\linewidth]{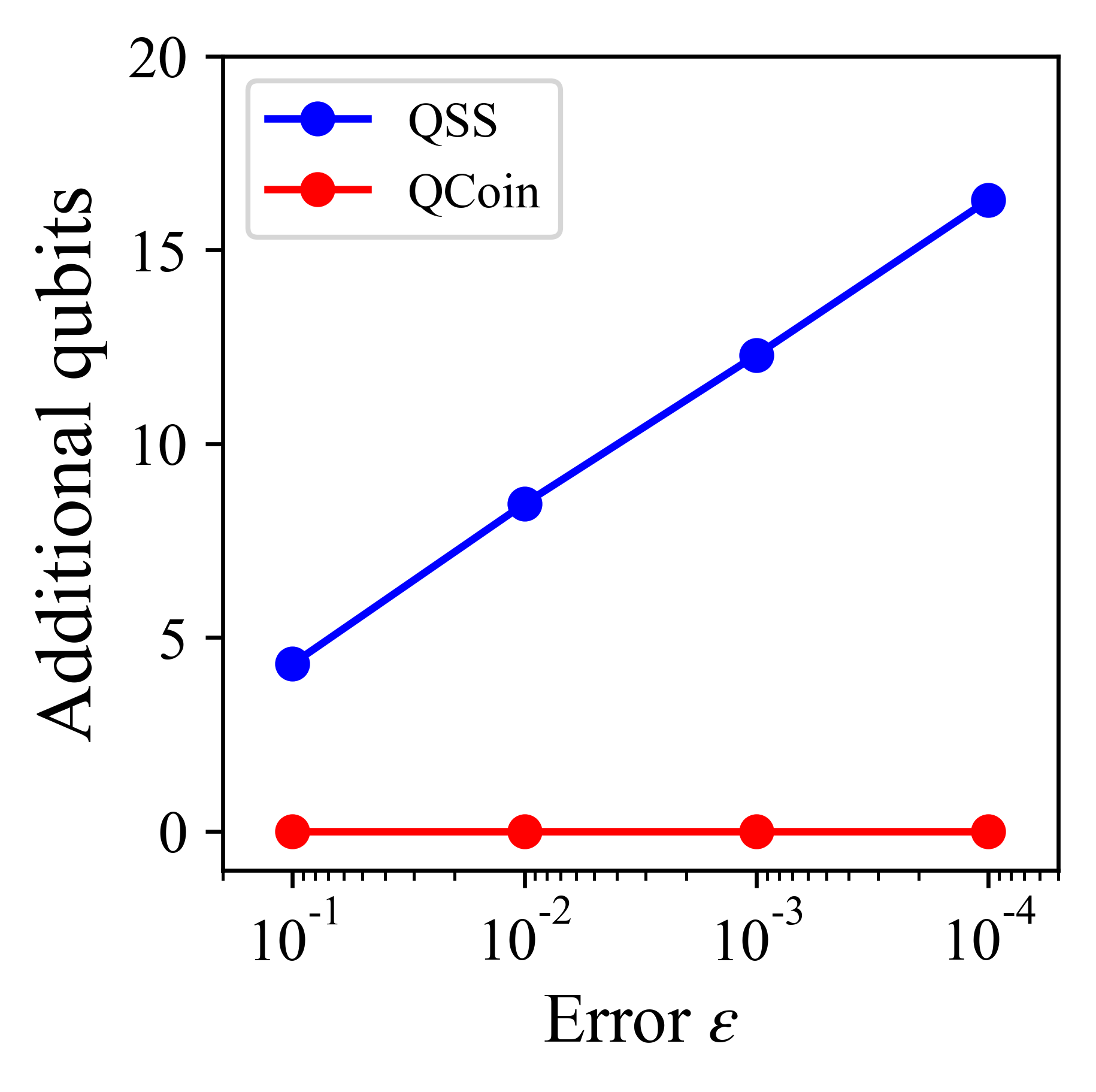}
  \includegraphics[height=0.47\linewidth]{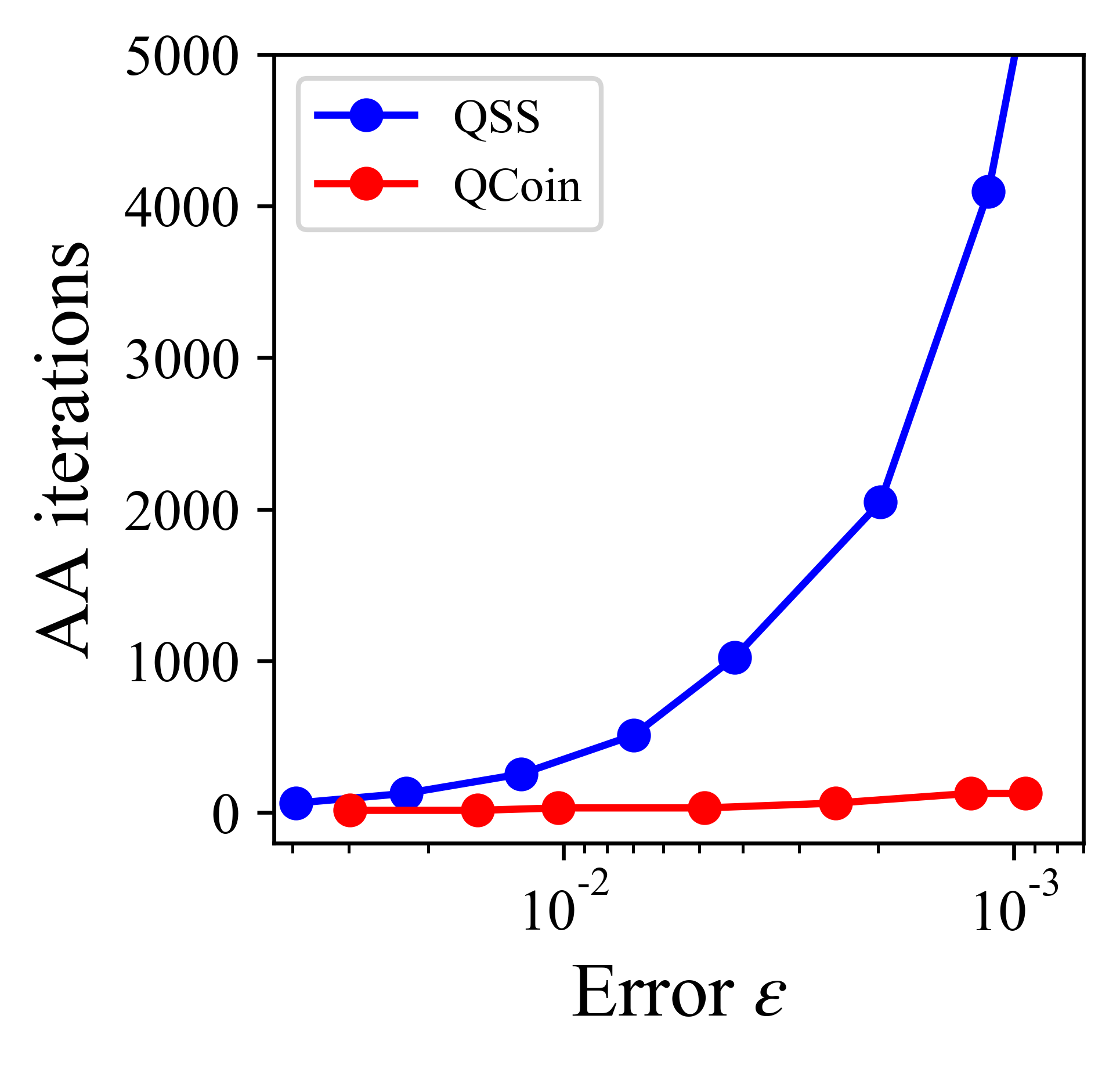}
  \mbox{}
  \caption{\label{Dis} Relation plots between the number of additional register qubits (AA iterations) and error $\epsilon$ in left (right).}
  \vspace{-1EM}
\end{figure}

\paragraph*{Required number of qubits.}
Quantum computers at this moment have a limited number of usable qubits. As mentioned above, the maximum number of qubits we can have is $\sim50$ qubits~\cite{50qubits} at the moment. Let us consider that the size of input is $N$ and the maximum AA iterations is $P$. In this case, QSS needs $\log N + \log P$ qubits to run the algorithm, while QCoin needs only $\log N$ (left in Figure \ref{Dis}). The examples of quantum circuits for QSS and QCoin also verify this fact (Figure \ref{Qcircuit-QSS} and Figure \ref{Qcircuit-QCoin}). For example, when the input $N=2^{10}=1024$ is given, using the current architecture of quantum computers, the number of AA iterations in QSS is limited to less than only $P=2^5=32$ times, whereas QCoin can has no such limitation by construction. This severely limits the applicability of QSS, making QCoin an attractive alternative in practice.

\begin{figure}[t]
  \centering
  \mbox{}
  \includegraphics[width=0.49\linewidth]{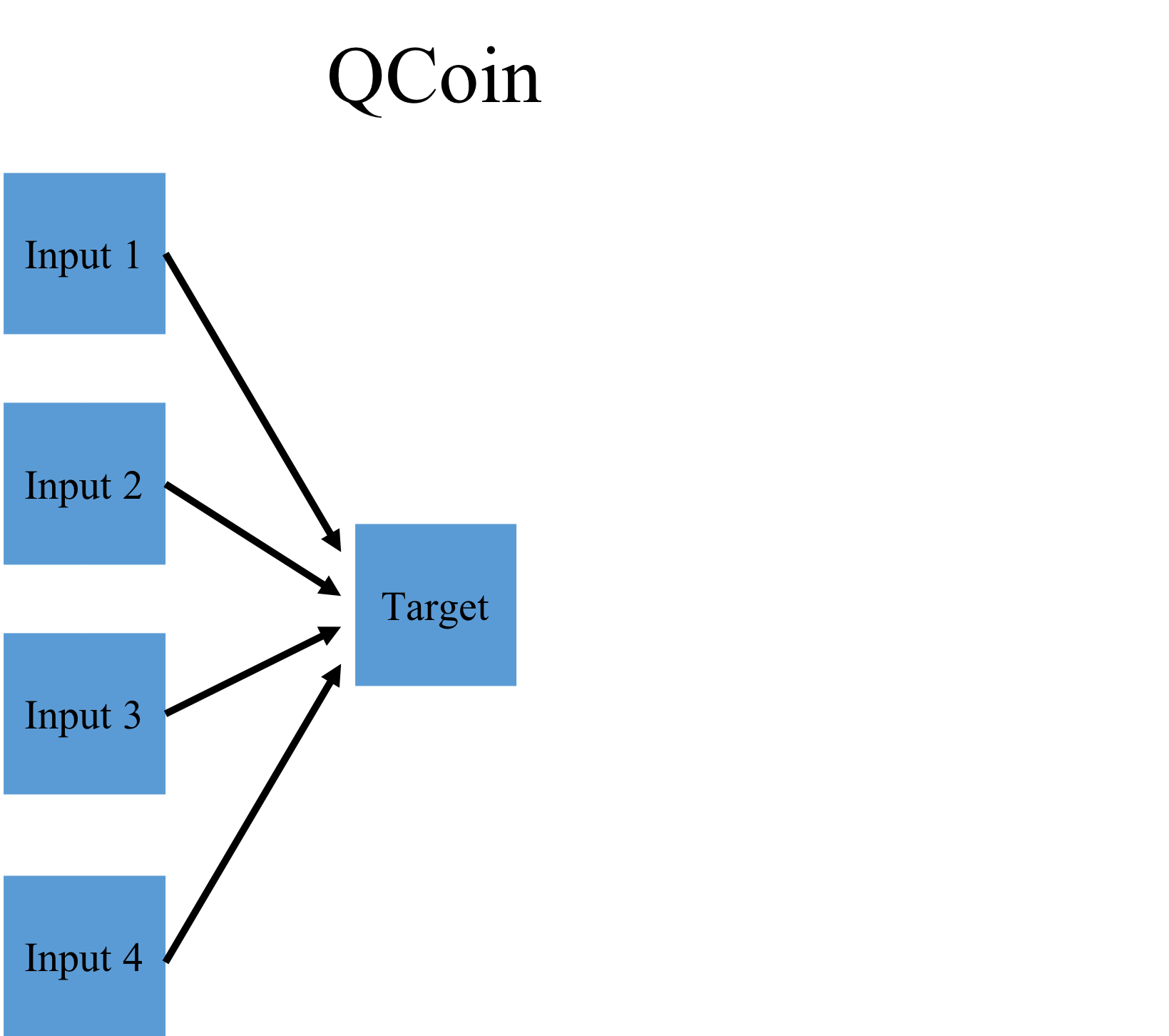}
  \includegraphics[width=0.49\linewidth]{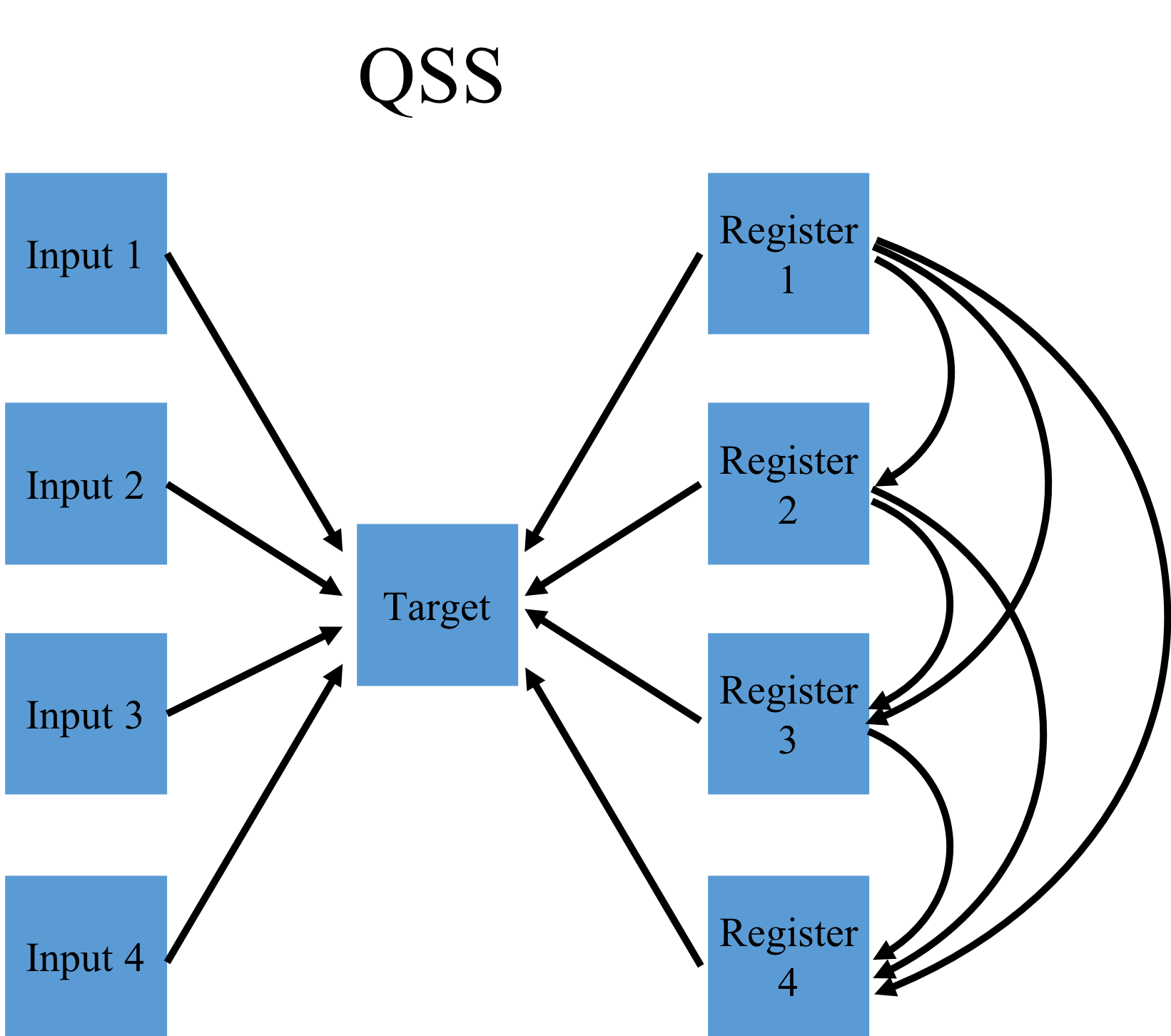}
  \mbox{}
  \caption{\label{Qarch} Minimum qubits' architectures for the quantum circuits in Figure 6 (QCoin) and Figure 5 (QSS). The arrow represents a connection with controlled gate; a root qubit at the arrow is a control qubit, and an end-qubit is a target.}
\end{figure}

\paragraph*{Connections among qubits.}
Another well-known limitation of quantum computer architecture is the number of connections among qubits. In many quantum algorithms, controlled-gates are important. However, it is currently difficult to prepare all interacting the qubits. For example, IBM Q20 Tokyo~\cite{IBMQ} has a total of 20~qubits, while the size of fully-connected qubits set is only up to 4.
%, this quantum computer can handle five sequences of 4~qubits at most, not a single 20~qubits sequence.

Due to the use of QFT, QSS needs a sequence of qubits which has a connection between target and the other qubits and full connections among the register (used for computation) qubits. On the other hand, QCoin needs connections only between target and input qubits. Figure~\ref{Qarch} shows the minimum qubits' architectures (connections) for the quantum circuits in Figure~\ref{Qcircuit-QSS} and Figure~\ref{Qcircuit-QCoin}. In general, we need not only more qubits, but also more connections among qubits for QSS than QCoin, which further prevents the use of QSS in real quantum computers.

\paragraph*{Quantum error correction and NISQs.}
Aside from errors due to the algorithm (i.e., noise due to a limited number of samples or iterations), the error in quantum computation arises from various factors; bit-flip errors, decoherence of superposition states, errors on logic gates, and so on. Note that simulators currently do not include such errors. While it is theoretically possible to perform error corrections~\cite{knill2000theory}, its implementation on current quantum computers is still considered challenging~\cite{reed2012realization}. It is thus generally assumed that one cannot perform calculation accurately as the scale of a quantum circuit (computation time, the number of qubits, and the number of gate operations) is becoming larger.

Such an "unscalable" quantum computer is called an ``NISQ" (Noisy Intermediate-Scale Quantum Computer)~\cite{NISQ}. On NISQs, quantum algorithms which require many qubits and a long computation time will not work due to the errors in actual quantum computers. However, NISQs are more realistic models for actual quantum computers in the near future. Therefore, researchers have been vigorously investigating a novel class of quantum algorithm called ``hybrid quantum-classical"~\cite{VQE}. In this class of algorithms, an algorithm alternately repeats quantum calculations in a small circuit and adjusting parameters of the quantum circuit based on the classical calculation. 
A hybrid quantum-classical algorithm generally needs fewer qubits and lower depth of quantum circuit, thus suitable to run on NISQs.
%. using the values observed from the quantum circuit to get the correct answer. 

In QSS, the relations of AA iterations and estimation error appear as in Figure~\ref{Dis} (right). It is obvious that much more time and larger circuit for one continuous quantum operation are required for QSS than for QCoin. Therefore, we infer that QCoin performs better than QSS due to its shorter computation time. In our experiments and the original experiments by Johnston~\shortcite{QSS}, QSS is not really performing well on NISQs. Based on its \hl{performance} on a simulator, we hypothesize that its inferior performance on an actual quantum computer is not owing to the limited number of AA iterations, but its use of many \hl{qubits} and controlled-gate operations. We, however, should mention that we could not run a large enough quantum circuit for QSS due to the restriction of the qubit architecture to fully confirm the influence of decoherence alone. 

On the other hand, our QCoin method performs well even on an actual quantum computer. We think that it is because QCoin is hybrid quantum-classical; the use of a quantum coin is done as in classic Monte Carlo integration, while shifting and scaling of the error interval are done by AA in quantum computation. Hybrid quantum-classical algorithms generally need fewer qubits and lower depth of quantum circuit, which is considered suitable to run on NISQs. While the idea of QCoin was invented a while ago~\cite{QCoin}, there has been no effort to investigate whether QCoin is executable on NISQs, and we think that this finding alone is novel in the field of quantum computing. 
%(Thus, we demonstrate the difference of the behaviors on a real quantum computer between QSS and QCoin to confirm the superiority of the QCoin method as the hybrid quantum-classical algorithm.)

%------------------------------- Limitations ----------------------------------------
\section{Limitations}
Aside from the limited complexity of integrands due to the current architecture of quantum computers, we have a few more limitations. In classic computers, by giving up the use of random numbers, it is possible to perform quasi Monte Carlo integration~\cite{morokoff1995quasi} to achieve the convergence rate of $O(\log(N)^s/N)$ for $s$-dimension integrands. While the QCoin's convergence rate of $O(1/N)$ is still better, it is unclear if and how we can incorporate quasi Monte Carlo to achieve an even better convergence rate in quantum computation, or whether it is possible. The classical part of QCoin is still limited to Monte Carlo integration.
Moreover, due to the constraints of hardware, our experiments for both QSS and QCoin on an actual quantum computer omitted the input circuit part, which generally involves controlled-gate operations. As such, if we could have included the omitted input part, errors in our experiments might potentially go up due to the use of more controlled-gate operations.

%------------------------------- Conclusion ------------------------------------------
\section{Conclusion}
We proposed a concrete algorithm of QCoin and performed numerical experiments for the first time after 20~years of its theoretical introduction~\cite{QCoin}. Our implementation of QCoin shows a faster convergence rate than that of classical Monte Carlo integration. This performance is equivalent to QSS. We formulated QCoin as a hybrid quantum-classical method and explained why QCoin is more stable than QSS in the presence of noise in actual quantum computers. We discussed hardware limitations of near-term quantum computers and concluded that QCoin needs fewer qubits and simpler architecture, thus being much more practical than QSS. Our experiments on a quantum computer confirmed this robustness against noise and faster convergence rate than classical Monte Carlo integration. We believe that QCoin is a practical alternative to QSS if we were to run rendering algorithms on quantum computers in the future.

%-------------------------------- References ----------------------------------------
% bibtex
\bibliographystyle{eg-alpha-doi} 
\bibliography{Reference}       
%-------------------------------------------------------------------------

\end{document}